\pdfoutput=1

\documentclass{article}
\usepackage{arxiv}

\usepackage{lipsum}
\usepackage{lineno}
\usepackage{amsmath}
\usepackage{amsfonts}
\usepackage{amsthm}
\usepackage{amsmath,amssymb}
\usepackage{bm}
\newcommand\norm[1]{\left\lVert#1\right\rVert}
\DeclareMathAlphabet\mathbfcal{OMS}{cmsy}{b}{n}
\DeclareMathOperator*{\argmin}{arg\,min}
\usepackage{algpseudocode}
\usepackage[ruled, vlined, linesnumbered]{algorithm2e}
\algnewcommand{\algorithmicgoto}{\textbf{go to}}%
\algnewcommand{\Goto}[1]{\algorithmicgoto~\ref{#1}}%
\makeatletter
\newcommand{\removelatexerror}{\let\@latex@error\@gobble}
\makeatother
\usepackage{bm}
\usepackage{stmaryrd}
\usepackage{subcaption}
\usepackage{graphicx}
\usepackage[toc,page]{appendix}
\usepackage{multirow}
\usepackage{hyperref}
\usepackage{cleveref}
 \usepackage{float}
\newtheorem{remark}{Remark}
\DeclareMathOperator{\Tr}{Tr}
\modulolinenumbers[5]
\usepackage{xcolor}
\usepackage{setspace}

\title{An integrated numerical model for coupled poro-hydro-mechanics and fracture propagation using embedded meshes}

\author{
  Guotong Ren\\
  Department of Petroleum Engineering\\
  University of Tulsa\\
  Tulsa, OK 740104 \\
  \texttt{guotong-ren@utulsa.edu} \\
   \And
 Rami M.~Younis \\
  Department of Petroleum Engineering\\
  University of Tulsa\\
  Tulsa, OK 74104 \\
  \texttt{rami-younis@utulsa.edu} \\
}

\begin{document}
\maketitle

\begin{abstract}
Integrated models for fluid-driven fracture propagation and general multiphase flow in porous media are valuable to the study and engineering of several systems, including hydraulic fracturing, underground disposal of waste, and geohazard mitigation across such applications. This work extends the coupled model multiphase flow and poromechanical model of  \cite{ren2018embedded} to admit fracture propagation (FP). The coupled XFEM-EDFM scheme utilizes a separate fracture mesh that is embedded on a static background mesh. The onset and dynamics of fracture propagation (FP) are governed by the equivalent stress intensity factor (SIF) criterion. A domain-integral method (J integral) is applied to compute this information. An adaptive time-marching scheme is proposed to rapidly restrict and grow temporal resolution to match the underlying time-scales. The proposed model is verified with analytical solutions, and shows the capability to accurately and adaptively co-simulate fluid transport and deformation as well as the propagation of multiple fractures.
\end{abstract}

\keywords{Fracture propagation \and Coupled hydro-mechanics \and Porous media \and Extended finite element method \and Embedded discrete fracture method}

\section{Introduction}
Numerical models are an important enabling technology towards the advancement of a number of engineered systems within the nexus of energy, water, and the environment (e.g. subsurface energy resource extraction, waste disposal or storage, and geological intermittent energy storage systems). The first-order response of such systems is often driven by coupled poro-thermo-hydro-mechanics, including fracturing and multiphase flow. While these underlying processes typically occur with local spatial and temporal support within the extent of the system-scale, their long-range interactions and causal relations cannot be ignored. Subsequently, ubiquitous (across engineering subprocesses) models are necessary in the context of joint assimilation of multiple types of observations (e.g. displacement, fluid flow, and thermodynamic state) acquired over the duration of multiple operational substeps (e.g. hydraulic fracturing, injection, and production). The focus of this work is the development of an efficient, robust, and ubiquitous numerical approximation for geological systems undergoing concurrent or consecutive types of operations. In particular, the target processes are fracture, displacement, and multiphase flow and transport at the system-scale.
\begin{figure}[ht]
	\centering
	\begin{subfigure}[t]{0.21\linewidth}
		\centering
		\includegraphics[width=0.9\linewidth]{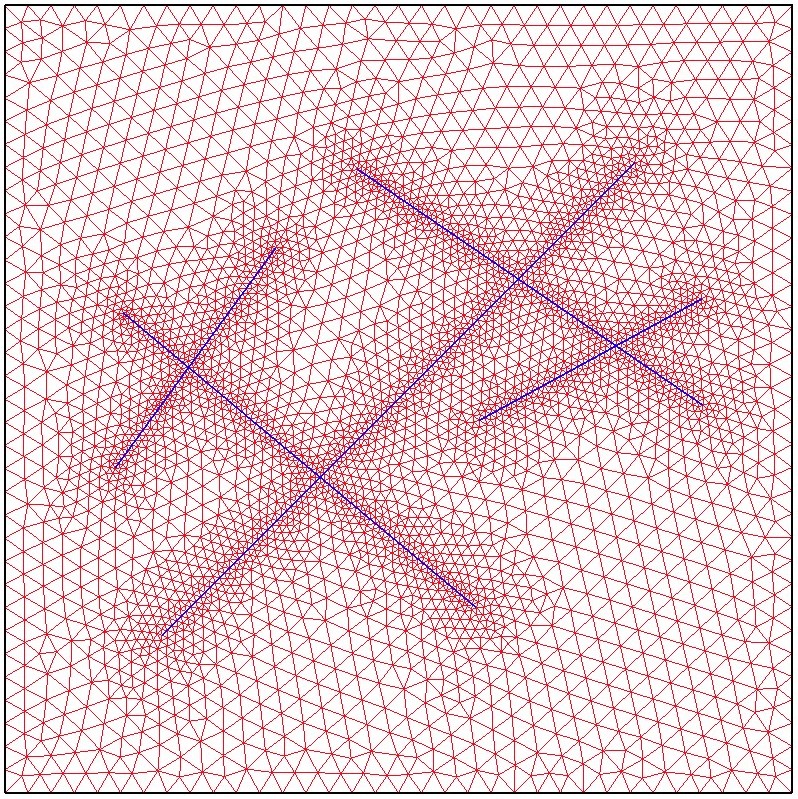}
		\caption{}
		\label{fig:udfm}
	\end{subfigure}
	\quad
	\begin{subfigure}[t]{0.21\textwidth}
		\centering
		\includegraphics[width=0.9\linewidth]{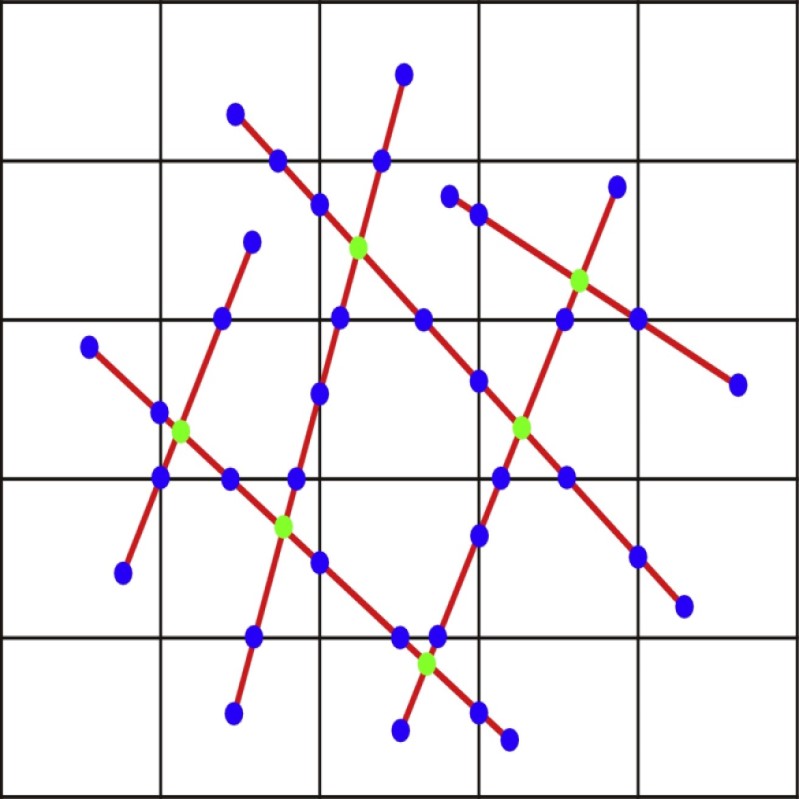}
		\caption{}
		\label{fig:edfm}
	\end{subfigure}
	\begin{subfigure}[t]{0.4\textwidth}
		\centering
		\includegraphics[width=0.9\textwidth]{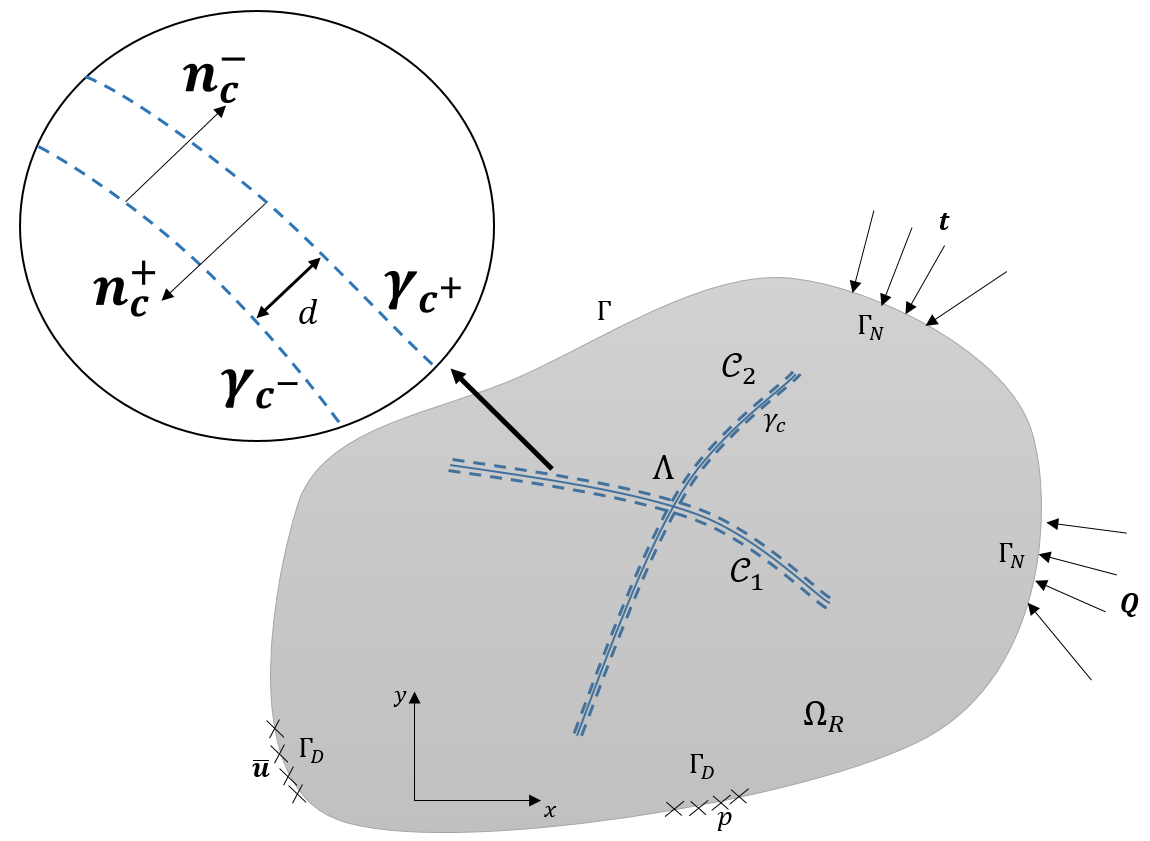}
		\caption{}
		\label{domainpic}
	\end{subfigure}
	\caption[Discrete fracture representations]{Examples of discrete fracture representations:~(\subref{fig:udfm}) a fitted triangular mesh;~(\subref{fig:edfm}) an embedded Cartesian mesh; and~(\subref{domainpic}) two intersecting fractures (dashed lines) represented using reduced-dimension immersed boundaries (solid lines).}
	\label{fig:discretization}
\end{figure}

{\color{black} An important branch of hydromechanical models with fracture seeks to homogenize fracture within the context of a continuum model. In terms of mechanics, continuum damage models (CDM) (e.g. continuum damage mechanics~\cite{kachanov1999rupture} and crack band theory~\cite{bavzant1983crack}) can describe the extent of rock failure within a representative element of the continuum by a damage tensor that depends on the underlying stress field. As a continuum model, there is naturally the requirement for sufficient scale-separation allowing such homogenization. In terms of hydrodynamics, continuum flow models in fractured rock are also well-studied. Models such as dual-porosity and dual-permeability (e.g.~\cite{warren1963behavior}) or multiple interacting continua MINC (e.g.~\cite{wu1988multiple}) homogenize sub-continuum damage to accommodate flow and transport. Continuum models abstract challenges in accommodating complex fracture intersections (e.g.~\cite{li2018recent,yun2019improved}) and can lead to computationally efficient simulation models. On the other hand, by their  very construction, such models may be inappropriate when there is no clear scale-separation in terms of damage or flow and the subscale heterogeneity is severe and perhaps non-isotropic (see for example~\cite{jiang2016hybrid} for flow). Additional considerations such as stress-locking when damage localizes into a fracture, and mesh-dependency have been reported in applications of CDM (\cite{jirasek1998analysis}). Recent work (\cite{roth2015combined,tamayo2019continuous}) proposed a CDM model in a multi-scale context whereby it is applied as a standalone propagation criterion, and is combined with a discrete fracture method (explicit representation of discontinuity). Additionally hybrid approaches that combine CDM models for diffuse and dense damage with explicit discrete models for large fractures have been proposed in the context of coupled flow and poromechanics (e.g.~\cite{ren2017fully,yan2019numerical}). One challenge ahead of applying such models in dynamic contexts is in representing the interaction between collocated CDM and explicit processes and in delineating the two.}

Discrete Fracture and Matrix methods approximate the continuity of mass and momentum directly on the fracture and porous media continua with transmission conditions across them. Within this class of approach, fractures may be represented using diffusive phase-field indicators (e.g. \cite{heister2015primal,lee2016pressure,chukwudozie2019variational}) or explicitly. In the latter approach, fracture aperture and geometry-dependent friction and stick-slip conditions may be modeled directly. In explicit approaches, fractures may be modeled as lower-dimensional, evolving, sharp-interfaces, or immersed boundaries with an implied aperture field along the reduced dimension. Within this class of model, \textit{fitted} methods (e.g. Figure \ref{fig:udfm}) utilize unstructured meshes to represent the matrix continuum (rock skeleton), and edges, or faces to represent fractures with implied aperture (e.g. \cite{carrier2012numerical, settgast2017fully, garipov2016discrete}). Using fitted representations, Fracture Propagation (FP) requires that the mesh be adapted dynamically. On the other hand, \textit{embedded} representations (e.g. Figure \ref{fig:edfm}) offer a convenience in the choice of matrix-mesh topology since it, in principle, need not be constrained to the fracture geometry. Subsequently, using embedded representations, the background mesh can remain static as the fracture propagates locally. Examples of numerical models using embedded meshes include the extended finite element method (XFEM) (e.g. \cite{melenk1996partition, moes1999finite, flemisch2018benchmarks, odsaeter2019simple}) and the embedded discrete fracture method (EDFM) (e.g. \cite{li2008efficient, jiang2017improved, ctene2017projection}). 

While embedded models have been proposed to approximate coupled multiphase flow and poromechanics (e.g. \cite{deb2017modeling, ren2018embedded,yan2019numerical}), and more recently to also include hydraulic FP (e.g. \cite{mohammadnejad2013extended, salimzadeh2015three,gupta2018coupled,gordeliy2013coupling,gordeliy2013implicit}), several challenges remain. Firstly, the structure of the resulting algebraic systems involve local and variable degrees-of-freedom that can hinder the efficiency of preconditioned iterative solution methods (e.g. \cite{ren2018embedded}). This aspect is universal to both monolithic and partitioned discretizations, and nonlinear and linear solution methods. A second challenge concerns coupling models between multiphase hydromechanics in the matrix and fractures to the criteria for failure under FP, particularly involving branching or intersection. Finally, adaptive methods must be developed to accommodate transitions into, and out of periods during which fractures are to propagate. This work develops a mixed XFEM-EDFM embedded model with FP, with a focus on addressing i) solution efficiency and ii) temporal adaptivity to accommodate the onset of propagation at arbitrary times during the simulation process. {\color{black} In this work, we consider that fractures are open to flow as in the conditions believe to prevail during a hydraulic-fracturing operation. That is, we neglect the enforcement of contact conditions as well as associated stick and frictional slip contact models. 
}

In Section~\ref{PD}, the initial boundary value problem is formulated along with the fracture propagation constraints. In Sections \ref{ND} and \ref{FPSOLVER}, the mixed discretization scheme and solution methods are developed. Section \ref{NE} presents several numerical results that verify correctness and efficiency, accuracy using reference problems, empirical consistency, as well as computational examples of consecutive and concurrent propagation and hydromechanics.

\section{Problem Formulation}\label{PD}

\subsection{Preliminaries}
As illustrated in Figure~\ref{domainpic}, we consider a spatial domain $\Omega\subset \mathbb{R}^2$ with external boundary $\Gamma$ and its associated outward-oriented unit-normal $\bm{n}_{\Gamma}$. Dirichlet and Neumann boundaries for fluid flow are denoted as $\Gamma_{p}$ and $\Gamma_{Q}$ respectively, while $\Gamma_{u}$ and $\Gamma_{t}$ are the counterparts for mechanics. The boundary segments are disjoint ($\Gamma_{p}\bigcap  \Gamma_{Q} =  \Gamma_{u} \bigcap \Gamma_{t} = \varnothing$), and $\Gamma_{p} \cup \Gamma_{Q} = \Gamma_{u} \cup \Gamma_{t} = \Gamma$ . 

We consider a collection of fractures $\Omega_F = \{ \mathcal{C}_c \subset \Omega$, $c=1,\ldots,N_F \}$, each of which is parameterized over a real open interval $I=\left(a_1, a_2\right)$ by a sufficiently smooth mapping, $\boldsymbol{\gamma}_c: I \times \mathbb{R}^+ \rightarrow \Omega$. Subsequently, a fracture's tips are $\mathcal{D}_c = \{ \boldsymbol{\gamma}_c \left( \eta, t \right) : \eta \in \partial I, t \geq 0 \}$, and their instantaneous velocities are $ \frac{\partial \boldsymbol{\gamma}_c}{\partial t}\left( \partial I ,t \right) $. Geometrically, the instantaneous unit-tangent at $\eta \in I$ is (\textcolor{black}{\cite{guggenheimer1977applicable}}),
{\color{black}
\begin{equation} 
\bm{t}_{c} = \frac{1}{ \| \frac{ \partial \boldsymbol{\gamma}_c } { \partial \eta} \|_2 } \frac{ \partial \boldsymbol{\gamma}_c } { \partial \eta}, 
\end{equation}}
and the oriented unit-normal to one side of the fracture is subsequently defined as,
{\color{black}
\begin{equation} \bm{n}_{c}^{+}=  \frac{1}{\|\frac{\partial\bm{t}_c}{\partial\eta} \|_2} \bigg[ \frac{\partial ^2 \boldsymbol{\gamma}_c }{\partial \eta ^2} - \left( { \frac{\partial ^2 \boldsymbol{\gamma}_c }{\partial \eta ^2} \cdot \bm{t}_c } \right) \bm{t}_c\bigg], 
\end{equation}
}
and on the other side (orientation) as,
{\color{black}
\begin{equation}
 \bm{n}_{c}^{-}= - \bm{n}_{c}^{+}.
\end{equation}}
A scalar aperture field is defined on fractures $\omega_c\left(\mathcal{C}_c\right)$ and it is assumed that $\omega_c << 1$. With this assumption, it is reasonable to define a matrix domain as $\Omega_M := \Omega \setminus \Omega_{F}$. State fields defined on matrix or fracture are distinguished by the subscripts $M$ and $F$ respectively. 

Fields defined on the matrix, $\mathcal{Z}_M$, are assumed to take two limiting values at each fracture location and across its aperture; these are denoted $ \mathcal{Z}_M^{+}$ and $\mathcal{Z}_M^{-}$ and may be distinct (resulting in a discontinuous field). The associated jump across fractures is denoted as,
{\color{black}
\begin{equation} \llbracket \mathcal{Z}_M \rrbracket = \mathcal{Z}_M^{+} - \mathcal{Z}_M^{-}. 
\end{equation}
}
The projection matrices $\bm{N}_c:= \bm{n}_{c} \otimes \bm{n}_{c}$ and $\bm{T}_c:= \bm{I} - \bm{N}_c$ are introduced such that the tangential gradient and divergence operators on fractures are defined as, {\color{black}$\nabla_{c} i := \bm{T}_c \nabla i$ and $\nabla_{c} \cdot i:= \bm{T}_c : \nabla i$ respectively.}

\subsection{Fluid Flow}
{\color{black} We consider the flow of two immiscible phases; wetting and non-wetting fluid phases denoted by subscript $\kappa \in \{ w, n \} $ respectively. We define the following field variables, supported independently on the matrix and fracture: $p_\kappa$ is the phase pressure; $S_\kappa$ the saturation; $\rho_\kappa$ the mass density; and $\hat{Q}_{\kappa}$ as point-sources of mass. Additionally, in the matrix, we have the Lagrangian porosity field denoted as $\phi^{*}$. The Lagrangian porosity augments the volumetric strain field and thereby introduces a coupling to the deformation as is described in a subsequent section. Continuity equations are posed for each of the two phases within $\Omega_M$ and $\Omega_F$. In the matrix and for each phase, we have,}
{\color{black}
\begin{equation}\label{governingM}
\int_{\Omega_M}\partial_{t} \left( \rho_{\kappa} S_{\kappa,M} \phi^{*} \right) d\bm{x} +  \int_{\Gamma} \rho_{\kappa}\bm{v}_{\kappa,M} \cdot \bm{n}_{\Gamma}d\bm{x} = \int_{\Omega_{F}}\llbracket \rho_{\kappa}\bm{v}_{\kappa,M}\cdot \bm{n}_{c}\rrbracket d\bm{x} + \hat{Q}_{\kappa,M},
\end{equation}
}

\textcolor{black}{and in the fracture, for $c = 1,\ldots, N_F,$}
{\color{black}
\begin{equation}\label{governingF}
\partial_{t}\int_{\mathcal{C}_{c}}\rho_{\kappa} S_{\kappa,F} d\boldsymbol{x} + \int_{\mathcal{C}_{c}} \nabla_{c} \cdot (\rho_{\kappa}\bm{v}_{\kappa,F})d\bm{x}  = -\int_{\mathcal{C}_{c}}\llbracket\rho_{\kappa} \bm{v}_{\kappa,M}\cdot \bm{n}_{c}\rrbracket d\bm{x} + \hat{Q}_{\kappa,F},
\end{equation}
}
\textcolor{black}{where $\bm{v}_{\kappa,M/F}$ are the fluid phase velocities. Effectively, these forms of continuity imply full coupling between mechanics and flow. In the matrix, the Lagrangian porosity integrates strain and we assume two-phase Darcy flow, whereas in fracture, aperture is dictated by displacement and it effects the fluid velocity. We assume Poiseuille flow in fracture so that the fluid occupies the entire fracture volume. The constitutive relations applied in our computational examples are described in a later section. The problem is closed by enforcing pressure continuity across fracture and fixed fluid velocity across the outer-boundary; i.e.,
\begin{subequations}
	\begin{align}
	&p_{\kappa,M} = p_{\kappa,F} \quad &\text{on} \quad \mathcal{C}_{c} \\
	&\bm{v}_{\kappa,M} \cdot \bm{n}_{\Gamma} = 0 \quad &\text{on} \quad \Gamma_{Q}.
	\end{align}
\end{subequations}
}

\subsection{Geomechanics}
We define the average pore pressure in the matrix or fracture using the saturation weighted average; i.e. $p_{M/F} = p_{w,M/F}S_{w} + p_{n, M/F}S_{n}$. The independent variable in the geomechanical system is the displacement field $\bm{u}: \Omega \times \mathbb{R}^+ \to \mathbb{R}^2$. Under the assumption of infinitesimal deformation ( $\norm{\nabla \bm{u}} \ll 1$), the quasi-static continuity of momentum requires that,
\begin{equation}
\label{GeomechanicsSF}
\nabla \cdot \left(\bm{\sigma} \left( \bm{u},p_{M}\bm{I} \right) \right) + \rho_{b}\bm{f}=0, \quad \text{on}\quad \Omega_{M},
\end{equation}
where $\bm{\sigma}$ is the total stress; $\bm{I}$ is the identity matrix; $\rho_{b}$ is the average density of the rock and fluid; and $\bm{f}$ is the body force per unit volume. On the outer boundary, Neumann (force) and Dirichlet (displacement) conditions are considered,
\begin{subequations}
	\begin{align}
	\bm{\sigma}\cdot\bm{n}_{\Gamma}= \bm{t}  \quad &\text{on}\quad\Gamma_{t}\\
	\bm{u} = \bm{\hat{u}}   \quad  &\text{on}\quad \Gamma_{u},
	\end{align}
\end{subequations}
while on immersed fracture boundaries, the fluid pressure $p_{F}$ is imposed onto the oriented surfaces of the fracture,
\begin{subequations}
	\begin{align}
	\bm{\sigma}\cdot \bm{n_{c}^{+}} = -\bm{\sigma}\cdot \bm{n_{c}^{-}}=p_{F}\bm{I}\cdot\bm{n_{c}} \quad &\text{on}\quad\mathcal{C}_{c},
	\end{align}
\end{subequations}

\subsection{Constitutive Laws}
Adopting Biot's single phase poroelasticity theory \cite{biot1957elastic} the effective stress law is written as,
\begin{equation}
\bm{\sigma} = \bm{\sigma}^{'} - \alpha p_{M} \bm{I},
\end{equation}
where $\bm{\sigma}$ is the total stress tensor; $\bm{\sigma}^{'}$ is the effective stress tensor; and $\alpha \in (0,1]$ is a  Biot coefficient. The effective stress $\bm{\sigma}^{'}$ is modeled using linear elasticity theory,
\begin{equation}
\bm{\sigma}^{'} = \lambda (\nabla \cdot \bm{u}) \bm{I} + 2G\bm{\varepsilon}(\bm{u}),
\end{equation}
where $\lambda, G >0$ are Lam$\acute{\text{e}}$ coefficients evaluated from properties of the matrix skeleton, and the strain $\bm{ \varepsilon}$ is a second order tensor. Under infinitesimal deformation the strain tensor is a function of displacement as,
\begin{equation}\label{strain}
\bm{\varepsilon} =  \nabla^{sym} \bm{u} := \frac{1}{2}(\nabla^{\text{T}} \bm{u} + \nabla \bm{u}),
\end{equation}
and the volumetric strain, $\epsilon$ is equal to the trace of the strain tensor, $\Tr{\left(\bm{\varepsilon}\right)}$. The Lagrangian porosity in the matrix under linear poroelastic infinitesimal deformation is modeled after \cite{dean2006comparison} as,
\begin{equation}
\phi^{\ast} = \phi_{0} + \alpha(\epsilon - \epsilon_{0}) + \frac{1}{M}(p_{M} - p_{M0}),
\end{equation}
where $\epsilon_{0}$ and $ p_{M0}$ are the reference states of volumetric strain and matrix pressure, respectively, both of which are  chosen as $0$ in this work; $\phi_{0}$ is the porosity at the state of $\epsilon_{0}$ and $ p_{M0}$; and $\frac{1}{M}$ is a Biot coefficient. 

{\color{black}
The multiphase extension to the Darcy velocity in the matrix is modeled as,
\begin{equation}
\bm{v}_{\kappa, M} = -\bm{k}_{M} \cdot \frac{k_{r\kappa,M}}{\mu_{\kappa}}(\nabla p_{\kappa, M} - \rho_{\kappa}\bm{g} \frac{dh}{dz}).
\end{equation}
where $ \bm{k}_{M}$ is the second order permeability tensor; $\bm{g}$ is the gravitational force;  and $\frac{dh}{dz}$ is the gradient of elevation with respect to gravity. The relative permeability for the wetting and non-wetting phases in matrix is calibrated by the Corey relationship,
\begin{subequations}
	\begin{align}
	k_{rn,M} = k_{rn}^{end} (\frac{S_{n} - S_{n,r}} {1 -  S_{n,r} - S_{w,r}})^{n_{n}},\\
	k_{rw,M} = k_{rw}^{end}(\frac{S_{w} - S_{w,r}} {1 -  S_{n,r} - S_{w,r}})^{n_{w}}.
	\end{align}
\end{subequations}
where $Kr_{w}^{end}$, $Kr_{n}^{end}$ are the end points of the wetting and non-wetting phase relative permeability, $S_{w,r}$ and $S_{n,r}$ are the residual saturation of the wetting and non-wetting phases $n_{n}$ and $n_{w}$ are the exponential numbers.

In fracture, fluid velocity is modeled according to Poiseuille's law,
\begin{equation}
\bm{v}_{\kappa, F} = -\frac{\omega_{c} ^ {2}S_{\kappa,F}}{12\mu_{\kappa}}(\nabla_{c} p_{\kappa,F} - \rho_{\kappa}\bm{g} \frac{dh}{dz}),
\end{equation}
where $\omega _{c} = \llbracket\bm{u}\rrbracket \cdot \bm{n}_{c}$ is fracture aperture.
}
\subsection{Rock failure and propagation}\label{RF}
The macroscopic concepts of Irwin's law are adopted to formulate the model for fracture growth rate. In particular, we assume the availability of an (empirical) critical fracture toughness, $K_c$, that is independent of fracture growth rate. Given a fracture tip $\bm{a}\left( t \right) \in \mathcal{D}_{c}$, we affix a local instantaneous frame, and consider the equivalent stress intensity factor $K_{I}^{eq}$  which is defined as a function of its mode I and II counterparts {\color{black} \cite{moes1999finite}},
\begin{equation}\label{eqKI}
K_{I}^{eq} = \frac{1}{2}\cos(\theta / 2) \{K_{I}(1+\cos(\theta)) - 3K_{II}\sin(\theta)\},
\end{equation}
and $\theta$ is the direction of maximum tensile stress
\begin{equation}\label{theta}
\theta = 2 \arctan \frac{1}{4}\bigg(K_{I}/K_{II} - \text{sign}(K_{II})\sqrt{(K_{I}/K_{II})^{2} +8}\bigg) .
\end{equation}
{\color{black}
\begin{figure}[!htb]
\centering
\includegraphics[width=0.6\textwidth]{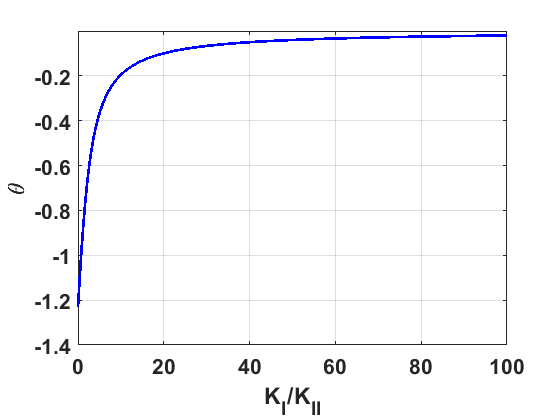}
\caption{\textcolor{black}{$\theta$ vs. $K_{I}/K_{II}$}}
\label{theta_K}
\end{figure}
The relationship between $\theta$ and the ratio $K_{I}/K_{II}$ is illustrated in Figure \ref{theta_K}, which suggests that the absolute value of $\theta$ decreases monotonically as $K_{I}/K_{II}$ increases.
}

 {\color{black}
The J integral adopted here to extract SIF is (\cite{walters2005interaction}),
\begin{equation}
J = \lim_{\Gamma_{I}\rightarrow 0}\int_{\Gamma_{I}}\left[ -\sigma_{ij}\frac{\partial u_{i}}{\partial x_{1}} + \sigma_{ij}\varepsilon_{ij}\delta_{1j} \right] n_{j}d\Gamma,
\label{contour_Int}
\end{equation}
where Einstein summation rules are adopted; Domain $\mathcal{B}_{\rho}$ is confined by a circle around the crack tip of radius $\rho$, and then $\Gamma_{I} = \partial \mathcal{B}_{\rho} \cup (\mathcal{B}_{\rho} \cap \mathcal{C}_c)$; $\bm n$ is the outward unit-normal to the neighborhood.

In order to model $K_{I}$ and $K_{II}$, the following relationship holds (\cite{moes1999finite})
\begin{equation}
    J = \frac{1}{E'}(K_{I}^2 + K_{II}^2).
\end{equation}
where $E'$ is equal to $E$ under the plane stress condition and equal to $E/(1-\tilde{\nu}^2 )$ under the plane strain condition. $\tilde{\nu}$ is the Poisson's ratio. 
}

In this work, the creation of initial fracture or defect is not modeled; rather, it is assumed that initially, there are preexisting fractures represented by the mappings $\bm{\gamma}_{c}$. Under linear elastic fracture mechanics theory, equilibrium (or static) FP requires that the strain energy release rates are less than or equal to the critical \cite{bavzant2014fracking}, i.e. $K^{eq}_{I} \leq K{c}$.
Subsequently, during water injection, pressure build-up within the fracture leads to an increase in $K_{I}^{eq}$. If $K_{I}^{eq}$ achieves a critical threshold $K_{c}$, the tip is assumed to advance, relieving the local build-up in $K_{I}^{eq}$. In a quasi-static manner, this process may continue while the local relief due to advancement and the build-up due to injection are stable and balanced. This quasi-static process terminates when the build-up due to pressure causes $K_{I}^{eq}$ to drop below the threshold criterion. Finally, a stable propagation scheme of linear elastic fracture that is regulated by the $K_{I}^{eq}$ reads,
\begin{equation}\label{propagationrule}
\begin{cases}
	\text{Stable Propagation} & K_{I}^{eq}=  K_{c} \\
    \text{Static} & K_{I}^{eq} < K_{c}  \\
\end{cases}.
\end{equation}

\section{Numerical Discretization}\label{ND}
We consider the mesh $\mathcal{T}_{h}$ as a subdivision of the domain $\Omega_{M}$ into disjoint elements $\Omega^{e}$ (quadrilaterals) in two dimensions. Subsequently, we define $\bar{\Omega} = \cup_{\Omega^{e}\in \mathcal{T}_{h}} \Omega^{e}$ as the union set of all disjoint matrix elements. The diameter of each element $\Omega^{e} \in \mathcal{T}_{h}$ is denoted by $h$. We denote partitioned boundaries of the domain $\bar{\Omega}$ as $\Gamma_{t}^{e}, \Gamma_{u}^{e}, \Gamma_{p}^{e}, \Gamma_{q}^{e}$. Another set of meshes denoted by $\hat{\mathcal{T}}_{h}$ and corresponding to fracture $\bm{\gamma}_{c}$ is attached to the base mesh resulting in a set of disjoint elements  $\bm{\gamma}_{c}^{e}$. The fracture elements $\bm{\gamma}_{c}^{e}$ are simply chosen as the portions that are partitioned by matrix elements $\Omega^{e}$. Thereafter, $\bar{\bm{\gamma}}_{c} = \cup_{\bm{\gamma}_{c}^{e}\in \hat{\mathcal{T}}_{h}}\bm{\gamma}_{c}^{e}$ is defined as the union set of all disjoint fracture elements. {\color{black} Generally, the fracture mesh (choice of segmentation of fracture) can be independent of the choice of matrix mesh. Accommodating this may be achieved by a preprocessing of connections and transmissibility in the context of EDFM, and by a treatment of enrichment in the XFEM context. While this feature may be desirable in practice, in this work we assume that the fracture segments conform to the edges of the matrix mesh in which they are embedded.} We also introduce a partition of the time interval, $\mathcal{I}_{T}^{n} = (t^{n}, t^{n+1})$ with $\Delta t = \left | \mathcal{I}_{T}^{n} \right |$, such that $\bar{\mathcal{I}}_{T} = \cup_{n}\mathcal{I}_{T}^{n} $ for $n \in {0,...,N_{t}}$.

{\color{black}
Flow equations are discretized using a finite-volume approximation whereas mechanical equations are approximated using a finite element method. The unknowns are staggered; pressure $p_{\kappa,M/F}$ and saturation $S_{\kappa,M/F}$ are cell-centered within $\Omega^{e}$ and $\gamma^{e}_{c}$, and displacement nodes are located at the vertices of grid $\Omega^{e}$. The current setting simultaneously ensures both local mass conservation and eliminates spurious spatial instability at early times for compressible systems (see for example, \cite{vermeer1981accuracy,murad1994stability,jha2007locally,kim2011stability}).
}

\subsection{XFEM approximation to poromechanics}

\subsubsection{Weak forms}
	First, we follow the Bubnov--Galerkin approach that the test ($\mathcal{S}_{0}$) and trial functional space ($\mathcal{S}$) share the same functional space, 
     \begin{equation}
		\mathcal{S}_{0}(\Omega) = \left \{ \delta \bm{u} | \delta \bm{u} \in \bm{H}^{1}_{0}(\Omega): \delta \bm{u} = 0 \  \text{on} \  \Gamma_{u}\right \}
	\end{equation}
	\begin{equation}
		\mathcal{S}(\Omega) = \left \{ \bm{u} |  \bm{u} \in \bm{H}^{1}(\Omega): \bm{u} = \hat{\bm{u}} \  \text{on} \  \Gamma_{u}\right \}.
	\end{equation}
 Furthermore, the test and trial functional space of admissible strain field $\delta \bm{\varepsilon}$ and $\bm{\varepsilon}$ shall be denoted by $\mathcal{E}_{0}$ and $\mathcal{E}$, respectively. Subsequently, variational forms of the displacement $\delta \bm{u}$ and strain $\delta \bm{\varepsilon}$ are decomposed into standard and enhanced parts in order to account for the discontinuity,
\begin{equation}
\begin{aligned}
	&\delta \bm{u} = \delta \bar{\bm{u}} + \delta \tilde{\bm{u}}, \quad \delta \bm{u} \in \mathcal{S}_{0}\\
	&\delta \bm{\varepsilon} = \delta \bar{\bm{\varepsilon}} + \delta \tilde{\bm{\varepsilon}}, \quad \delta \bm{\varepsilon} = \nabla^{sym} \delta \bm{u} \in \mathcal {E}_{0},
\end{aligned}
\end{equation}
where the $\delta \bar{\bm{u}}$ and $\delta \bar{\bm{\varepsilon}}$ are corresponding standard parts of the displacement and strain fields while  $\delta \tilde{\bm{u}}$ and $ \delta \tilde{\bm{\varepsilon}}$ are enhanced parts of the displacement and strain fields.  {\color{black}Here, we propose the following sets of weak forms of \cref{GeomechanicsSF}} combining the Biot poroelastic theory:
{\color{black}
\begin{equation}\label{weakform1}
\begin{aligned}
&\int_{\Omega}  \delta \bar{\bm{\varepsilon}} : \bm{\sigma}' d\bm{x} - \int_{\Omega} \delta \bar{\bm{\varepsilon}}:\alpha p_{M}\bm{I} d\bm{x}- W^{ext}(\delta \bar{\bm{u}}) = 0  \quad \forall  \delta \bar{\bm{u}} \in \mathcal{S}_{0} \\
&\int_{\Omega}  \delta \tilde{\bm{\varepsilon}} : \bm{\sigma}' d\bm{x} - \int_{\Omega} \delta \tilde{\bm{\varepsilon}}:\alpha p_{M}\bm{I}d\bm{x} -  W^{ext}(\delta \tilde{\bm{u}}) = 0 \quad \forall  \delta \tilde{\bm{u}} \in \mathcal{S}_{0},
\end{aligned}
\end{equation}
where external virtual work terms can be expressed as,
\begin{equation}\label{weakform2}
\begin{aligned}
	&W^{ext}(\delta \bar{\bm{u}})  = \int_{\Gamma_{t}}\delta \bar{\bm{u}} \cdot \bm{t}d\bm{x}\quad \forall  \delta \bar{\bm{u}} \in \mathcal{S}_{0} \\ 
      &W^{ext}(\delta \tilde{\bm{u}})  = \int_{\bm{\gamma}_{c}}\llbracket \delta \tilde{\bm{u}}\rrbracket  \cdot p_{F}\bm{n}_{c}d\bm{x} \quad \forall  \delta \tilde{\bm{u}} \in \mathcal{S}_{0},
\end{aligned}
\end{equation}
}
Note that $p_{M}$ is applied in the strain energy term $\delta \bm{\varepsilon} : \bm{\sigma}'$ while $p_{F}$ is used in the external virtual work term $W^{ext}(\delta \tilde{\bm{u}})$.  These contribute to coupling between mechanics and fluid flow. 

\subsubsection{Displacement interpolation functions and discretization}
\begin{figure}[!htb]
		\centering
		\includegraphics[width=0.5\textwidth]{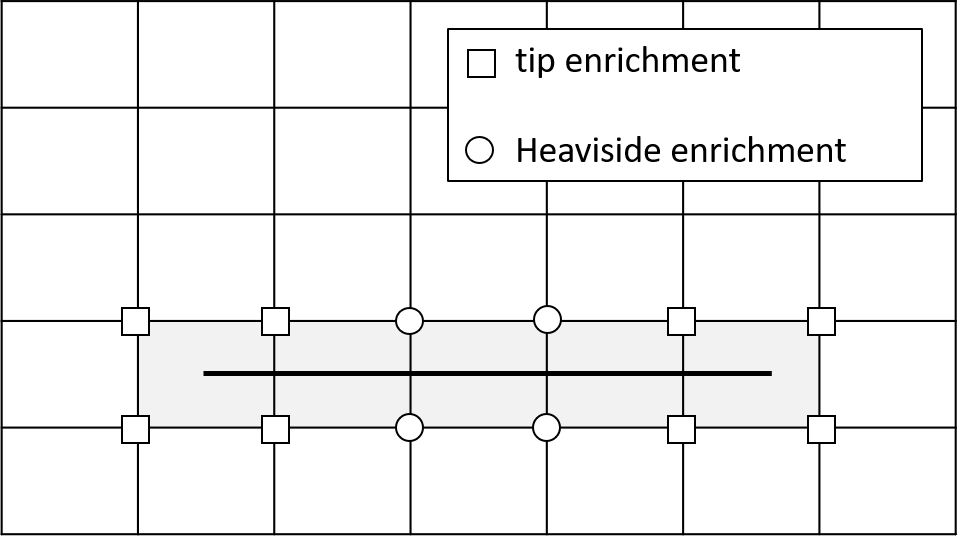}
		\caption{\textcolor{black}{Illustration of enrichment nodes; Black line represents the fracture; All grid vertices are included in the node set $I$; Squares consist of subset $K$ while circles consist of subset $L$.}}
		\label{enrichillustration}
\end{figure}
Following the approach of \cite{moes1999finite}, crack surfaces are modeled by step functions and tips by asymptotic near-tip fields. Three clusters of nodes shown in Figure \ref{enrichillustration} are defined:
\begin{itemize}
	\item $I$: the set of all nodes in the region $\bar{\Omega}$.
	\item $L$: the sub set of $I$ that are enriched by the Heaviside function, $H_{\bm{\gamma}_{c}}$, and $L$ are additional nodes that are used to capture the displacement discontinuity of the fracture body.
	\item $K$: the sub set of $I$ that are enriched by the asymptotic branch functions, {\color{black}	\begin{equation}
F_{l} = \left\{\begin{matrix}
\sqrt{r}\sin \frac{\theta}{2}& l = 1\\ 
\sqrt{r}\cos \frac{\theta}{2}& l = 2\\ 
\sqrt{r}\sin \frac{\theta}{2}\sin \theta& l = 3\\ 
\sqrt{r}\cos \frac{\theta}{2}\sin \theta &  l =4
\end{matrix}\right.
	\end{equation},} and nodes in $K$ are used to capture the stress singularity at the fracture tip.
\end{itemize}
  Therefore, the standard and enhanced terms and their variational counterparts are expressed as,
 {\color{black}
\begin{equation} \label{interpolation1}
\begin{aligned}
	\bar{\bm{u}} &= \sum_{i\in I}N_{i}\bar{\bm{u}}_{i} \\
	 \delta\bar{\bm{ u}}  &= \sum_{i\in I}N_{i}\delta\bar{\bm{u}}_{i}\\
	\tilde{\bm{u}} &=  \sum_{i\in L}N_{i}(H_{\bm{\gamma}_{c}} - H^{i}_{\bm{\gamma}_{c}})\tilde{\bm{a}}_{i}  + \sum_{i \in K}\sum_{l = 1}^{4}N_{i}(F_{l} - F^{i}_{l})\tilde{\bm{b}}^{l}_{i} \\ 
        \delta\tilde{\bm{u}} & =  \sum_{i\in L}N_{i}(H_{\bm{\gamma}_{c}} - H^{i}_{\bm{\gamma}_{c}})\delta\tilde{\bm{a}}_{i}  + \sum_{i \in K}\sum_{l = 1}^{4}N_{i}(F_{l} - F^{i}_{l})\delta\tilde{\bm{b}}^{l}_{i},
\end{aligned}\\
\end{equation}}
\begin{equation}\label{interpolation2}
\begin{aligned}
\bar{\bm{\varepsilon}} &= \sum_{i\in I}\bm{B}^{u}_{i}\bar{\bm{u}}_{i}  \\
	  \delta\bar{\bm{ \varepsilon}} &= \sum_{i\in I}\bm{B}^{u}_{i}\delta\bar{\bm{u}}_{i}\\
	\tilde{\bm{\varepsilon}} &=  \sum_{i\in L}\bm{B}^{a}_{i}\tilde{\bm{a}}_{i}  + \sum_{i \in K}\sum_{l = 1}^{4}\bm{B}^{b^{l}}_{i}\tilde{\bm{b}}^{l}_{i}  \\
         \delta\tilde{\bm{\varepsilon}} &=  \sum_{i\in L}\bm{B}^{a}_{i}\delta\tilde{\bm{a}}_{i}  + \sum_{i \in K}\sum_{l = 1}^{4}\bm{B}^{b^{l}}_{i}\delta\tilde{\bm{b}}^{l}_{i},
\end{aligned}
\end{equation}
where $N_{i}$ are finite element linear basis functions for quadrilateral elements; {\color{black}Enrichment functions are shifted by the function values evaluated at vertices $i$, $H_{\bm{\gamma}_{c}}^i$ and $F_{l}^i$  respectively.} $\bar{\bm{u}}_{i}$ is the displacement for standard nodes; $\tilde{\bm{a}}_{i}$ is the displacement for Heaviside enriched nodes; $\tilde{\bm{b}}^{l}_{i}$ is the displacement for asymptotic branch function enriched nodes, and $\bm{B}^{u}_{i}$, $\bm{B}^{a}_{i}$, $\bm{B}^{b^{l}}_{i}$ can be derived correspondingly by substitution of \cref{interpolation1} into \cref{strain}.

{\color{black}
The discretized version of terms in \cref{weakform1,weakform2} is expressed in the \cref{discrge}. The subscripts $i$ and $j$ identify the nodal number,  $D$ is the elastic matrix. The numerical integration in elements cut by fractures is performed using triangulation and coordinate transformation of the tip element (\cite{ren2018embedded}).
\begin{table}[!htb]
	\centering
	\caption{\textcolor{black}{Discretized forms of the geomechanics equation}}
		\label{discrge}
	\begin{tabular}{ll}  \hline 
		$\int_{\Omega}  \delta \bar{\bm{\varepsilon}} : \bm{\sigma}' d\bm{x}$ and $ \int_{\Omega}  \delta \tilde{\bm{\varepsilon}} : \bm{\sigma}' d\bm{x}$ & $ \int_{\Omega^{e}} (\bm{B}_{i}^{r})^{\rm{T}}D\bm{B}_{j}^{s}d\bm{x}\quad(r,s = u,a,b^{l})$\\  \hline 
		$\int_{\Omega} \delta \bar{\bm{\varepsilon}}:\alpha p_{M}\bm{I} d\bm{x}$ and $ \int_{\Omega} \delta \tilde{\bm{\varepsilon}}:\alpha p_{M}\bm{I}d\bm{x} $    &              $\int_{\Omega^{e}}(\bm{B}^{r}_{i})^{\rm{T}} \alpha p_{M}(1,1,0)^{T}d\bm{x} \quad (r = u,a,b^{l})$                                                                                                                           \\   \hline 
$\int_{\bm{\gamma}_{c}}\llbracket \delta \tilde{\bm{u}}\rrbracket  \cdot p_{F}\bm{n}_{c}d\bm{x}$	&  $\int_{\bm{\gamma}^{e}_{c}}N_{i}\llbracket F_{l}\rrbracket p_{F}\bm{n}_{c}d\bm{x}$  and $\int_{\bm{\gamma}^{e}_{c}} N_{i}\llbracket H_{\bm{\gamma}_{c}}\rrbracket p_{F} \bm{n}_{c}d\bm{x}$                    \\   \hline 
$\int_{\Gamma_{t}}\delta \bar{\bm{u}} \cdot \bm{t}d\bm{x}$ & $\int_{\Gamma^{e}_{t}}N_{i}\bm{t}d\bm{x}$     \\                     \hline                                 
	\end{tabular}
\end{table}
}

\subsection{EDFM approximation of fluid flow and transport.}
A first order backward Euler fully implicit scheme is adopted for the time discretization of flow equations.  Fluxes between different connections, i.e. matrix-matrix (M-M), matrix-fracture (M-F), fracture-fracture (F-F), are approximated using a two-point flux approximation  (TPFA).  A standard finite-volume scheme is employed as the spatial discretization for M-M and F-F connections. We follow the strategy proposed by \cite{lee2001hierarchical}  for M-F flux calculation. The fully-implicit discretization for \cref{governingM} and \cref{governingF} are listed in Table \ref{flowDiscretized},
\begin{table}[!htb]
\caption{The discretization of flow equations}
\label{flowDiscretized}
\centering
\begin{tabular}{ll}
\hline
Continuous form                                                                                     & Discrete form for a single cell $\Omega^{e}$ or $\bm{\gamma}^{e}_{c}$                                                                                                                                                  \\ \hline
$\int_{\Omega_{M}}\partial_{t}\big(\rho_{\kappa} S_{\kappa,M} \phi^{*}\big) d\bm{x}$   & $1/\Delta t \big[ \big((\phi_{0} + \alpha\epsilon + \frac{1}{M}p_{M})\rho_{\kappa} S_{\kappa, M}\big)_{i}^{n+1} - \big((\phi_{0} + \alpha\epsilon + \frac{1}{M}p_{M})\rho_{\kappa} S_{\kappa,M}\big)_{i}^{n}\big]$ \\ \hline
$\partial_{t}\int_{\bm{\gamma}_{c}(t)}\rho_{\kappa} S_{\kappa,F} d\bm{x}$           &  $L_{i}/\Delta t \big[ \big(\omega_{c}\rho_{\kappa} S_{\kappa, M}\big)_{i}^{n+1} - \big(\omega_{c}\rho_{\kappa} S_{\kappa,M}\big)_{i}^{n}\big]$                                                                                                                                                                                               \\ \hline
$\int_{\Gamma}\rho_{\kappa}\bm{v}_{\kappa,M} \cdot \bm{n}_{\Gamma}d\bm{x}$                            & $\sum_{j\in adjMM(i)}T_{MM}(k_{r\kappa} \rho_{\kappa}/\mu_{\kappa})_{(i+j)/2}^{n+1}(p_{\kappa,M_{i}} - p_{\kappa,M_{j}})^{n+1}$                                                                                       \\ \hline
$\int_{\bm{\gamma}_{c}(t)} \nabla_{c} \cdot (\rho_{\kappa}\bm{v}_{\kappa,F})d\bm{x}$                       & $\sum_{j\in adjFF(i)}T_{FF}( k_{r\kappa}\rho_{\kappa}/\mu_{\kappa})_{(i+j)/2}^{n+1}(p_{\kappa,F_{i}} - p_{\kappa,F_{j}})^{n+1}$                                                                                       \\ \hline
$\int_{\bm{\gamma}_{c}(t)} \llbracket\rho_{\kappa}\bm{v}_{\kappa,M}\cdot \bm{n}_{c}\rrbracket d\bm{x}$ & $\beta\sum_{j\in adjMF(i)}T_{MF}( k_{r\kappa}\rho_{\kappa}/\mu_{\kappa})_{(i+j)/2}^{n+1}(p_{\kappa,M_{i}} - p_{\kappa,F_{j}})^{n+1}$                                                                                       \\ \hline
\end{tabular}
\end{table}\\
where the superscripts $n+1$ and $n$ represent the current and previous time-steps, and $L_{i}$ is the length of fracture segment $i$. We also assume that while fracture porosity is one,  the volume of the fracture is affected by its aperture $\omega_{c}$. The connectivity routines, $adjMM(i)$  and $adjFF(i)$ return matrix or fracture cell index $j$ adjacent to matrix or fracture cell index $i$ or given matrix cell index $i$, $adjMF(i)$ returns the fracture cell index $j$. The definitions of transmissibility terms $T_{MM}$, $T_{FF}$ and $T_{MF}$ are as proposed in \cite{jiang2016hybrid}. First order phase potential upwinding (PPU) is employed to complete the $k_{r\kappa}$ on the interface, while $\rho_{\kappa}$ and $\mu_{\kappa}$ are interpolated from two adjacent cells.  
\begin{remark}
	If FP does not occur at a given time-step, then $\bm{v} = \bm{0}$ and the computational domain remains static. Therefore, 
\begin{equation}
   \partial_{t}\int_{\bm{\gamma}_{c}(t)}\rho_{\kappa} S_{\kappa,F}d\bm{x}  \approx \int_{\bm{\gamma}_{c}(t^{n+1})}1/\Delta t\big[\big(\rho_{\kappa} S_{\kappa,F}\big)^{n+1} - \big(\rho_{\kappa} S_{\kappa,F}\big)^{n}\big] d\bm{x}.
\end{equation}
Under FP however, a new fracture segment is to be added into the system at time $t^{n+1}$, and subsequently,
\begin{equation}
\partial_{t}\int_{\bm{\gamma}_{c}(t)}\rho_{\kappa} S_{\kappa,F} d\bm{x} \approx  \frac{1}{\Delta t} \left[ {\int_{\bm{\gamma}_{c}(t^{n+1})}\big(\rho_{\kappa} S_{\kappa,F}\big)^{n+1} d\bm{x} - \int_{\bm{\gamma}_{c}(t^{n})} \big(\rho_{\kappa} S_{\kappa,F}\big)^{n} d\bm{x} } \right].
\end{equation}
\end{remark}

\begin{remark}
\begin{figure}[!ht]
	\centering
	\includegraphics[width=0.5\textwidth]{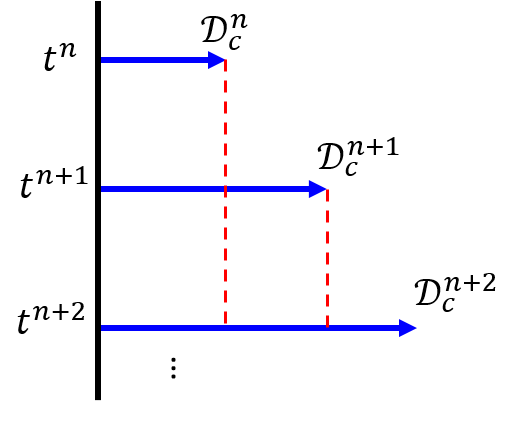}\\
	\caption{Schematics of fluid interaction between fracture and matrix during FP}
	\label{Fig::correctionfactor}
\end{figure}
Correction for M-F fluid transfer under propagation is necessary. As illustrated in Figure \ref{Fig::correctionfactor}, consider a newly established tip element that propagates from length $\mathcal{D}_{c}^{n}$ to length $\mathcal{D}_{c}^{n+1}$ over a time-step $\Delta t = t^{n+1} - t^{n}$. The total fluid losses from the fracture to matrix is,
\begin{equation}
	Q_{MF} = \int_{t^{n}}^{t^{n+1}}\int_{\mathcal{D}_{c}^{n}}^{\mathcal{D}_{c}^{n+1}}q_{MF}H(t-\tau)dxdt,
\end{equation}
where $H(t-\tau)$ is the Heaviside function and $\tau$ is the arrival time of the fracture tip at position $x$. Assuming a constant propagation velocity $c$ and a uniform flux $q_{MF}$ over the time period $t^{n+1}-t^{n}$,  and utilizing a trapezoidal rule for the integration, we have,
\begin{equation}
	Q_{MF} = \int_{t^{n}}^{t^{n+1}}\int_{\mathcal{D}_{c}^{n}}^{\mathcal{D}_{c}^{n}+ct}q_{MF}^{n+1}dxdt = \frac{1}{2}q_{MF}^{n+1}(\mathcal{D}^{n+1}_{c} - \mathcal{D}_{c}^{n})(t^{n+1}-t^{n}).
\end{equation}
This model is applied to M-F source terms, and the $\beta$ appearing in the Table \ref{flowDiscretized} equals $1/2$ in this model.
\end{remark}
\subsection{The fully coupled system}
After we assemble the Jacobian system resulted from Table \ref{discrge} and Table \ref{flowDiscretized}, the fully coupled system solved by the Newton-Raphson method is,
{\color{black} 
\begin{equation}\label{fcsytem}
\begin{pmatrix}
\frac{\partial R_{1}}{\partial p_{n, M}} & \frac{\partial R_{1}}{\partial S_{w, M}} & \frac{\partial R_{1}}{\partial p_{n, F}} &  \frac{\partial R_{1}}{\partial S_{w, F}}  &  \frac{\partial R_{1}}{\partial \bm{u}} \\ 
\frac{\partial R_{2}}{\partial p_{n, M}} &  \frac{\partial R_{2}}{\partial S_{w, M}} & \frac{\partial R_{2}}{\partial p_{n, F}} & \frac{\partial R_{2}}{\partial S_{w, F}} &\frac{\partial R_{2}}{\partial \bm{u}}  \\
\frac{\partial R_{3}}{\partial p_{n, M}} &  \frac{\partial R_{3}}{\partial S_{w, M}} & \frac{\partial R_{3}}{\partial p_{n, F}} & \frac{\partial R_{3}}{\partial S_{w, F}} &\frac{\partial R_{3}}{\partial \bm{u}}  \\
\frac{\partial R_{4}}{\partial p_{n, M}} &  \frac{\partial R_{4}}{\partial S_{w, M}} & \frac{\partial R_{4}}{\partial p_{n, F}} & \frac{\partial R_{4}}{\partial S_{w, F}} &\frac{\partial R_{4}}{\partial \bm{u}}  \\
\frac{\partial R_{5}}{\partial p_{n, M}} &  \frac{\partial R_{5}}{\partial S_{w, M}} & \frac{\partial R_{5}}{\partial p_{n, F}} & \frac{\partial R_{5}}{\partial S_{w, F}} &\frac{\partial R_{5}}{\partial \bm{u}}  \\
\end{pmatrix}\begin{pmatrix}
\delta p_{n, M}\\ 
\delta S_{w, M}\\ 
\delta p_{n, F}\\ 
\delta S_{w, F}\\ 
\delta \boldsymbol{u}\\
\end{pmatrix}=-\begin{pmatrix}
R_1\\ 
R_2\\ 
R_3\\ 
R_4\\ 
R_5\\
\end{pmatrix},
\end{equation}
where gradients can be derived by using the formulations in Table 1 and 2.  }
\section{Fracture Propagation}\label{FPSection}
A direct local solution for FP speed via \cref{propagationrule} is intractable. Rather, at a given time-step, the condition in 
\cref{propagationrule} is tested for each tip within the domain. FP tip-advancement algorithms are required to indicate the onset of propagation over a time-step, $\Delta t$, and to approximate a length, $\Delta a$ with orientation, $\theta$, of the propagation, assuming a linear trajectory.

\begin{figure}[ht]
	\centering
	\begin{subfigure}[b]{0.3\textwidth}
		\centering
		\includegraphics[width=0.9\textwidth]{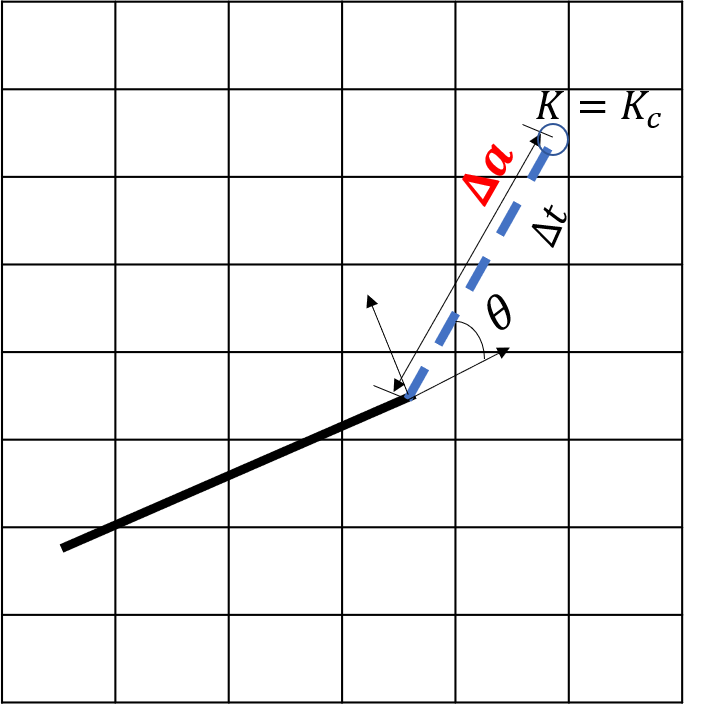}\\
		\caption{Scheme A}
		\label{Fig::FPAlgo1}
	\end{subfigure}
	\begin{subfigure}[b]{0.3\textwidth}
		\centering
		\includegraphics[width=0.9\textwidth]{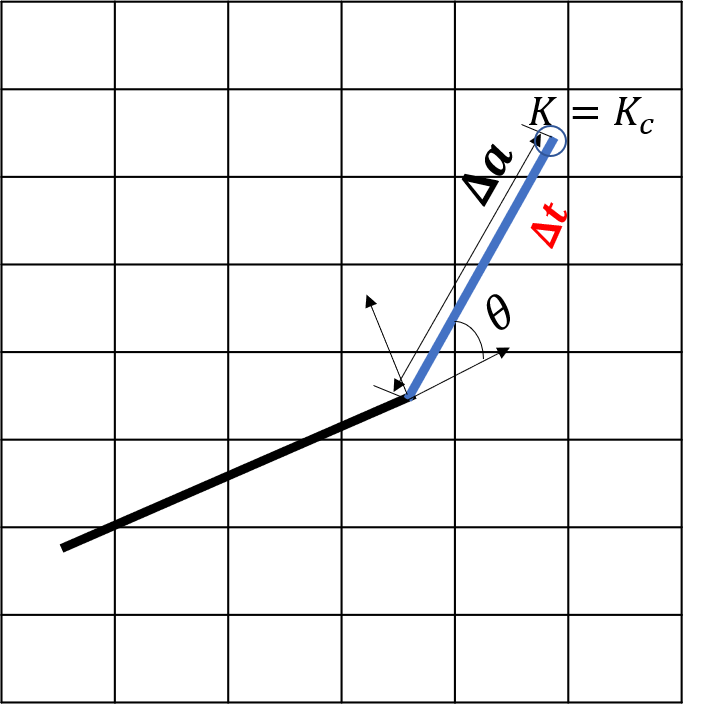}\\
		\caption{Scheme B}
		\label{Fig::FPAlgo2}
	\end{subfigure}
	\caption{Comparison of two schemes for fracture propagation. Black line: old fracture segment, Blue dashed/solid line: new fracture segment; Red color stands for the parameter that needs to be determined.}
	\label{fig:schemes}

\end{figure}

Broadly, there are two classes of tip-advancement algorithm differing in the assumptions applied to the tip advancement $\Delta a$ over time-step $\Delta t$, and their relation to the propagation criterion as illustrated in Figure~\ref{fig:schemes}. 

\paragraph{Scheme A} As illustrated in Figure \ref{Fig::FPAlgo1}, given a fixed target time-step, $\Delta t$, fracture advancement $\Delta a$ that will satisfy the criterion \cref{propagationrule} is determined. One class of algorithm requires frequent reordering of enriched nodes and reassembly of elastic stiffness matrices associated with enriched nodes until a suitable step-size is found. Example of methods in this category include, \cite{hunsweck2013finite,zeng2018fully}. The unknown, $\Delta a$, must satisfy \cref{propagationrule}, and can be determined by solving,
\begin{equation} \Delta a =  
	\begin{cases}
	0 & K_{I}^{eq}\left( t \right) -  K_{c} <0 \\
	\argmin{| {K_{I}^{eq}-  K_{c}} | } & \text{otherwise} \\
	\end{cases}.
\label{ConstrainEqn}
\end{equation}

\paragraph{Scheme B} Alternately, as illustrated in Figure \ref{Fig::FPAlgo2}, a fixed propagation length $\Delta a$ is specified, and the corresponding time-step size $\Delta t$ and its associated state variables ($p_{\kappa, M/F}, S_{\kappa, M/F}, \bm{u}$) are determined to satisfy the constraint in \cref{propagationrule}, e.g. \cite{gordeliy2013coupling}. To obtain such a step, the coupled problem \cref{ConstrainEqn} can be solved for the time-step. In this approach however, the modeling of multiple, simultaneously propagating fractures is challenging since the various tip propagation speeds may differ widely. 

An alternative combining ideas from both approaches entails selection of a target time-step size and utilizing an approximate corresponding advancement step. Since the approximate step may not satisfy the constraint in \cref{ConstrainEqn}, a sequence of subsequent solution substeps are computed such that the terminal substep satisfies the equality constraint in \cref{propagationrule}. Hence, the time of this final substep can be considered as the correct time-step for the given advancement step. In \cite{gupta2018coupled} and \cite{shauer2019improved} a regularized form of the constraint equation is applied to identify approximate advancement increments and at each substep in \cite{gupta2018coupled}, the mesh is adapted and the coupled system is resolved. In that work, the fluid flow and its coupling to mechanics within the matrix are neglected. Therefore, intermittent periods of flow simulation processes are not modeled. This is critical for the application of such models in data assimilation of combined flow and fracture systems.

In this work, we adapt this general approach to the embedded method under general two-phase flow and mechanics in fracture and matrix. We address the following four technical components to achieve this:
\begin{enumerate}
\item Accurate estimation of the SIF in embedded fractures through a numerical evaluation of J integral,
\item Geometric algorithms to update fracture geometry $\bm{\gamma}_{c}$ and insertion of new $\bm{u}$ into numerical discretization, 
\item Reliable extrapolations of state variables ($p_{\kappa}, S_{\kappa} \text{ and } \bm{u}$) following propagation in order to improve subsequent nonlinear algebraic solution,
\item Adaptive time-step selection criteria to improve computational performance at the onset of end of FP, and to improve performance within the internal FP iteration. 
\end{enumerate}

\subsection{Proposed methods}
The general solution procedure is listed in Algorithm \ref{FPSOLVER}, and Algorithms \ref{DU}-\ref{DF} are sub-routines to complete the algorithm. The input $I_{m}$ in Algorithm \ref{DU} is a Boolean record of whether tips meet the propagation criterion. For an arbitrary element $I_{m}^{i} \in I_{m}$,
\begin{equation}
I_{m}^{i}=\left\{\begin{matrix}
\text{TRUE} & 0<(K_{I}^{eq, i} -K_{c})/K_{c} <\epsilon_{SIF} \\ 
\text{FALSE} & (K_{I}^{eq, i} -K_{c})/K_{c} < 0
\end{matrix}\right.,
\end{equation}
where $\epsilon_{SIF}$ is a user specified tolerance. At the start of each time-step solve from \cref{l1} and \cref{l2} in Algorithm \ref{FPSOLVER}, we first update the domain based on the $I_{m}$, then solve for $\{p_{\kappa, M/F}, S_{\kappa, M/F}, \bm{u}\}$. In the post processing, from \crefrange{l3}{l4},  we either update $I_{m}$ if FP is stable or restart the solve, otherwise. Sections \ref{NJINT} - \ref{dins} develop the numerical approximation to the SIF, the proposed geometric updates, and the state-variable initial guess and updates.

\begin{figure}[!thb]
 \removelatexerror
\begin{algorithm}[H]\label{DU}
\caption{Domain\_Update}
\textbf{Input} $I_{m}$ \\
\textbf{Output} $p_{\kappa}, S_{\kappa},\bm{u}$\\
\For{$i \in I_{m}$}{
		\If{$i=$\upshape{TRUE}}{
            Follow the routine in Section \ref{fgu}
		}
}
\If{$\exists i \in I_{m} = $\upshape{TRUE}}{
Algorithm \ref{FPInitAlgo}: Initialization() in Section \ref{dins}\\
}
\end{algorithm}
\end{figure}

\begin{figure}[!thb]
 \removelatexerror
\begin{algorithm}[H]
\caption{Check\_StablePropagation}\label{CS}
\textbf{Input} $K_{I}^{eq}$ \\
\textbf{Output} IS\_STABLE\\
$\text{IS\_STABLE} \gets \text{TRUE}$;\\
\For{$k \in \mathcal{D}_{c}$}{
		\If{$(K_{I,k}^{eq} - K_{c})/K_{c} >\epsilon_{SIF}$}{
		IS\_STABLE $\gets \text{FALSE}$ \\
		Break
		}
}
\end{algorithm}
\end{figure}

\begin{figure}
 \removelatexerror
\begin{algorithm}[H]\label{DF}
\caption{Do\_Flagging}
\textbf{Input} $K_{I}^{eq}$ \\
\textbf{Output} $I_{m}$\\
$I_{m} \gets \{\}$;\\
\For{$k \in \mathcal{D}_{c}$}{
		\If{$0 < (K_{I,k}^{eq} - K_{c})/K_{c} < \epsilon_{SIF}$}{
		$I_{m} \gets I_{m} \cup \text{TRUE}$ \\
		}
		\ElseIf{$K_{I,k}^{eq} - K_{c} < 0$}{
		$I_{m} \gets I_{m} \cup \text{FALSE}$\\
		}
}
\end{algorithm}
\end{figure}

\begin{figure}[!thb]
 \removelatexerror
\begin{algorithm}[H]\label{FPSOLVER}
	\caption{FP solver}
    \textbf{Input}  $\Delta t, I_m$ \\
    \textbf{Output} $p_{\kappa, M/F}, S_{\kappa, M/F}, \bm{u}$\\
        Algorithm \ref{DU}: Domain\_Update($I_{m}$)\label{l1}\\
		Solve the fully coupled system \cref{fcsytem}\label{marker}\label{l2}\\ 
		Algorithm \ref{interactionInt}: $K_{I}^{eq} \gets$\upshape{SIF\_Calculation()}\label{l3} \\
		\If{\upshape{Algorithm \ref{CS}: Check\_StablePropagation}$(K_{I}^{eq}) = \text{TRUE}$}{
		     Algorithm \ref{DF}: $I_{m}\gets$  Do\_Flagging($K_{I}^{eq}$)
		 }
		 \Else{
		    	Apply \cref{tscut} to update $\Delta t$\\
				\Goto {marker} and restart\label{l4}
		}
\end{algorithm}
\end{figure}

\subsection{Proposed numerical approximation of Interaction Integral}\label{NJINT}
\begin{figure}[!ht]
	\begin{subfigure}[b]{0.49\textwidth}
	\centering
	\includegraphics[width=0.65\textwidth]{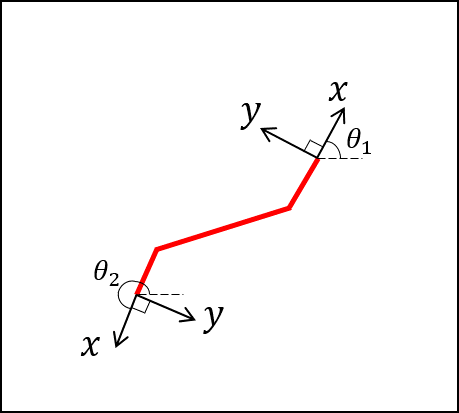}\\
	\caption{}
	\label{Fig::tiplocalcoord}
	\end{subfigure}
	\begin{subfigure}[b]{0.49\textwidth}
	\centering
	\includegraphics[width=0.6\textwidth]{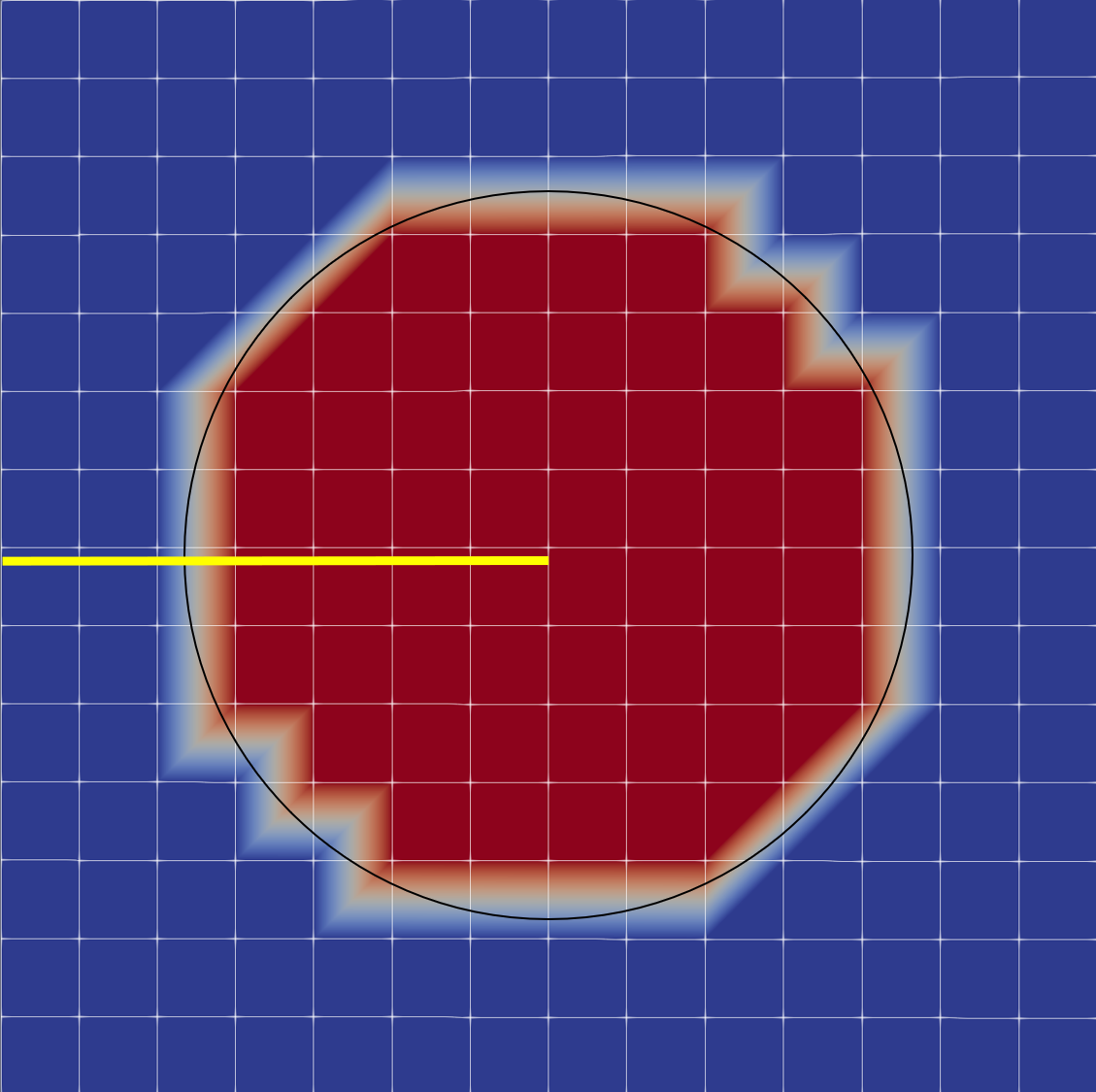}\\
	\caption{}
	\label{Fig::Jint}
	\end{subfigure}
	\caption{(a) Fracture tip local coordinate, (b) $q$ value distribution, red and blue colors correspond to value 1 and 0, respectively.}
\end{figure}

 An interaction-integral approach is adopted that superimposes the actual field with an auxiliary state within the expression for energy release rate. In this work, the auxiliary is chosen as the asymptotic tip analytical solution of mode I and II fractures. Denoting the solution fields (stress, strain and displacements) with superscript $(1)$, and the auxiliary fields under modes I and II with superscripts $(2) I$ and $(2) II$ respectively. For the sake of the computational convenience, the contour integral of \cref{contour_Int} can be converted into an equivalent area integral as (the derivation is shown in \ref{app:jint}),

{\color{black}
	\begin{figure}[!htb]
		\centering
		\includegraphics[width=0.5\textwidth]{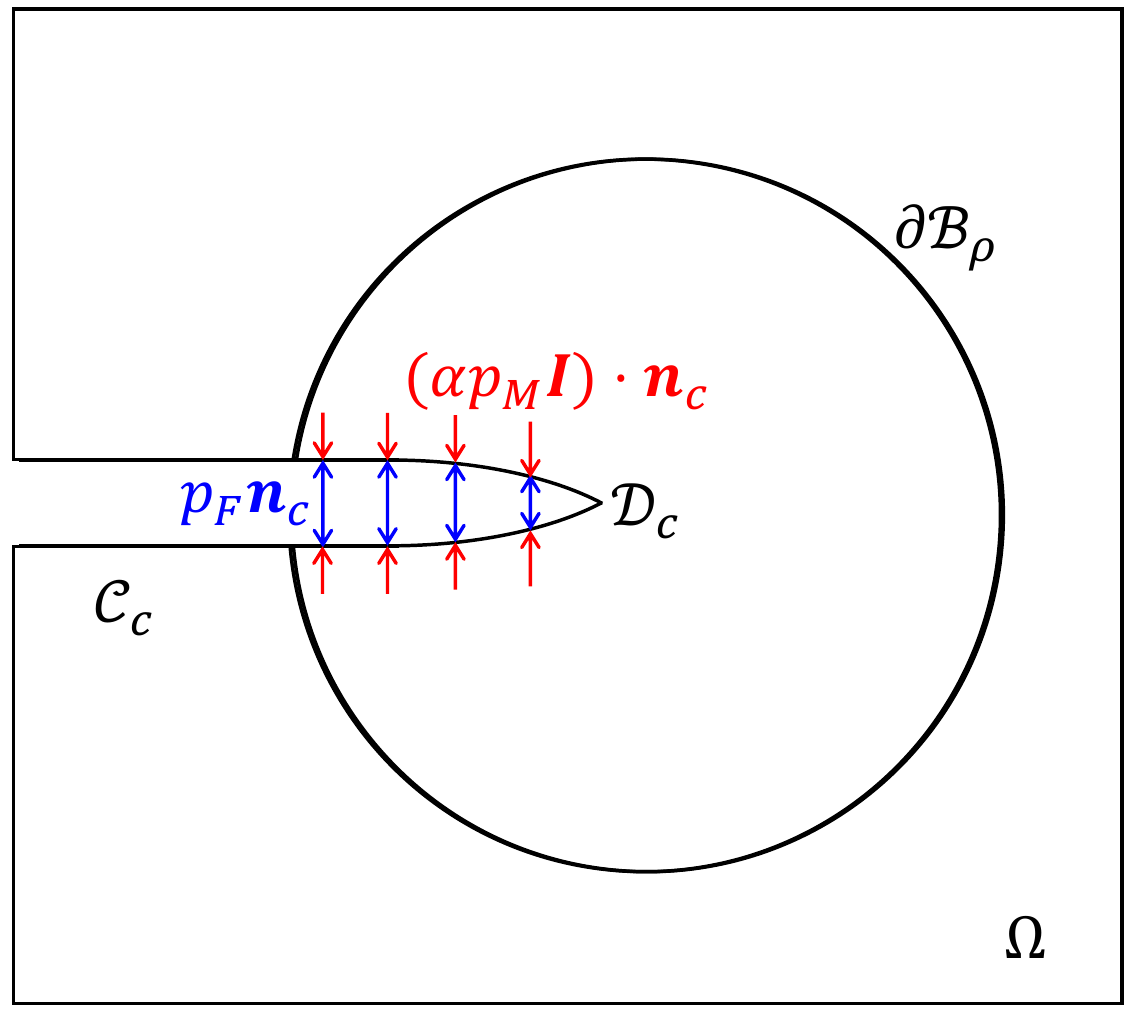}
		\caption{\textcolor{black}{J integral graphical illustration.}}
		\label{JIGI}
	\end{figure}
	\begin{equation}
	I_{I,II} = \int_{\mathcal{B}_{\rho}}\left[ \sigma_{ij}^{(1)}\frac{\partial u_{i}^{(2)I,II}}{\partial x_{1}} + \sigma_{ij}^{(2)I,II}\frac{\partial u_{i}^{(1)}}{\partial x_{1}} - \sigma_{ij}^{(1)}\varepsilon_{ij}^{(2)I, II}\delta_{1j} \right]\frac{\partial q}{\partial x_{j}}d\bm{x} + \int_{\Gamma_{L}} p^{*}_{F}q\frac{\partial u_{2}^{(2)I,II}}{\partial x_1}d\Gamma,
	\label{area_Jint}
	\end{equation}
	where $\Gamma_{L} =(\bm{\gamma}^{+}_{c} \cup \bm{\gamma}^{-}_{c})\cap\mathcal{B}_{\rho} $; $p_{F}^{*}:= p_{F} - \bm{n}^T_c\left(\alpha p_{M}\bm{I}\right)\bm{n}_c = p_{F} - \alpha p_{M}$ is the net pressure applied on the fracture surface. The graphical illustration to the line integral of \cref{area_Jint} is shown in Figure \ref{JIGI}. Along the fracture line, while $p_F$ offers tensile forces to open $\mathcal{C}_c$, the effective pore pressure $\alpha p_{M}$ applies compression to the surface.
}
The weight function $q: \mathcal{B}_{\rho} \rightarrow [0,1] $ is
\begin{equation}
q = \left\{\begin{matrix}
1 & \norm{\bm{x} - \bm{x}_{tip}^{*}} \leqslant \rho  \\ 
0 & \norm{\bm{x} - \bm{x}_{tip}^{*}} > \rho 
\end{matrix}\right..
\end{equation}
The value of scalar field $q$ is depicted in Figure \ref{Fig::Jint}.  In addition to the contribution from the bulk material, the second term in \cref{area_Jint} considers contribution of fracture pressure to the interaction integral. Note that the tensors in \cref{area_Jint} require projection onto the local reference coordinates located at the fracture tip. The transformation matrix $\mathcal{T}$ is defined as,
\begin{equation}
\mathcal{T} = \begin{bmatrix} \cos \theta &   \sin\theta \\ -\sin \theta & \cos \theta \end{bmatrix}, 
\end{equation}
and, the coordinate transform from the global frame $\bm{x}$ to the local frame $\bm{x^{*}}$ is defined as,
\begin{equation}
\bm{x}^{*}=\mathcal{T}(\bm{x} - \bm{x}_{\text{Tip}}).
\end{equation}
Subsequently, the tensor and vector transformation of an arbitrary field $\mathcal{Z}$ and $\mathcal{X}$ between the two coordinates are computed as,
\begin{align}
\mathcal{Z} (\bm{x^{*}}) &= \mathcal{T} \mathcal{Z} (\bm{x})  \mathcal{T}^{T}\\
\mathcal{X} (\bm{x^{*}}) &= \mathcal{T} \mathcal{X} (\bm{x}). 
\end{align}

\begin{algorithm}[!ht]
\caption{SIF\_Calculation}\label{interactionInt}
    \textbf{Output} $K_{I}^{eq}$\\
Define domain $\mathcal{B}_{\rho}$ and calculate scalar field $q$ and transformation matrix $\mathcal{T}$\\
Define two element sets, $\Omega^{\mathcal{B}_{\rho}}$ and $\bm{\gamma}^{\mathcal{B}_{\rho}}_{c}$  that satisfy $\{ \Omega^{\mathcal{B}_{\rho}} | \Omega^{\mathcal{B}_{\rho}} \cap  \partial \mathcal{B}_{\rho} \neq \varnothing \}$ and  $\{ \bm{\gamma}^{\mathcal{B}_{\rho}}_{c} | \bm{\gamma}_{c}^{\mathcal{B}_{\rho}} \cap  \mathcal{B}_{\rho} \neq \varnothing \}$\\
\For {$\Omega^{e} \in \Omega^{\mathcal{B}_{\rho}}$}{
	\For{\normalfont{each quadrature point}  $\bm{x_{G}}$ \normalfont{with a weight} $\mathcal{W}_{G}$ \normalfont{in each elements or sub-triangles}}{
		compute $\bm{\sigma}^{(1)}(\bm{x}^{*})$, $ \frac{\partial \bm{u}^{(1)}}{\partial \bm{x}} (\bm{x}^{*})$ and $\frac{\partial q}{\partial\bm{x}} (\bm{x}^{*})$ in the local coordinate\\
		use mode I or II fracture analytical solutions to get $\sigma_{ij}^{(2)}(\bm{x}^{*})$ and $\varepsilon_{ij}^{(2)}(\bm{x}^{*})$\\
		integrand is calculated $I \gets I + \big(\sigma_{ij}^{(1)}\frac{\partial u_{i}^{(2)}}{\partial x_{1}} + \sigma_{ij}^{(2)}\frac{\partial u_{i}^{(1)}}{\partial x_{1}} - \sigma_{ij}^{(1)}\varepsilon_{ij}^{(2)}\delta_{1j}\big)\partial q / \partial x_{j}\mathcal{W}_{G}$
	}	
}
\For { $\bm{\gamma}_{c}^{e} \in \bm{\gamma}_{c}^{\mathcal{B}_{\rho}}$ } {
\For{\normalfont{each quadrature point} $\bm{x_{G}}$ \normalfont{with a weight} $\mathcal{W}_{G}$}{
	compute $\frac{\partial q}{\partial\bm{x}} (\bm{x}^{*})$ in the local coordinates\\
	$I \gets I + \big(2 q\frac{\partial p^{*}_{F}}{\partial x_{1}}u_{2}^{(2)} + 2 p^{*}_{F}\frac{\partial q}{\partial x_{1}}u_{2}^{(2)}\big)\mathcal{W}_{G}$
}
}
Apply \cref{eqKI} to get $K_{I}^{eq}$
\end{algorithm}

Algorithm~\ref{interactionInt} develops the process applied to perform the numerical approximation. The accuracy of the computed estimate is naturally dependent not only on the quadrature approximation in Algorithm~\ref{interactionInt}, but also on the accuracy of the underlying estimates of the state variables.

\subsection{Fracture geometry updates} \label{fgu}
We consider piece-wise linear approximations to curved fracture paths. A given element may contain multiple fracture segments. To accommodate the XFEM approximation, a modification of the signed-distance function is first introduced, followed by proposed updates to the EDFM discretization.
\subsubsection{XFEM}
\begin{figure}[ht]
	\centering
	\begin{subfigure}[b]{0.49\textwidth}
		\centering
		\includegraphics[width=0.6\textwidth]{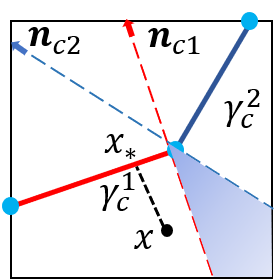}\\
		\caption{ }
		\label{Fig::SignDisXFEM}
	\end{subfigure}
	\begin{subfigure}[b]{0.49\textwidth}
		\centering
		\includegraphics[width=1\textwidth]{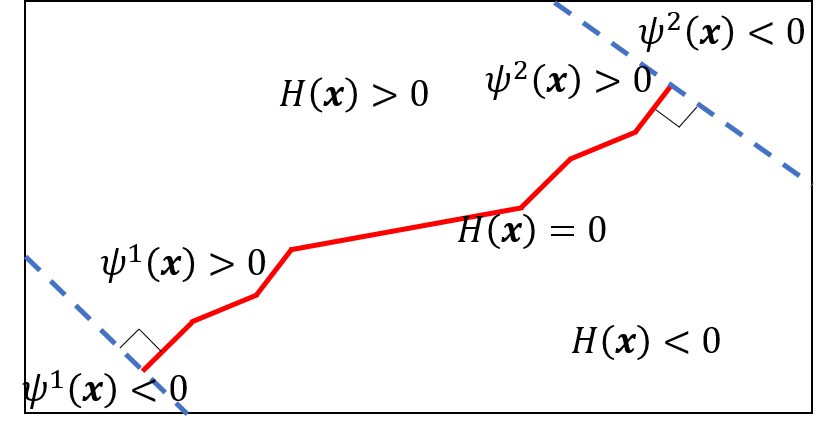}\\
		\caption{}
		\label{Fig::IntXFEM}
	\end{subfigure}
	\caption{A kinked fracture in an element, (a) $\bm{x}_{*}$ is closest point to $\bm{x}$ on the crack, $\bm{e_{1}}, \bm{e_{2}}$ are unit norm vectors of $\bm{\gamma}^{1}_{c}$ and $\bm{\gamma}^{2}_{c}$, (b) a fracture is described by $H(\bm{x}) = 0,\psi^{1}(\bm{x}) > 0 \ \text{and}\  \psi^{2}(\bm{x}) > 0$.}
	\label{XFEMDiscr}
\end{figure}

{\color{black}
In order to constrain the fracture in both perpendicular and lateral directions, functions, $H({\boldsymbol{x}})$, $\psi^{1}(\bm{x})$ and $\psi^{2}(\bm{x})$, are used here.  $\psi^{1}(\bm{x})$ and $\psi^{2}(\bm{x})$ are descriptions of lines in 2-D that are orthogonal to the fracture tip as illustrated in Figure \ref{Fig::IntXFEM}. Subsequent updates of these functions are necessary during FP. 
In a matrix cell containing a piece-wise linear fracture $\mathcal{C}_c$, we rely on a signed normal distance function as follows. Given an arbitrary point $\bm{x}$ in the cell, let $\bm{x}_* \in \mathcal{C}_c$ be the closest point $\bm{x}$ on the fracture. Then the signed normal distance function $\phi(\bm{x})$ is defined as, 
\begin{equation}
\phi(\bm{x}) = \text{sgn} \left(\left(\bm{x} - \bm{x}_{*}\right) \cdot \bm{n}_c\left(\bm{x}_{*}\right)\right) \|\bm{x} - \bm{x}_*\|_2.
\end{equation}
where $\|\text{ }\|_2$ calculates the length of a segment.
As illustrated in Figure \ref{Fig::SignDisXFEM}, this leads in an ambiguity whenever $\bm{x}_*$ coincides with a vertex in the piece-wise linear construct. This is characterized by the condition that $\bm{x}$ lies within the cone formed by the two unit normal vectors. This is illustrated by the shaded region in the figure formed by $\bm{n}_{c1}$ and $\bm{n}_{c2}$. As a convention we apply the choice $\bm{n}_{c1}$ in the signed distance function.

Additionally, the Heaviside function is defined as,
\begin{equation}
H_{\gamma_c}(\bm{x}) := \begin{cases} 
1 & \phi(\bm{x}) > 0 \\ 
0 & \phi(\bm{x}) = 0  \\ 
-1 & \text{otherwise} 
\end{cases}
\end{equation}

}

\subsubsection{EDFM}
\begin{figure}[ht]
	\centering
	\begin{subfigure}[b]{0.35\textwidth}
		\centering
		\includegraphics[width=0.7\textwidth]{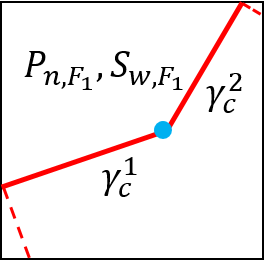}\\
		\caption{ Scheme 1}
	\end{subfigure}
	\begin{subfigure}[b]{0.35\textwidth}
		\centering
		\includegraphics[width=0.7\textwidth]{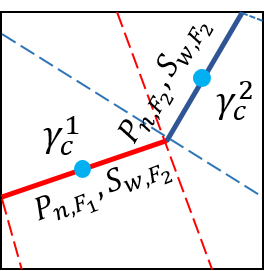}\\
		\caption{Scheme 2}
		\label{scheme2}
	\end{subfigure}
	\caption{Two approaches to model multiple fracture segments in an element; Red and blue solid lines represent fracture segments, dashed lines confine the integration region for $ \langle  d \rangle$ calculations.}
	\label{EDFMDiscr}
\end{figure}

Two possible schemes are proposed here as illustrated in Figure \ref{EDFMDiscr}. In the scheme 1, the two fracture segments ($\bm{\gamma}_{c}^{1}, \bm{\gamma}_{c}^{2}$) are treated as a whole, and hence, there is only one set of degrees of freedom ($p_{\kappa}$ and $S_{\kappa}$) assigned to the grid. The integration region is confined by the dashed lines in the rectangular region, and the average distance between a fracture is computed as,
\begin{equation}
 \langle  d \rangle= \frac{\int_{V_{*}} \norm{ \bm{x} - \bm{x}_{*}}_{2} dv}{V_{*}},
\label{avdDis}
\end{equation}
where $V_{*}$ is the integration region. A constrained triangulation is applied to partition the domain, and numerical integration is performed on each sub triangle. This scheme facilitates computation of the average distance in cases with multiple fracture segments in an element while limiting the total number of degree of freedoms introduced in the system. However, once a new fracture segment is introduced during FP, the properties of the old segment $\bm{\gamma}^{1}_{c} $ has to be mapped to the combined segment ($\bm{\gamma}^{1}_{c} + \bm{\gamma}^{2}_{c}$).

A second approach is to treat $\bm{\gamma}^{1}_{c} $ and $\bm{\gamma}^{2}_{c}$ separately with two distinct sets of dofs as illustrated in Figure \ref{scheme2}. The average normal distances are computed using \cref{avdDis} to each segment. All computational examples in this work apply the second approach.
\subsection{Domain re-initialization \& nonlinear solution}\label{dins}
\begin{figure}[ht]
	\centering
	\begin{subfigure}[b]{0.49\textwidth}
		\centering
		\includegraphics[width=0.6\textwidth]{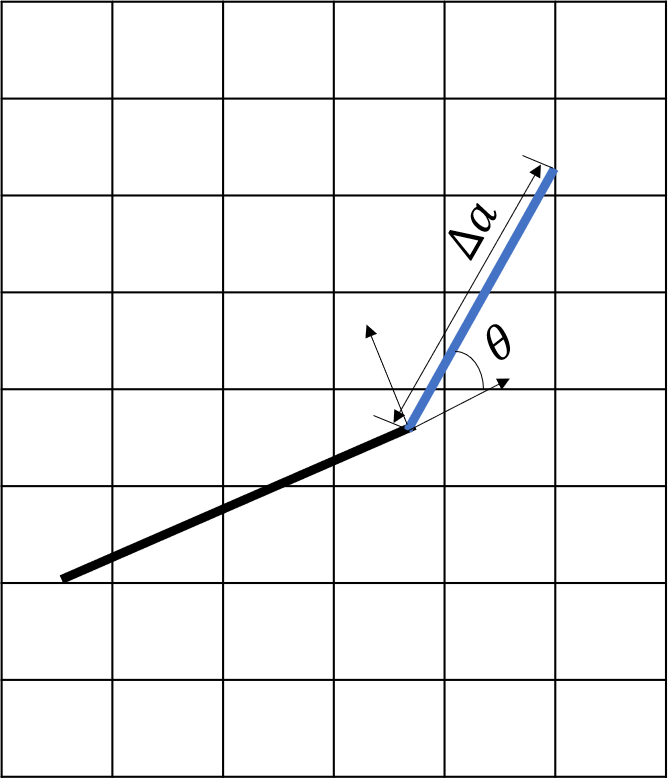}\\
		\caption{ }
		\label{Fig::FPAlgo}
	\end{subfigure}
	\begin{subfigure}[b]{0.49\textwidth}
		\centering
		\includegraphics[width=1.0\textwidth]{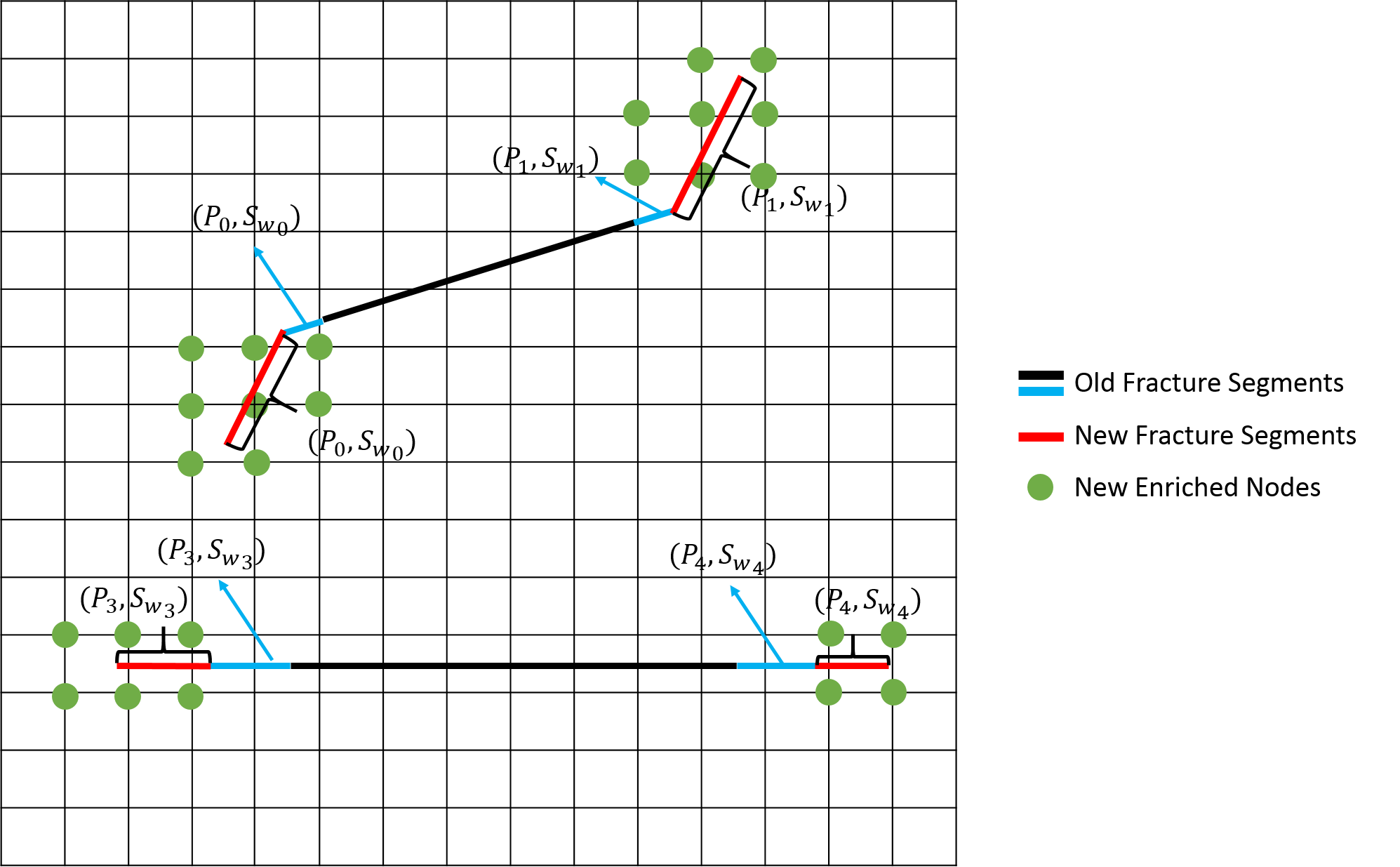}\\
		\caption{}
		\label{Fig::FPInit}
	\end{subfigure}
	\caption{(a) Fracture propagation algorithm schematics; (b) Initialization of newly added fracture segments}
\end{figure}
As illustrated in Figure \ref{Fig::FPAlgo}, once the stress intensity factor exceeds the FP criterion, fracture tips are to be advanced by a fixed advancement length $\Delta a$ along a direction $\theta$. Pressure and saturation initialization within the newly introduced fracture segments is necessary for the solution of the resulting nonlinear algebraic system. As illustrated in Figure \ref{Fig::FPInit}, the pressure and saturation initialization can be performed without any further geometric processing since the connection-list data structure provides constant time look-up of the face and its two connected cells; subsequently the update is,
\begin{equation}\label{UDFlow}
p_{\kappa, F_{j}} := p_{\kappa, F_{adjFF(j)}}.
\end{equation}
Note that, while this update serves as an initial guess after propagation, the initial fluid volume prior to the addition of the segment is assumed to be zero. On the mechanics side, a drained condition computation is performed to assign displacements of the newly introduced enriched nodes. This step is critical since the fracture aperture is directly associated to the displacement of enriched nodes, and a poor initial guess results in severely degraded nonlinear convergence behavior. By fixing $p_{\kappa,M/F}$ and $S_{\kappa,M/F}$, the discretized equations of mechanics in \cref{discrge} result in a linear problem. The algorithm for initialization of new fracture segments $p_{\kappa, M/F}, S_{\kappa, M/F} \text{ and } \bm{u}$ is listed in Algorithm\ref{FPInitAlgo}.

\begin{figure}[ht]
	\centering
	\includegraphics[width=0.7\textwidth]{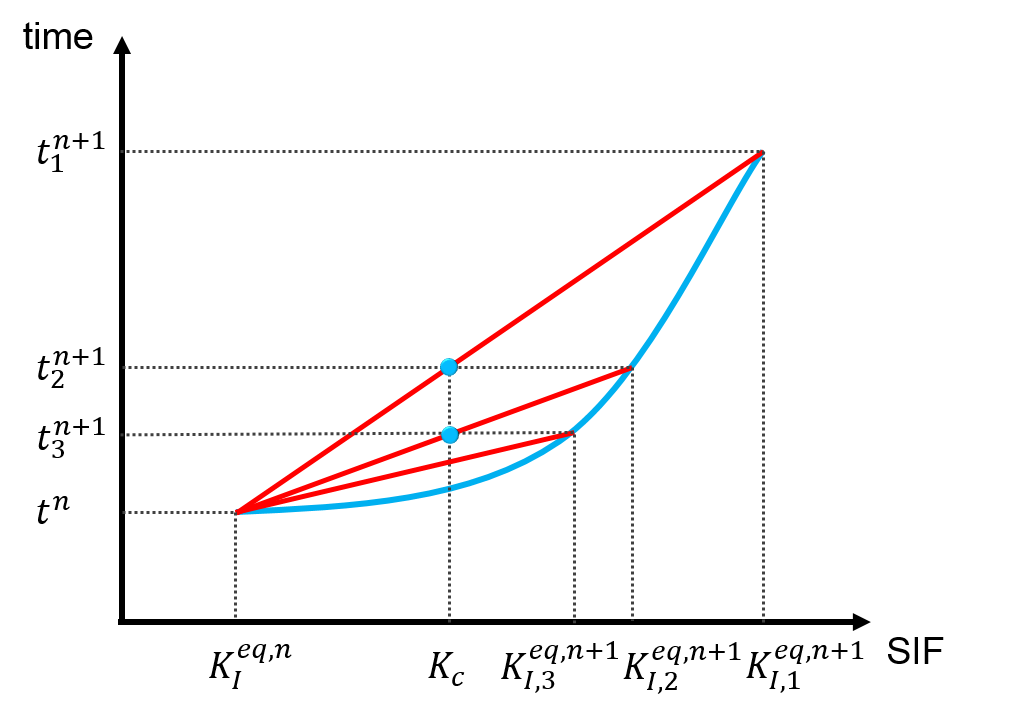}\\
	\caption{Graphical illustration of the time-step selection strategy. Blue curve: the relationship between time and SIF; Red lines: interpolation lines.}
	\label{ATS}
\end{figure}

\begin{algorithm}[H]
	\caption{Initialization}\label{FPInitAlgo}
	Apply \cref{UDFlow} to assign initial value to the newly added $p_{\kappa,F}$ and $S_{\kappa, F}$\\
	Solve \cref{weakform1,weakform2} for $\bm{u}$ by fixing  $p_{\kappa,M/F}$ and $S_{\kappa,M/F}$.
\end{algorithm}
Once the initialization is complete, the nonlinear system at time-step $t^{n+1}$ is solved by means of an Inexact Newton method. After primary unknowns are obtained, a time-step controller will decide the next taken time-step. However, if the propagation condition is violated, a linear interpolation is applied for the calculation of the time-step cut, 
\begin{equation}\label{tscut}
\Delta t := \alpha_{SIF}\frac{K_{c} ( 1 + \epsilon_{SIF})-K_{I}^{eq,old}}{K_{I}^{eq} - K_{I}^{eq,old}}\Delta t,
\end{equation} 
where $\alpha_{SIF}$ is a damping factor, $K_{I}^{eq,old}$ and $K_{I}^{eq}$ are previous and current equivalent SIF respectively, and $K_{c}$ are perturbed by $ 1 + \epsilon_{SIF}$ to ensure the positive coefficient. Figure \ref{ATS} illustrates the chord iterative process of \cref{tscut}. Starting from an initial guess $t_{1}^{n+1}$, the corresponding SIF falling on the blue curve equals $K_{I,1}^{eq,n+1}$ that is over the critical, $K_{c}$. An interpolation is made to find the next time-step $t_{2}^{n+1} - t^{n} = \alpha_{SIF}\frac{K_{c} ( 1 + \epsilon_{SIF})-K_{I}^{eq,n}}{K_{I, 1}^{eq, n+1} - K_{I}^{eq,n}} (t_{1}^{n+1} - t^{n})$. The procedure is performed iteratively until the criterion is satisfied.

The proposed time-step strategy is not only suitable for FP simulation, but also for production processes. For instance, a hydraulic fracturing process followed by long-range fluid flow will lead to a geometric increase in time-step size under the same framework, if $\Delta t_{max}$ is set to the largest time-scale. This occurs since at the transition, the SIF will decrease. On the other hand, under a transition from fluid production to onset of FP, the proposed time-step~\cref{tscut} will geometrically be rescaled thereby avoiding excessive redundant solves.

\section{Results}\label{NE}
The proposed methods are implemented within an in-house software framework~\cite{jiang2016hybrid,jiang2017improved,ren2018coupled}. Validation results illustrating accuracy are presented for benchmark models of elastic mechanics with no fluid flow. This is followed by several numerical examples of propagation with coupled multiphase flow and poromechanics.

\subsection{Model accuracy and KGD verification under elastic mechanics}
The viscosity- and toughness-dominated propagation regimes of the KGD fracture model are considered. The dimensionless toughness variable $\mathcal{K}_{m}$  characterizing these regimes is defined as,
\begin{equation}
    \mathcal{K}_{m} = \frac{K^{'}}{(2E^{'3}\mu ^{'}q)^{1/4}},
\end{equation}
where $K^{'} = 8K_{C} / \sqrt{2\pi}$, $\mu^{'} = 12\mu$, $E'=E/(1-v^{2})$ is the equivalent Young's modulus in the plain strain condition, and $q$ is the flow rate into one wing of the fracture. If $\mathcal{K}_{m} < 1$, the flow lies in the viscosity dominated regime and the solution is called the M-vertex solution when $\mathcal{K}_{m} = 0$. Under this scenario, rock is very brittle ($K_{C}\rightarrow 0$) so that energy dissipation is dominant in viscous flow. On the other hand, if $\mathcal{K}_{m} > 4$, the process is toughness-dominated, and in the limit that $\mathcal{K}_{m}  \rightarrow  \infty$, the solution is called a K-vertex solution. In that case, energy is mostly utilized to break the rock and influences from either small aperture or highly viscous fluid could be neglected. Model instances of the two regimes are created using the parameters listed in Table~\ref{KGDModelProperty}.

\begin{table}[ht]
\caption{Rock and fluid properties for KGD model verification }
\label{KGDModelProperty}
\centering
\begin{tabular}{cccccccc}
\hline
                 & $E$(GPa) & $\tilde{\nu}$ & $K_{c}(\text{MPa}\sqrt{\text{m}})$ & $\epsilon_{SIF}$ & $\mu (\text{Pa}\cdot \text{s})$ & $q(\text{m}^{2}/\text{s})$ & $\mathcal{K}_{m}$ \\ \hline
toughness dominated  &  8.3        & 0.25    & 0.5           & 0.001           &  $1e-8$             & 0.001                 & 16.7                 \\ \hline
viscosity dominated    &  8.3        & 0.25    & 0.1          & 0.001            &  0.001                & 0.001                 & 0.158                  \\ \hline
\end{tabular}
\end{table}

 \begin{figure}[ht]
   \centering
   \includegraphics[width=0.3\textwidth]{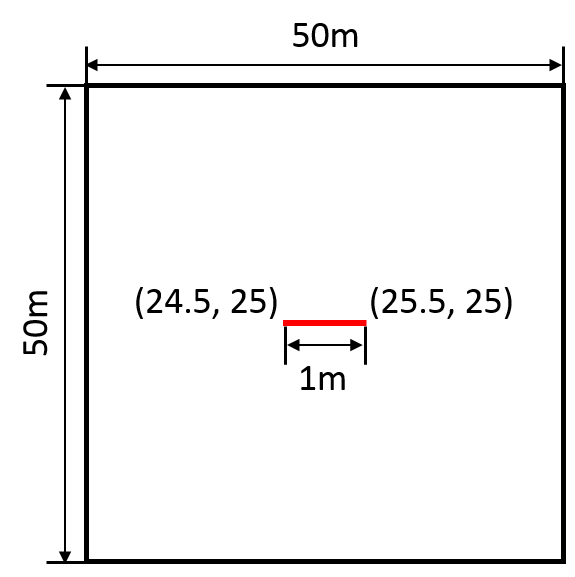}
   \caption{Model domain configuration for KGD verification tests. The red-line represents an initial fracture (not to scale).}
   \label{fig:KGDdomainsetup}
 \end{figure}

We consider the domain depicted in Figure~\ref{fig:KGDdomainsetup}, and in all tests, a uniform Cartesian mesh of dimensions $317 \times 123$ is applied unless noted otherwise. Incompressible fluid is injected into the center point of the fracture at a constant flow rate of $2q$ (refer to Table~\ref{KGDModelProperty}). In order to model the unbounded domain assumed in the analytical reference solutions, the mid-points on the left and right edges of the domain are restricted from displacement in the y-direction, while those on the top and bottom of the domain are restricted in x-direction. The remainder of the boundary is traction free. The initial fracture length ($1m$) is also small enough relative to the simulation domain size. Analytical solutions for either propagation regime are reviewed in~\ref{appendix:analytical}. 

With reference to the analytical solutions, the relative error in propagation length $L$ is,
\begin{equation}
\varepsilon_{L} =  \frac{\mid L(t) - L^{0}(t)\mid}{L^{0}(t)},
\end{equation}
where  $L^{0}$ is analytical solutions. Similarly, the relative error in aperture $\omega_{c}$ at the wellbore is defined as, 
\begin{equation}
\varepsilon_{\omega_{c}} =  \frac{\mid \omega_{c}(t) - \omega_{c}^{0}(t)\mid}{\omega_{c}^{0}(t)},
\end{equation}
where $\omega_{c}^{0}$ is the analytical solution.

\begin{figure}[!htb]
	\centering
       \begin{subfigure}[b]{0.49\textwidth}
		\centering
		\includegraphics[width=0.9\textwidth]{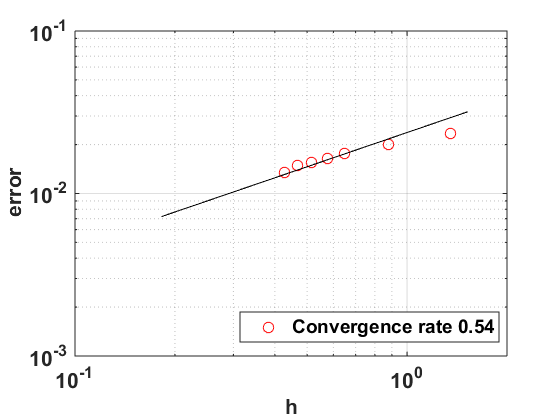}
		\caption{K vertex}
		\label{fig:meshrefine1}
	\end{subfigure}
	\begin{subfigure}[b]{0.49\textwidth}
		\centering
		\includegraphics[width=0.9\textwidth]{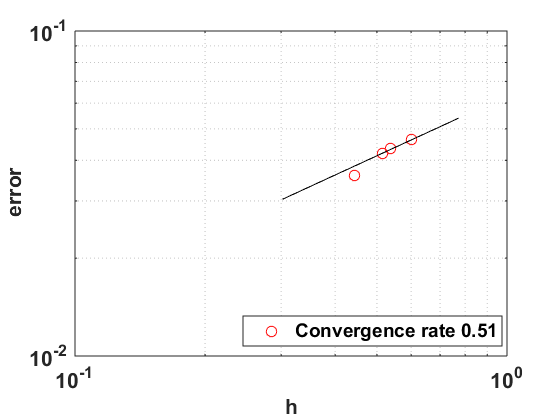}
		\caption{M vertex}
		\label{fig:meshrefine2}
	\end{subfigure}
	\caption{mesh refinement study of the K and M vertex solutions}
\end{figure}

Asymptotic mesh refinement results for the K- and M-vertex cases are presented in Figures~\ref{fig:meshrefine1} and~\ref{fig:meshrefine2} respectively. In these computations, finely resolved temporal and tip-length steps are fixed ($\Delta t = 1e-03 s$, $\epsilon_{SIF} = 2e-05$ for the K vertex case and $\epsilon_{SIF} = 1e-04$ for the M-vertex case) whereas a refinement path along $h$ is studied. The J-integral approximations apply radii of $3h$, and the tip element enrichment scheme is based on the topological geometry where only the element containing the tip is enriched by branch functions. The computed error compares the stable numerical solution for tip length in each simulation with the corresponding analytical solution. Convergence rates in both scenarios are approximately 0.5 order. Note that the propagation scheme relies not only on state variables, but also on the numerical approximation of the equivalent SIF using the converged state variables. Similar convergence rates applied to elastic problems have been reported previously~\cite{paul20183d}. To improve the accuracy of SIF estimate, the fixed-area tip enrichment scheme proposed in~\cite{wang2016numerical} could be applied here.

\subsubsection{Propagation in the toughness dominated regime}
\begin{figure}[!htb]
	\centering
       \begin{subfigure}[b]{0.49\textwidth}
		\centering
		\includegraphics[width=0.9\textwidth]{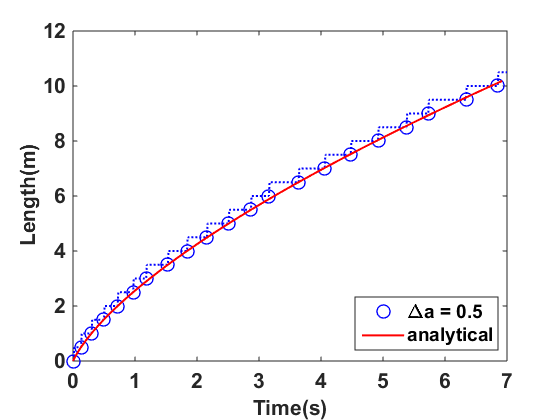}
		\caption{}
		\label{fig:toughnesslength05}
	\end{subfigure}
	\begin{subfigure}[b]{0.49\textwidth}
		\centering
		\includegraphics[width=0.9\textwidth]{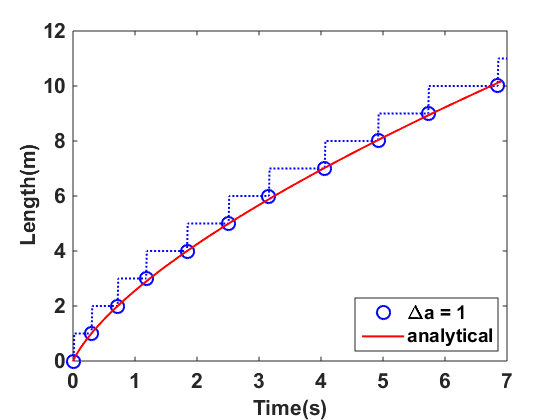}
		\caption{}
		\label{fig:toughnesslength10}
	\end{subfigure}
	\begin{subfigure}[b]{0.49\textwidth}
		\centering
		\includegraphics[width=0.95\textwidth]{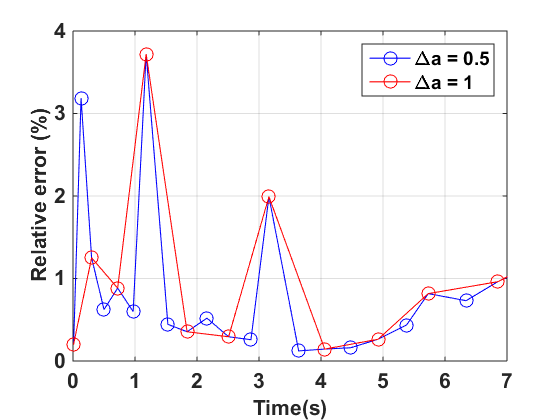}
		\caption{}
		\label{fig:toughnesslengtherror}
	\end{subfigure}
	\begin{subfigure}[b]{0.49\textwidth}
		\centering
		\includegraphics[width=0.9\textwidth]{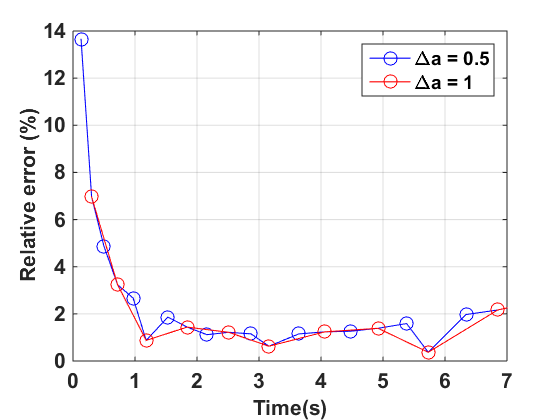}
		\caption{}
		\label{fig:toughnessapertureerror}
	\end{subfigure}
	   \begin{subfigure}[b]{0.49\textwidth}
		\centering
		\includegraphics[width=0.9\textwidth]{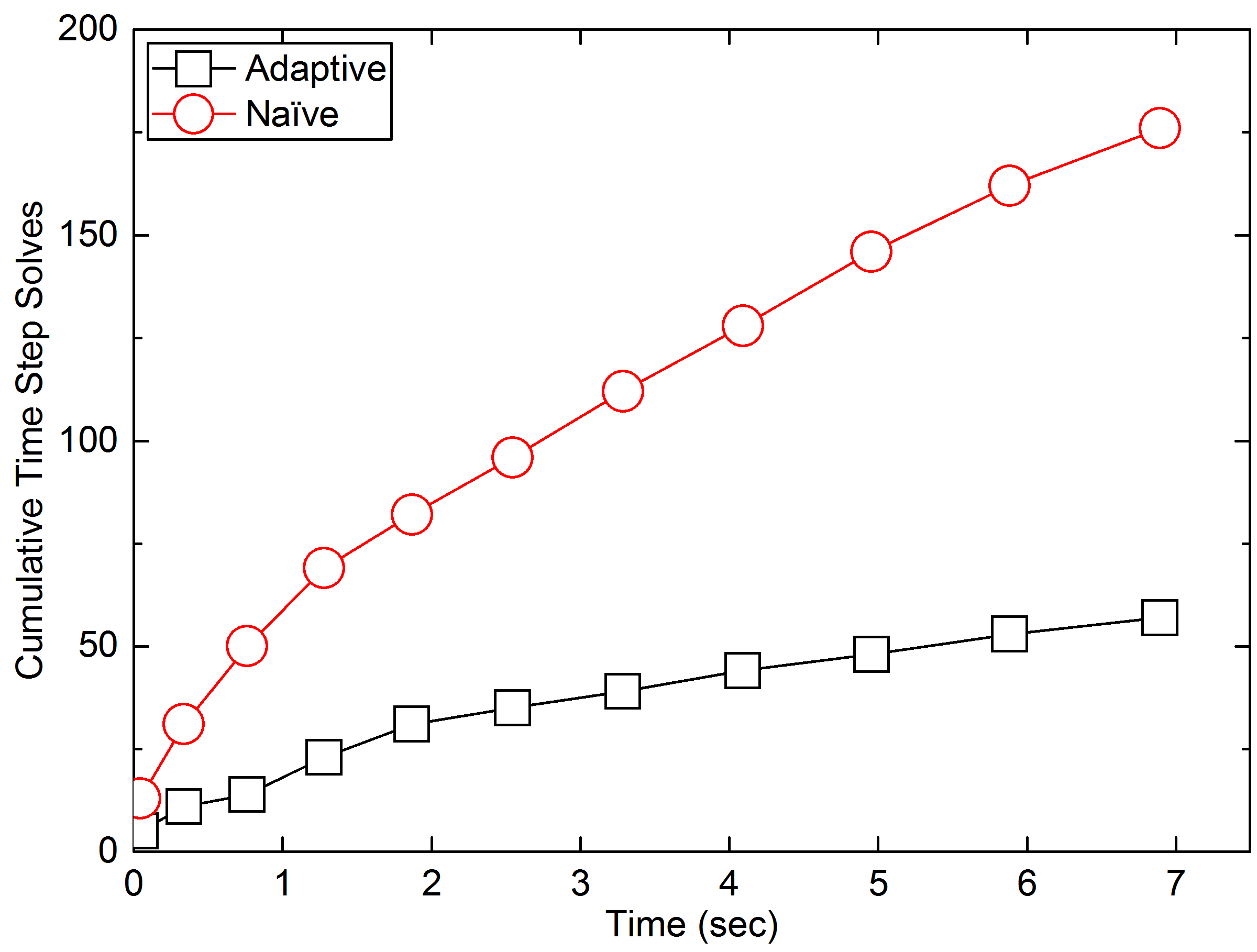}
		\caption{}
		\label{fig:cumulative_timestep_comparison}
	\end{subfigure}
	\begin{subfigure}[b]{0.49\textwidth}
		\centering
		\includegraphics[width=0.9\textwidth]{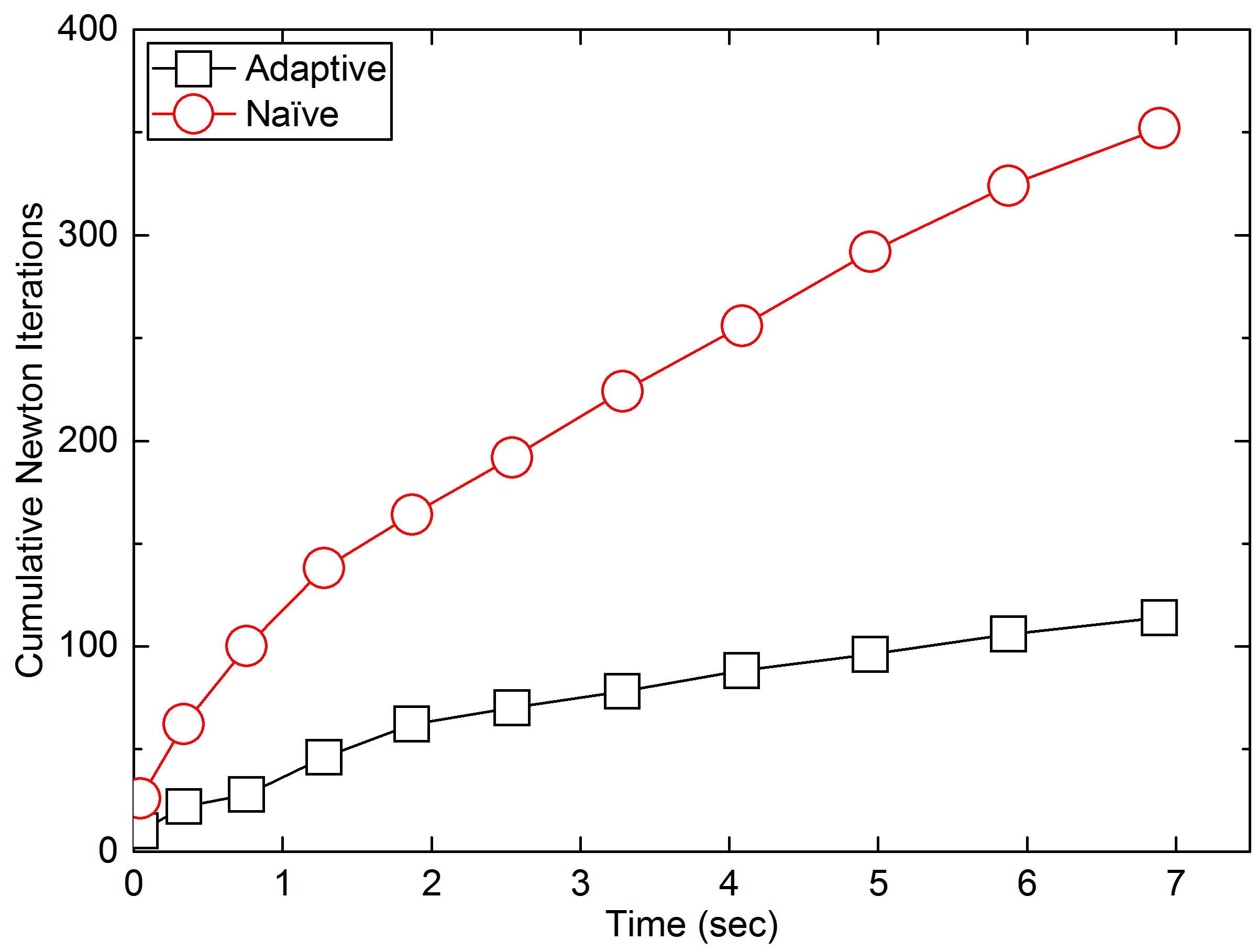}
		\caption{}
		\label{fig:cumulative_newton_comparison}
	\end{subfigure}
	\caption{Simulation results of toughness dominated FP. (a) $\Delta a = 0.5$: fracture length vs. time; (b) $\Delta a = 1.0$: fracture length vs. time; (c) relative error of the aperture at the wellbore. Dashed lines indicate intermediate states while circles are accepted solutions that satisfy the FP criterion. (d) relative error of the propagation length; (e) $\Delta a = 1.0$: cumulative time-step solves vs. time; (f) $\Delta a = 1.0$: cumulative nonlinear iterations vs. time;  }
\end{figure}

Applying the toughness dominated properties listed in Table \ref{KGDModelProperty}, Figures~\ref{fig:toughnesslength05} and~\ref{fig:toughnesslength10} show the time-series of fracture lengths computed (blue) using advancement step-lengths of $\Delta a = 0.5$ and $1.0$ respectively compared to the evolution of the analytical solution (red). Note that in these figures, the circular data markers represent solution-steps at which the propagation is stable as per Algorithm~\ref{CS}, whereas the stair-casing connectors indicate intermediate sub-steps of the algorithms. Clearly, selecting a larger advancement step-size for a given propagation velocity leads to a proportional increase in the time-separation between successive stable solutions. As is reflected by Figures~\ref{fig:cumulative_timestep_comparison} and~\ref{fig:cumulative_newton_comparison}, this comes at no extra cost in terms of computation in this case. Moreover, Figures~\ref{fig:toughnesslengtherror} and~\ref{fig:toughnessapertureerror} present the computed length and aperture relative-error time-series respectively using either step-size. These figures also compare the application of two different time-step algorithms. The Na\"ive scheme increases or cuts the time-step size by a constant factor, whereas the adaptive scheme applies the estimate in Equation~\ref{tscut}. In Figure \ref{fig:cumulative_timestep_comparison}, we compare the cumulative time-step solves over the course of the simulation for two different schemes. In this case, the proposed estimate can reduce computational cost by 70\% percent using $\epsilon_{SIF}=0.001$.

\subsubsection{Verification of propagation in the viscosity dominated regime}
\begin{figure}
	\centering
       \begin{subfigure}[b]{0.49\textwidth}
		\centering
		\includegraphics[width=0.9\textwidth]{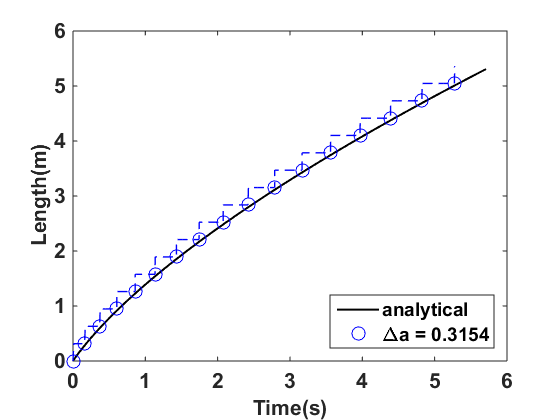}
		\caption{}
		\label{fig:viscouslength}
	\end{subfigure}
       \begin{subfigure}[b]{0.49\textwidth}
		\centering
		\includegraphics[width=0.9\textwidth]{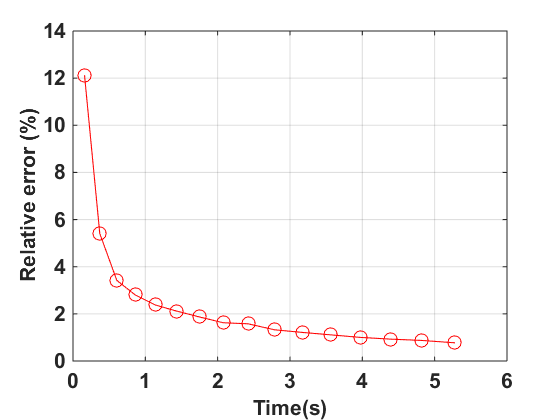}
		\caption{}
		\label{fig:viscouslengtherror}
	\end{subfigure}
	\begin{subfigure}[b]{0.49\textwidth}
		\centering
		\includegraphics[width=0.9\textwidth]{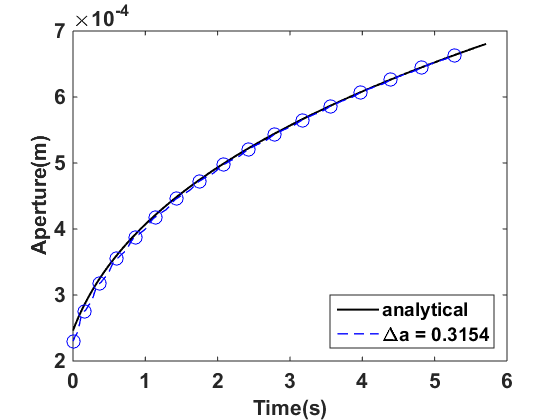}
		\caption{}
		\label{fig:viscousaperture}
	\end{subfigure}
	\begin{subfigure}[b]{0.49\textwidth}
		\centering
		\includegraphics[width=0.9\textwidth]{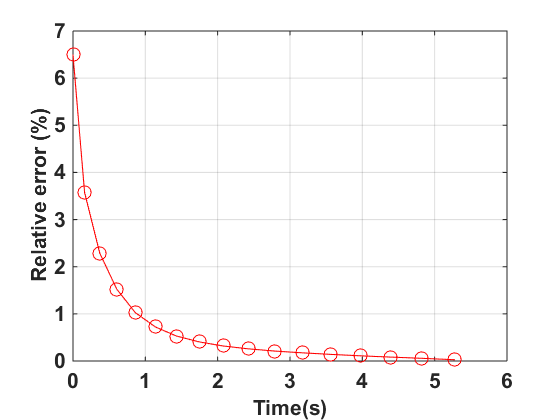}
		\caption{}
		\label{fig:viscousapertureerror}
	\end{subfigure}
	\begin{subfigure}[b]{0.49\textwidth}
		\centering
		\includegraphics[width=0.95\textwidth]{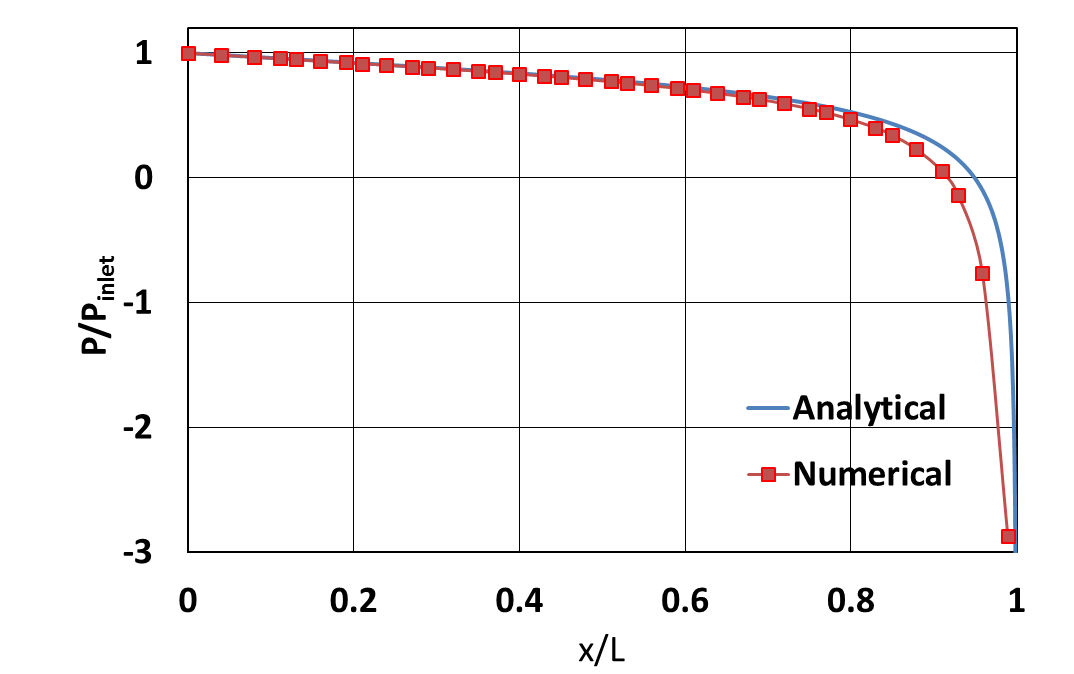}
		\caption{}
		\label{fig:viscousP}
	\end{subfigure}
	\caption{Simulation results of viscosity dominated FP. (a) fracture length vs. time; (b) relative error of the fracture length; (c) aperture at the wellbore vs. time; (d) relative error of the aperture at the wellbore; (e) pressure profile at FP length = 3.95; (f) Number of Nonlinear iterations at first 1 sec period time. Dashed lines indicate intermediate states while circles are accepted solutions that satisfy the FP criterion.}
\end{figure}

The computational performance and relative error obtained in the viscosity-dominated scenario are similar to those of the toughness-dominated case. They are summarized by Figures~\ref{fig:viscouslength} through~\ref{fig:viscousapertureerror}. The pressure profile along the fracture at a fixed time is extracted and plotted in Figure \ref{fig:viscousP}, where the axes are normalized by the value of inlet pressure and current length (3.95m). The computed pressure profile matches the analytical solution very well within the interior segment of the fracture, away from the regions near the tips. The analytical solution dictates that pressure exhibits a singular behavior near the tip, and obeys an asymptotic function $\mathcal{O}((1-\frac{x}{L})^{-1/3})$ in the M vertex solution (\cite{adachi2001}). In the numerical model, the pressure behavior is not explicitly constrained by this asymptotic function and linear interpolation is used instead.

\subsection{Flow, mechanics, and fracture in rock.}
A model problem is considered where an initial horizontal fracture of length $3 \text{ m}$ is centered within a square domain of sides $50 \text{ m}$ in length. The matrix permeability is $1e-15 \text{m}^{2}$ and porosity is 0.1. Two phase fluid flow in the porous media and fracture are considered; fluid density and viscosity are assumed to be equal (1000 $\text{kg/}\text{m}^{3}$) with isothermal compressibility of  $4.3e
9\text{Pa}^{-1}$ and $0.001$ $\text{Pa}\cdot\text{s}$, respectively. Corey-Brook relative permeability is applied to flow within the matrix, and linear models to flow within fracture. Initially, the pressure is $p_{M}=5 \text{ MPa}$; water saturation is $S_{w}=0.6$; the Biot coefficient is $\alpha = 0.8$; fracture toughness is $K_{c}=4e7 \text{Pa}\sqrt{\text{m}}$; and the tolerance is set at $\epsilon_{SIF}=$5\%. Water is injected into the center of the fracture at a rate of $0.01 \text{m}^3/\text{s}$. 

\begin{figure}[htb]
	\centering
	\begin{subfigure}[b]{0.45\textwidth}
		\centering
		\includegraphics[width=0.7\textwidth]{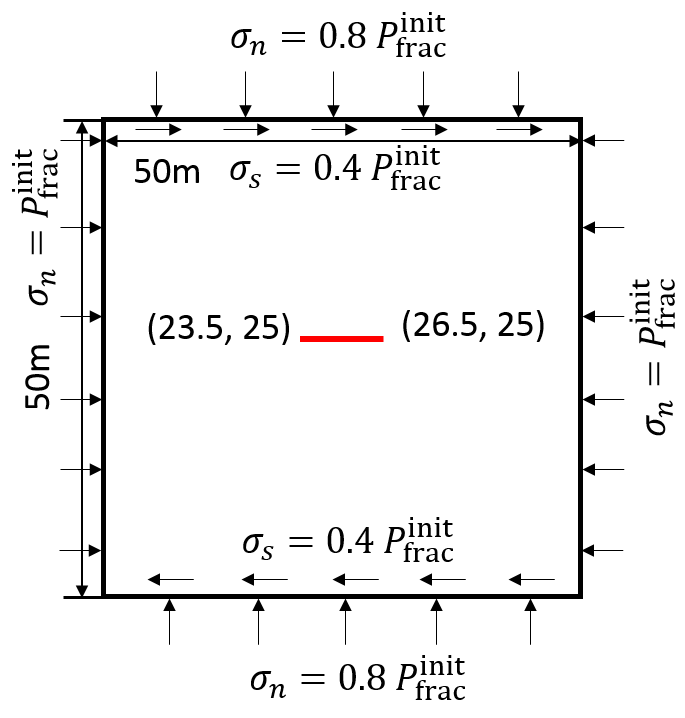}
		\caption{Mixed Mode type 1 (MMT-1)}
		\label{case4InitialPattern1}
	\end{subfigure}
	\begin{subfigure}[b]{0.45\textwidth}
		\centering
		\includegraphics[width=0.7\textwidth]{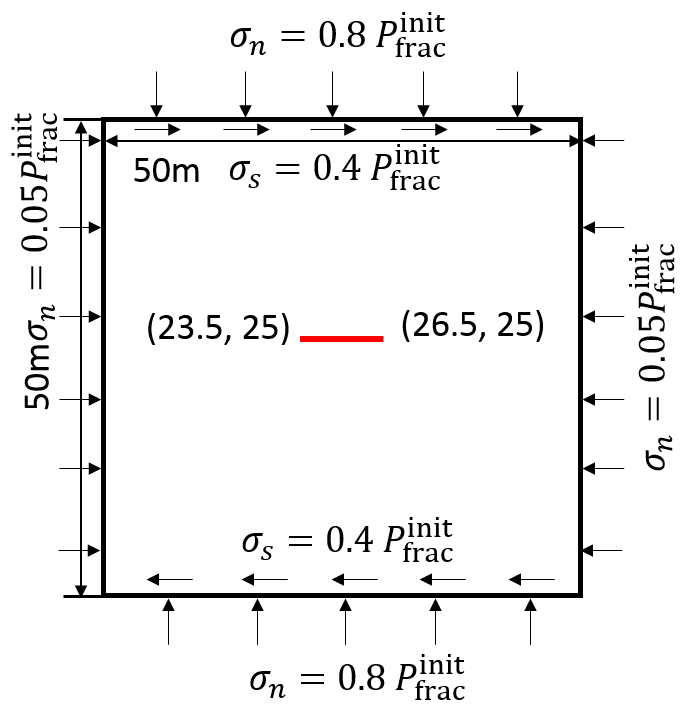}
		\caption{Mixed mode type 2 (MMT-2)}
		\label{case4InitialPattern2}
	\end{subfigure}
	\caption{Initial fracture configuration and boundary conditions, $y$ displacement is fixed at the mid point of left and right edges and $x$ displacement is fixed at the mid point of  top and bottom edges. }
	\label{fig:case4fractureSchematics}
\end{figure}

Two variations of the model problem are considered: mixed-mode type 1 (MMT-1) as depicted in Figure~\ref{case4InitialPattern1}, and mixed-mode type 2 (MMT-2) as depicted in Figure~\ref{case4InitialPattern2}. Under MMT-1 or 2, the shear stress will re-direct the fracture path along a direction that is perpendicular to the local maximum tensile stress. {\color{black}``Mixed mode" refers to load types including both opening and in-plane shear (\cite{mechanics1995fundamentals}). the difference between MMT-1 and MMT-2 is the magnitude of forces applied to the left and right boundaries.

In both MMT-1 and MMT-2, the shear force applied to the boundary alters the direction of the principal stresses and hence the FP path does not follow a straight line. Due to the different stress magnitudes acting on the left and right boundaries in the cases MMT-1 and MMT-2, the curvature of the path manifests differently.
}

\subsubsection{Empirical consistency tests}
The asymptotic convergence of the predicted fracture path is studied with reference to $\Delta a$ and $h$. In particular, defining a relative error as,
\begin{equation}
\varepsilon_{c} =  \frac{\left\Vert Y(x) - Y^{0}(x)\right\Vert _2}{\left\Vert Y^{0}(x)\right\Vert_2},
\end{equation}
where the mapping function $Y:\bm{\mathcal{R}}^+ \rightarrow \bm{\mathcal{R}}^+$ takes a vector of $x$ coordinates of reference points and yields a vector of $y$ coordinates of the same group of points on the propagation path. The reference state, $Y^{0}$ is selected as the prediction using the finest mesh in the refinement paths listed in Table~\ref{sensitivityParams}, where the minimum $\Delta a$ chosen on the finest mesh size $h$ is larger or equal to the radius $r$ of $\mathcal{B}_{\rho}$ to improve the accuracy of J-integral approximation.

\begin{table}[!htb]
\centering
\caption{Sensitivity analysis: parameters of mesh sizes ($h$) and propagation length step ($\Delta a$) in porous media}
\label{sensitivityParams}
\begin{tabular}{cccc}
\hline
\multirow{2}{*}{Sensitivity Tests} & \multirow{2}{*}{Meshes} & \multicolumn{2}{c}{$\Delta a$}       \\ \cline{3-4} 
                                   &                         & MMT-1 & MMT-2 \\ \hline
\multirow{4}{*}{$\Delta a$}        & 187 $\times$ 173        & 4.4               & 2.2              \\ \cline{2-4} 
                                   & 187 $\times$ 173        & 2.2               & 1.1              \\ \cline{2-4} 
                                   & 187 $\times$ 173        & 1.1               & 0.55             \\ \cline{2-4} 
                                   & 187 $\times$ 173        & 0.55              & 0.275            \\ \hline
\multirow{4}{*}{$h$}              & 37 $\times$ 37          & 2.2               & 2.2              \\ \cline{2-4} 
                                   & 87 $\times$ 83          & 2.2               & 2.2              \\ \cline{2-4} 
                                   & 137 $\times$ 127        & 2.2               & 2.2              \\ \cline{2-4} 
                                   & 187 $\times$ 173        & 2.2               & 2.2              \\ \hline
\end{tabular}
\end{table}

\begin{figure}[htb]
	\centering
	\begin{subfigure}[b]{0.45\textwidth}
	\includegraphics[width=1\textwidth]{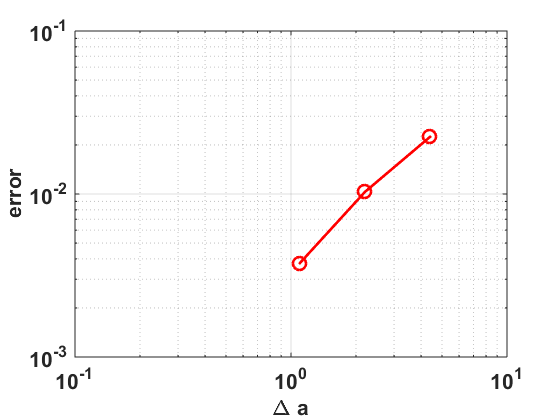}
	\caption{}
	\label{fig:ss_case4_error1}
	\end{subfigure}
	\begin{subfigure}[b]{0.45\textwidth}
	\includegraphics[width=1\textwidth]{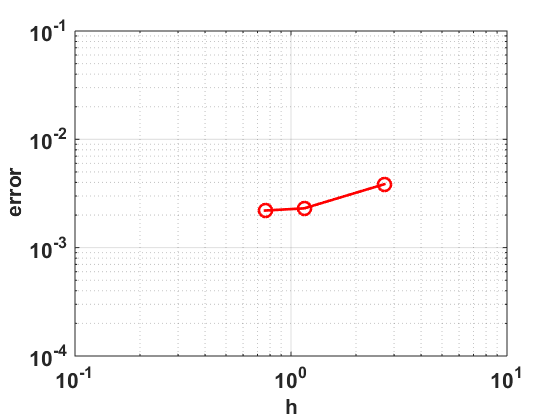}
	\caption{}
	\label{fig:ss_case4_error2}
	\end{subfigure}
	\begin{subfigure}[b]{0.45\textwidth}
	\includegraphics[width=1\textwidth]{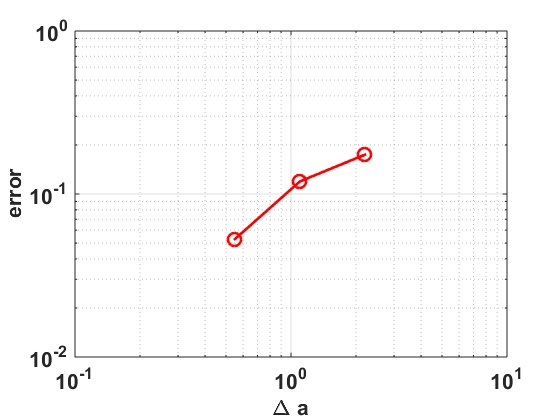}
	\caption{}
	\label{fig:ss_case4_error3}
	\end{subfigure}
	\begin{subfigure}[b]{0.45\textwidth}
	\includegraphics[width=1\textwidth]{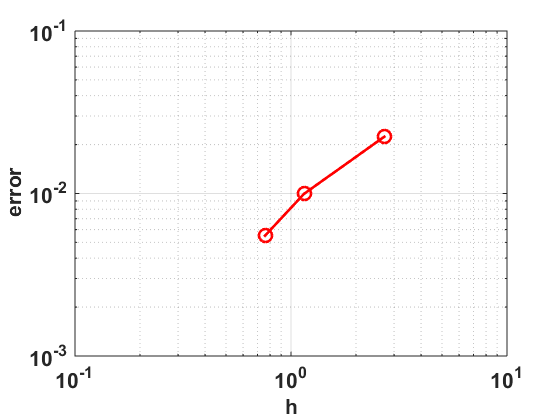}
	\caption{}
	\label{fig:ss_case4_error4}
	\end{subfigure}
	\caption{(a) $\Delta a$ consistency test in MMT-1; (b) $h$ consistency test in mixed MMT-1; (c) $\Delta a $ sensitivity test in mixed MMT-2; (d) $h$ sensitivity test in MMT-2.}
	\label{fig:ss_case4_error}
\end{figure}

As we can see from Figures \ref{fig:ss_case4_error1}, \ref{fig:ss_case4_error2}, \ref{fig:ss_case4_error3} and \ref{fig:ss_case4_error4}, all scenarios exhibit asymptotic decay in the relative error with respect to $\Delta a$ and $h$. Note that asymptotic decay in relative error is not sensitive to the higher curvature of the path in MMT-2 relative to MMT-1 in these cases.

Figures~\ref{fig:ss_case4_FPPath1} and~\ref{fig:ss_case4_FPPath2} present the paths and fracture length evolutions obtained for cases MMT-1 and MMT-2 respectively. Note that in the MMT-2 test, the initial fracture surface is not perpendicular to the minimum horizontal stress that is 0.05 times of the initial reservoir pressure $p_{i}$. This is intentionally designed to investigate the sensitivity to a highly curved fracture path. As expected, the fracture path exhibits a larger curvature than that in MMT-1. Despite distinct FP path curvature present in Figure \ref{fig:ss_case4_FPPath3}, differences in paths reduce with $\Delta a$. Therefore, in order to track the FP path closely, a finer mesh $h$ and smaller $\Delta a$ are required. The FP speed reflected in Figure \ref{fig:ss_case4_FPSpeed4} is not affected by $\Delta a$ significantly except that the speed associated with $\Delta a = 2.2$ becomes slower at late times. Mesh refinement $h$ does not have an significant impact on the propagation path and its speed shown from Figure \ref{fig:ss_case4_FPPath4} and \ref{fig:ss_case4_FPSpeed4}.
{\color{black}
	\begin{figure}[!htb]
		\centering
		\begin{subfigure}[b]{0.49\textwidth}
			\includegraphics[width=0.9\textwidth]{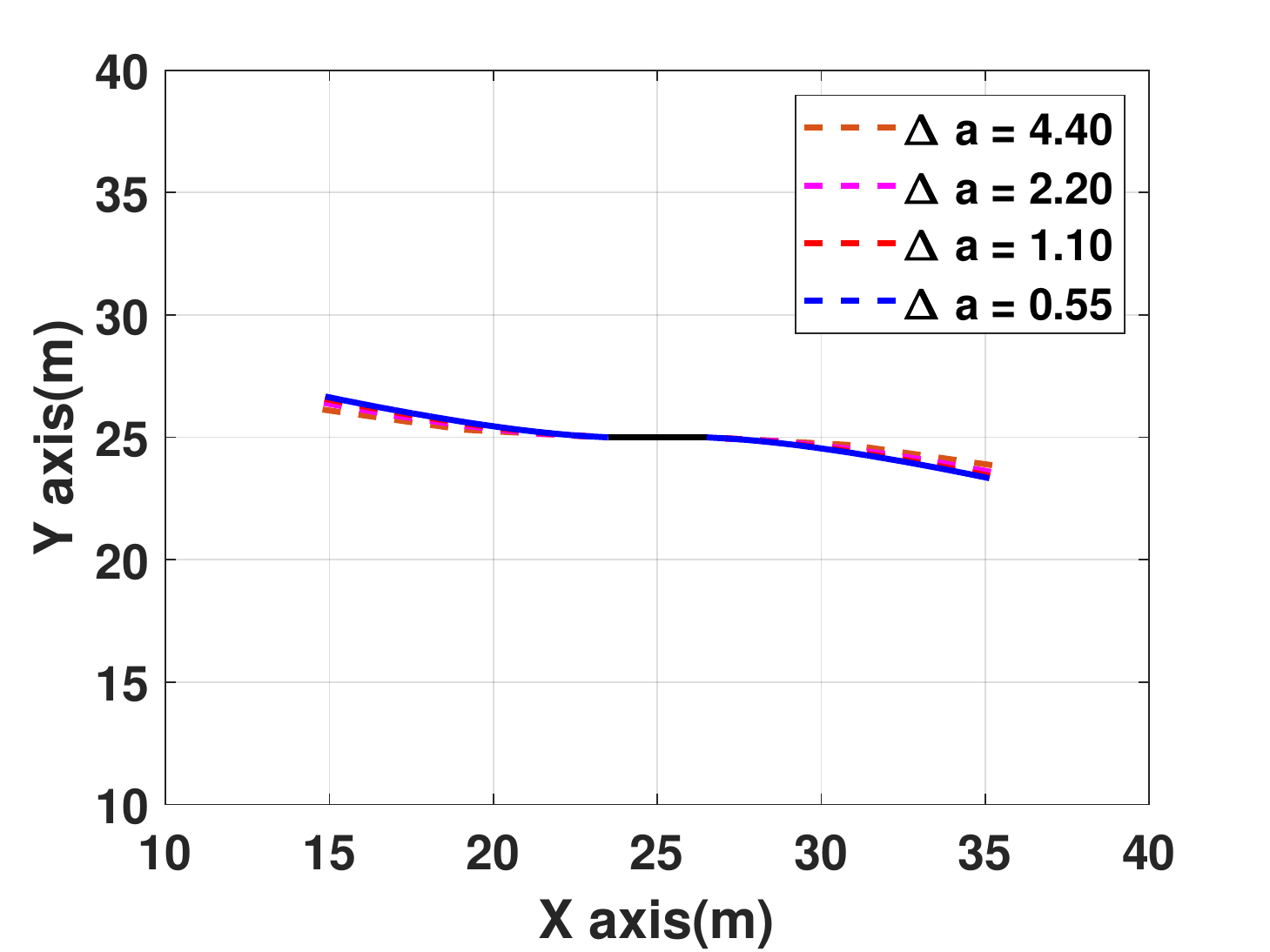}
			\caption{}
			\label{fig:ss_case4_FPPath1}
		\end{subfigure}
		\begin{subfigure}[b]{0.49\textwidth}
			\includegraphics[width=0.9\textwidth]{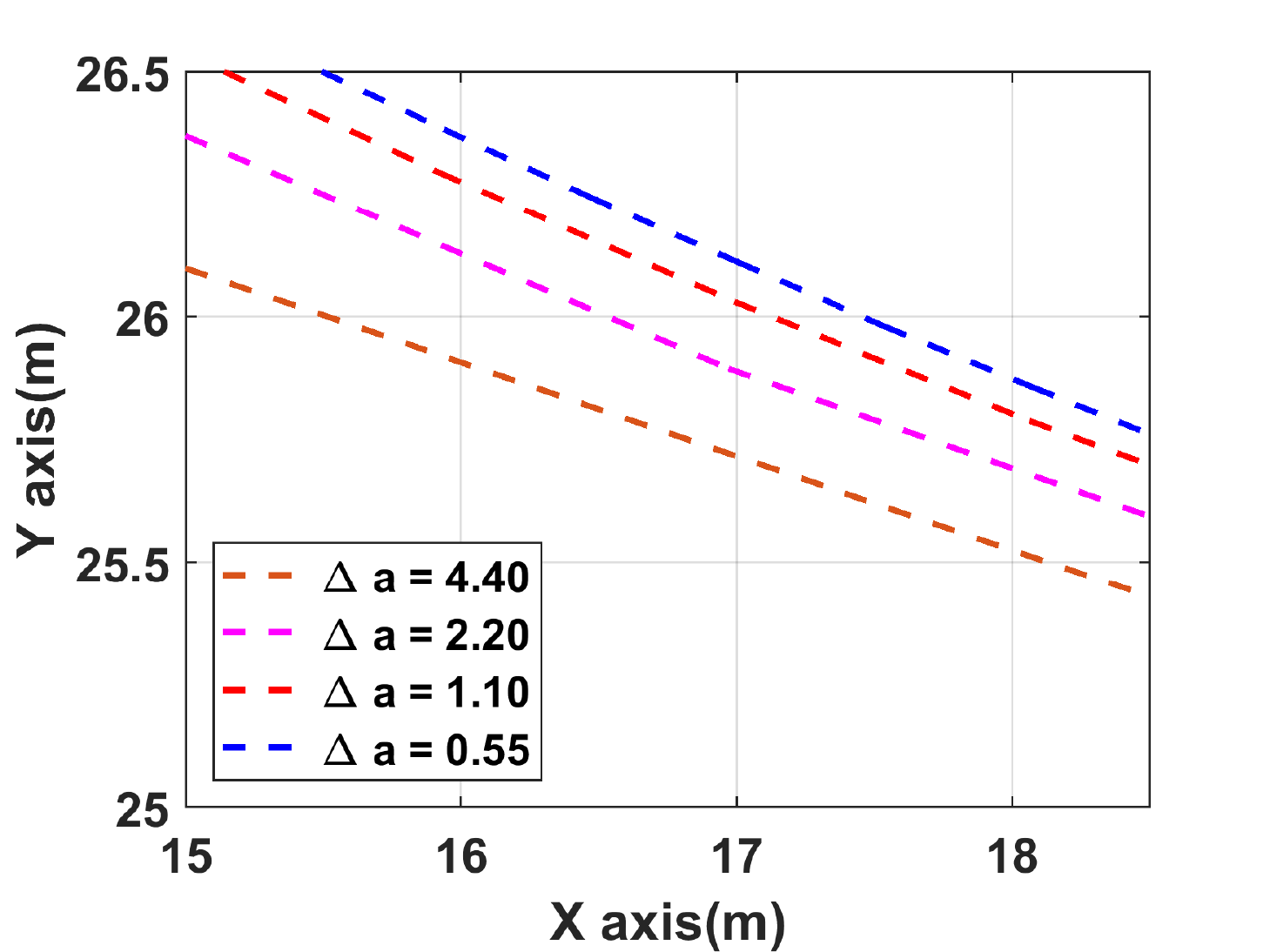}
			\caption{}
			\label{fig:ss_case4_FPPath1_zoom}
		\end{subfigure}

		\begin{subfigure}[b]{0.49\textwidth}
			\includegraphics[width=0.9\textwidth]{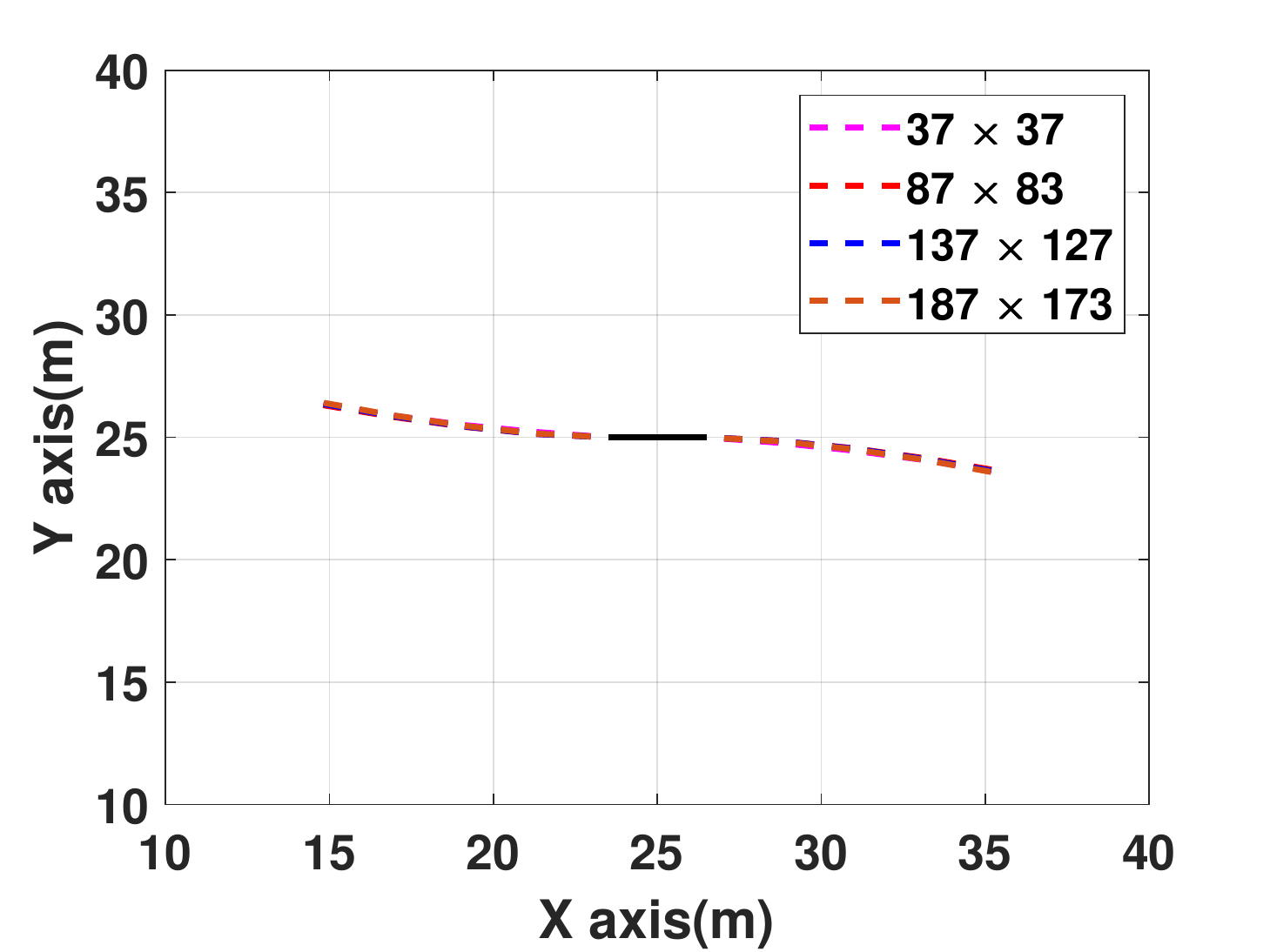}
			\caption{}
			\label{fig:ss_case4_FPPath2}
		\end{subfigure}
			\begin{subfigure}[b]{0.49\textwidth}
		\includegraphics[width=0.9\textwidth]{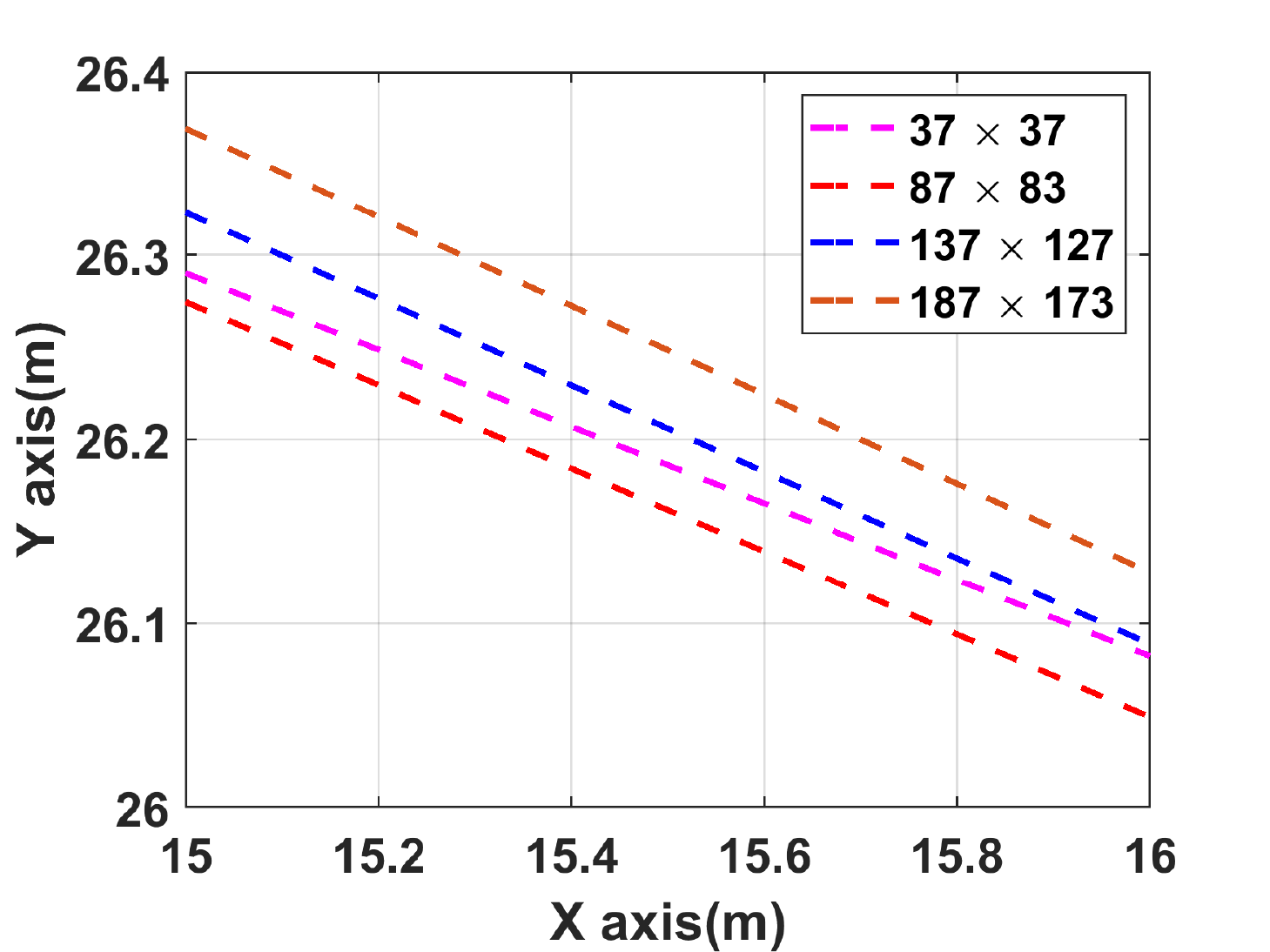}
		\caption{}
		\label{fig:ss_case4_FPPath2_zoom}
	\end{subfigure}
		\begin{subfigure}[b]{0.49\textwidth}
			\includegraphics[width=0.9\textwidth]{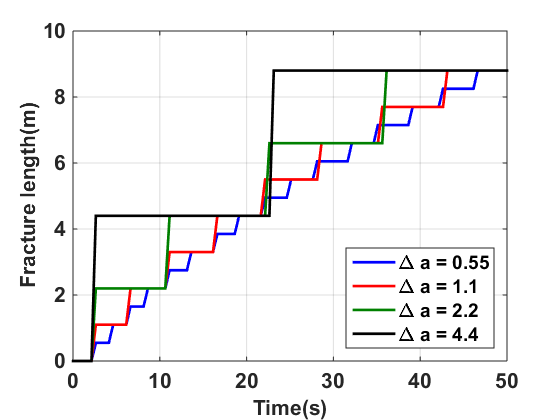}
			\caption{}
			\label{fig:ss_case4_FPSpeed1}
		\end{subfigure}
		\begin{subfigure}[b]{0.49\textwidth}
			\includegraphics[width=0.9\textwidth]{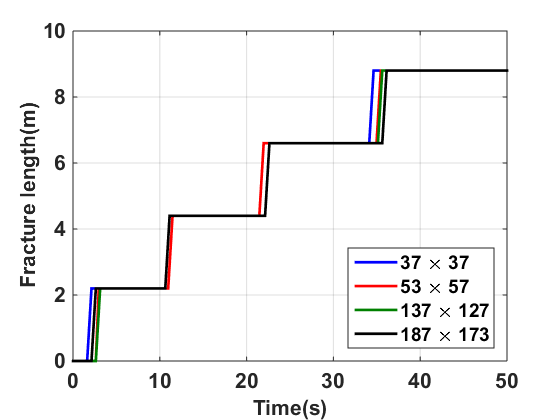}
			\caption{}
			\label{fig:ss_case4_FPSpeed2}
		\end{subfigure}
		\caption{Left column: fracture patterns at $t = 50 \text{s}$; right column: fracture length vs. time. (a), (b) and (e) $\Delta a$ sensitivity tests on MMT-1; (c), (d) and (f) $h$ sensitivity tests on MMT-1. (b) and (d) are zoomed counterparts of (a) and (c), respectively.}
	\end{figure}

\begin{figure}[!htb]
	\centering
	\begin{subfigure}[b]{0.49\textwidth}
		\includegraphics[width=0.9\textwidth]{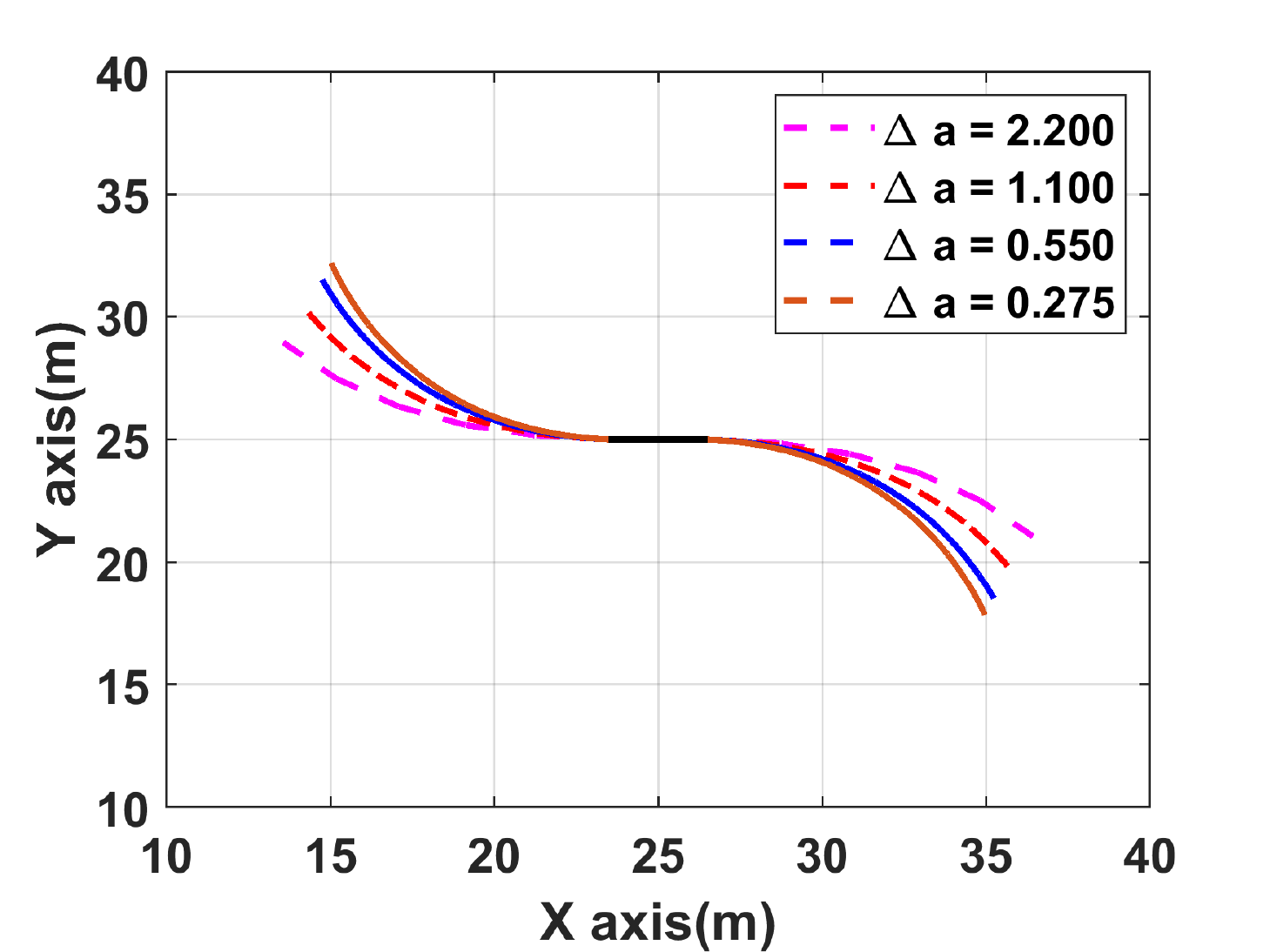}
		\caption{}
		\label{fig:ss_case4_FPPath3}
	\end{subfigure}
	\begin{subfigure}[b]{0.49\textwidth}
		\includegraphics[width=0.9\textwidth]{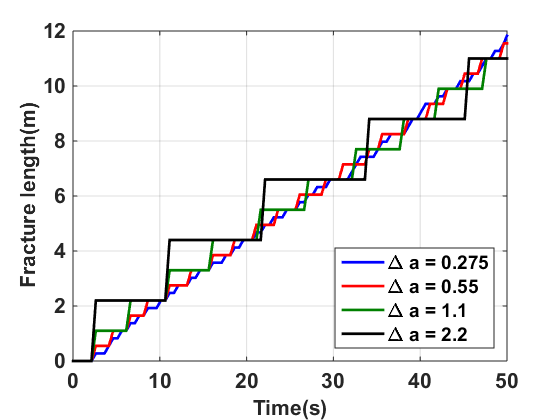}
		\caption{}
		\label{fig:ss_case4_FPSpeed3}
	\end{subfigure}
	\begin{subfigure}[b]{0.49\textwidth}
		\includegraphics[width=0.9\textwidth]{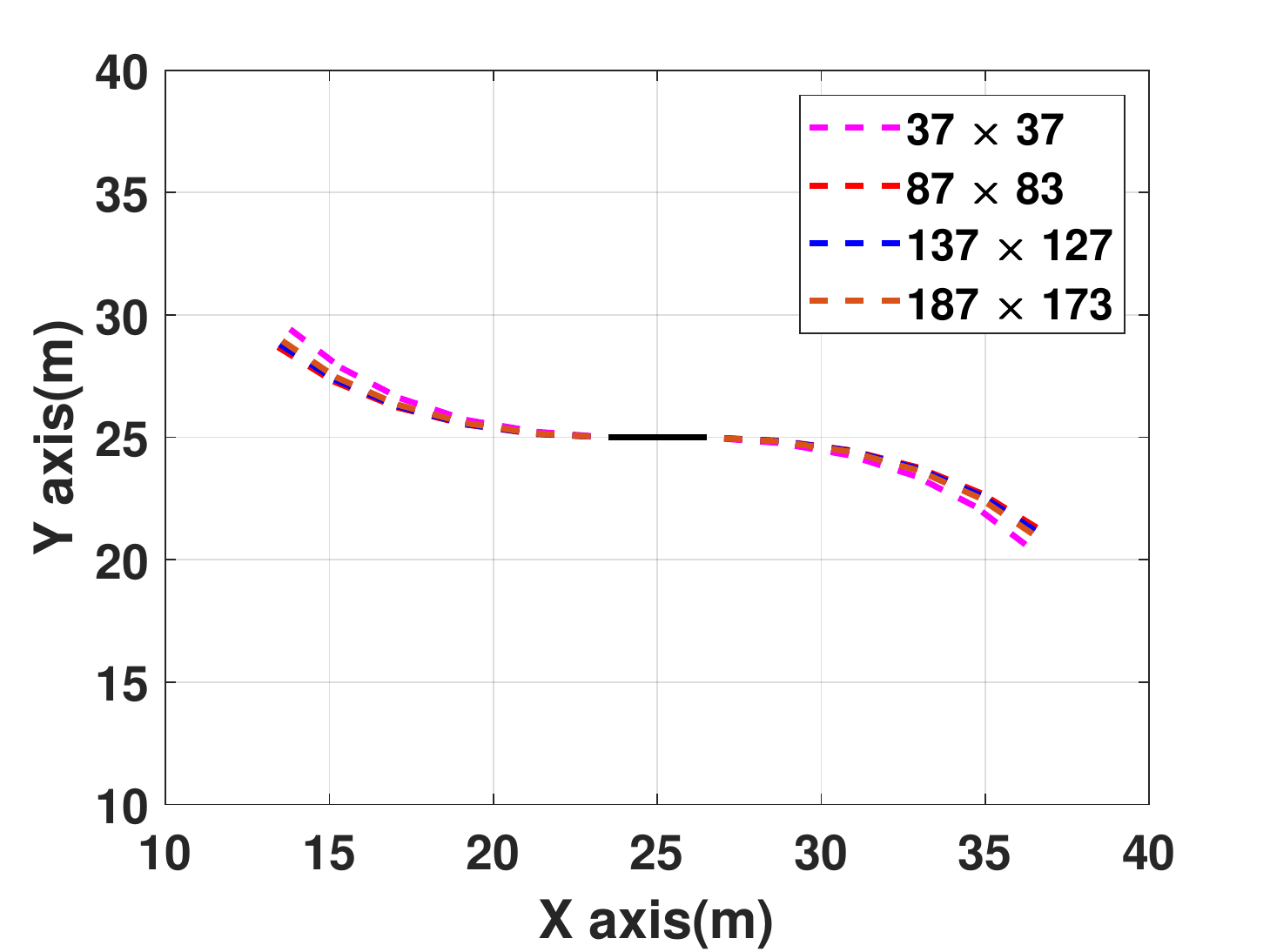}
		\caption{}
		\label{fig:ss_case4_FPPath4}
	\end{subfigure}
	\begin{subfigure}[b]{0.49\textwidth}
	\includegraphics[width=0.9\textwidth]{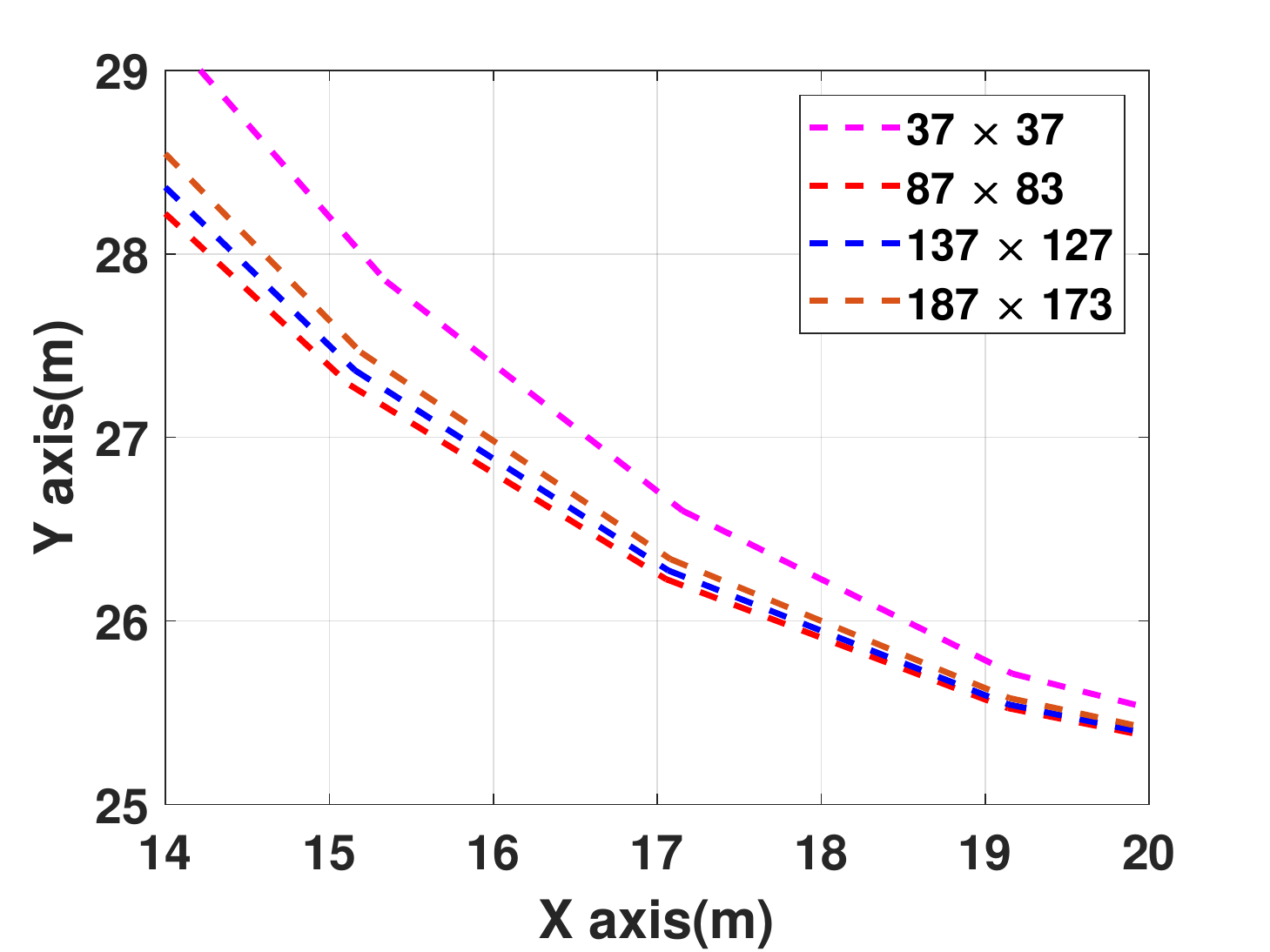}
	\caption{}
	\label{fig:ss_case4_FPPath4_zoom}
\end{subfigure}
	\begin{subfigure}[b]{0.49\textwidth}
		\includegraphics[width=0.9\textwidth]{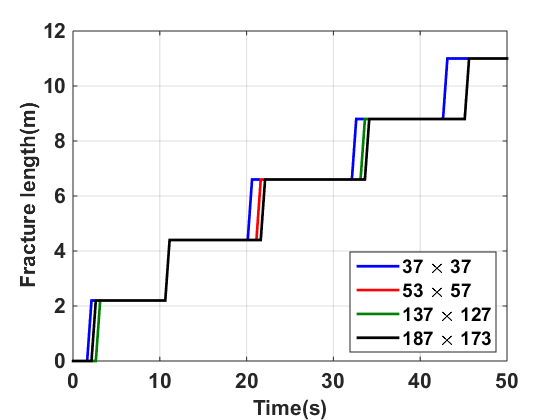}
		\caption{}
		\label{fig:ss_case4_FPSpeed4}
	\end{subfigure}
	\caption{Left column: fracture patterns at $t = 50 \text{s}$; right column: fracture length vs. time. (a) and (b) $\Delta a$ sensitivity tests on MMT-2; (c), (d) and (e) $h$ sensitivity tests on MMT-2. (d) is the zoomed counterparts of (c).}
\end{figure}
}

\begin{figure}[!htb]
	\centering
	\begin{subfigure}[b]{0.24\textwidth}
		\centering
		\includegraphics[width=0.8\textwidth]{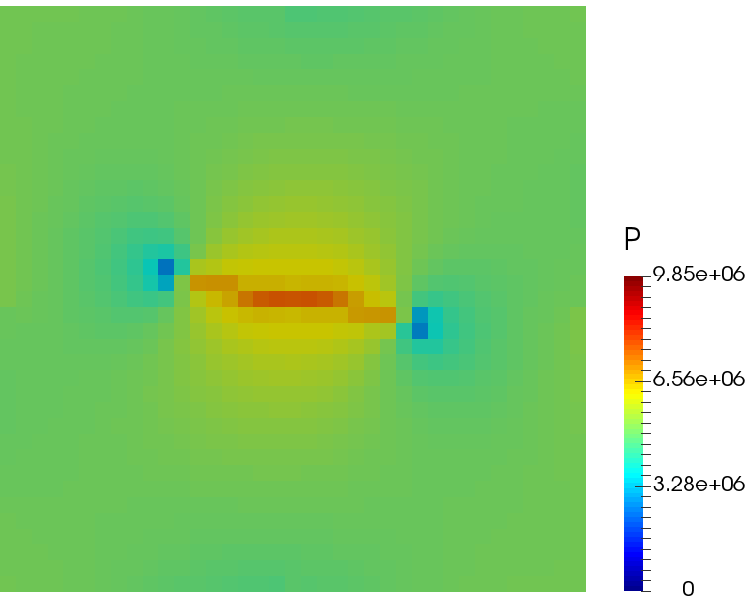}
		\caption{37$\times$37}
	\end{subfigure}
	\hspace{-0.5cm}
	\begin{subfigure}[b]{0.24\textwidth}
		\centering
		\includegraphics[width=0.8\textwidth]{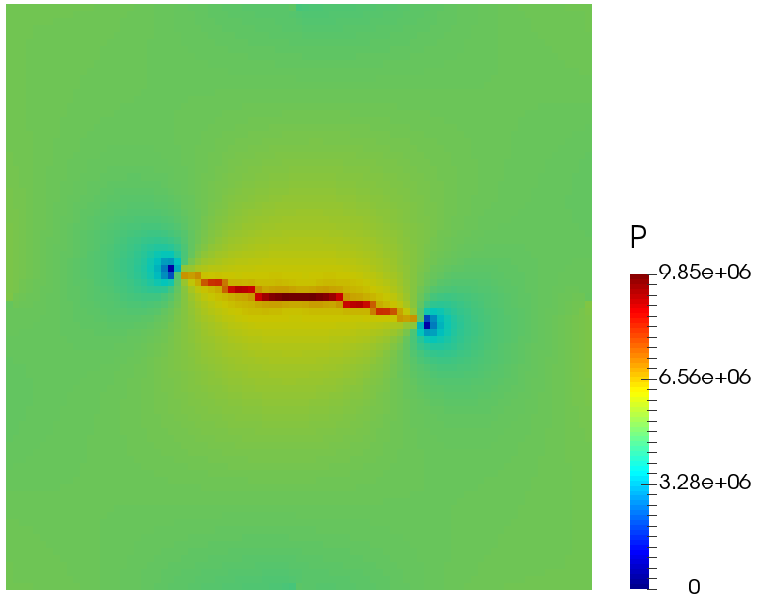}
		\caption{87$\times$83}
	\end{subfigure}
	\hspace{-0.5cm}
	\begin{subfigure}[b]{0.24\textwidth}
		\centering
		\includegraphics[width=0.8\textwidth]{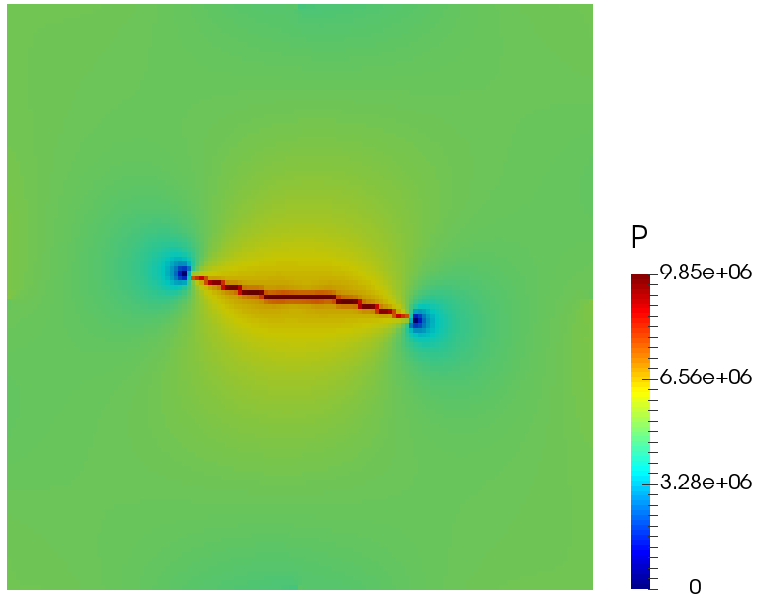}
		\caption{137$\times$123}
	\end{subfigure}
	\hspace{-0.5cm}
	\begin{subfigure}[b]{0.24\textwidth}
		\centering
		\includegraphics[width=0.8\textwidth]{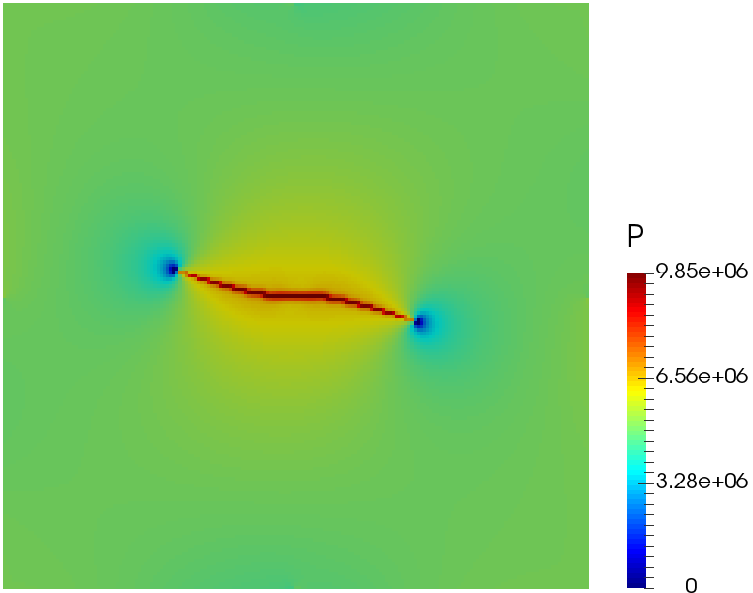}
		\caption{187$\times$173}
	\end{subfigure}

	\begin{subfigure}[b]{0.24\textwidth}
		\centering
		\includegraphics[width=0.8\textwidth]{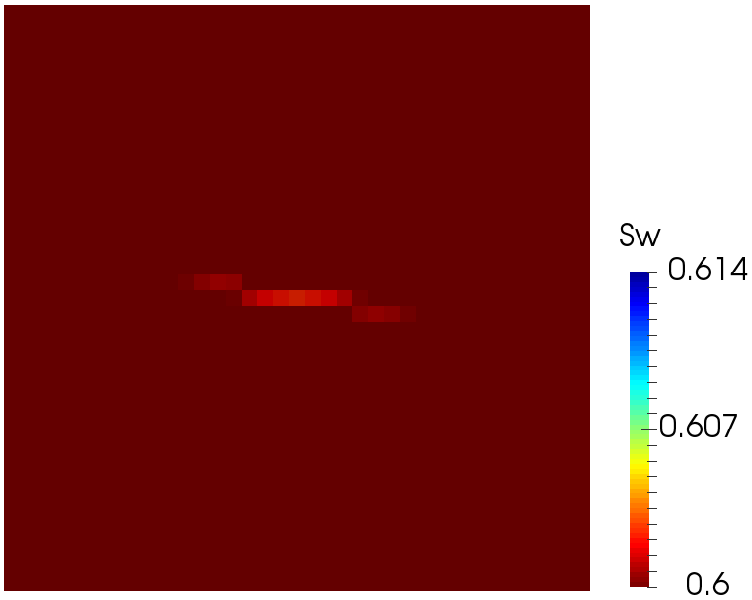}
		\caption{37$\times$37}
	\end{subfigure}
	\hspace{-0.5cm}
	\begin{subfigure}[b]{0.24\textwidth}
		\centering
		\includegraphics[width=0.8\textwidth]{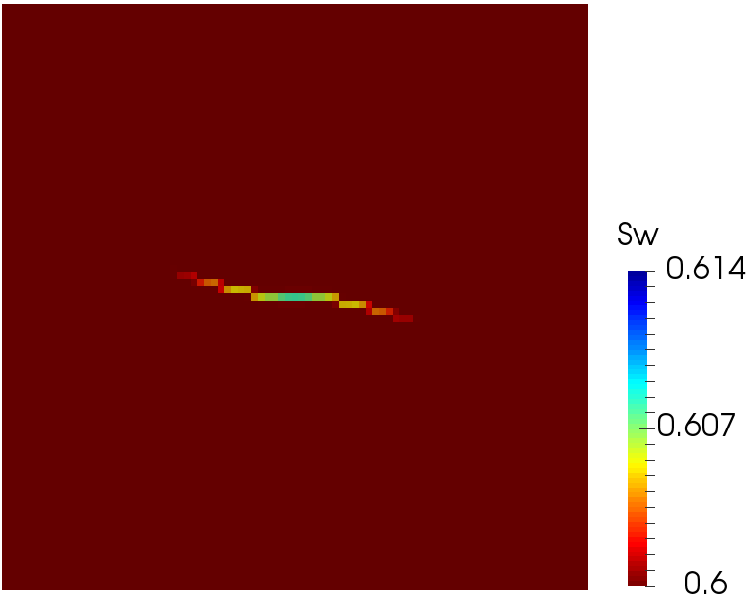}
		\caption{87$\times$83}
	\end{subfigure}
	\hspace{-0.5cm}
	\begin{subfigure}[b]{0.24\textwidth}
		\centering
		\includegraphics[width=0.8\textwidth]{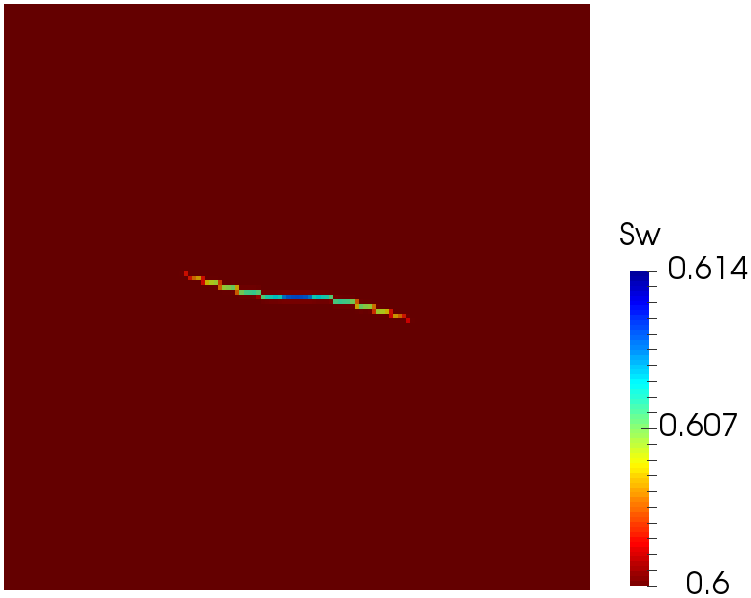}
		\caption{137$\times$123}
	\end{subfigure}
	\hspace{-0.5cm}
	\begin{subfigure}[b]{0.24\textwidth}
		\centering
		\includegraphics[width=0.8\textwidth]{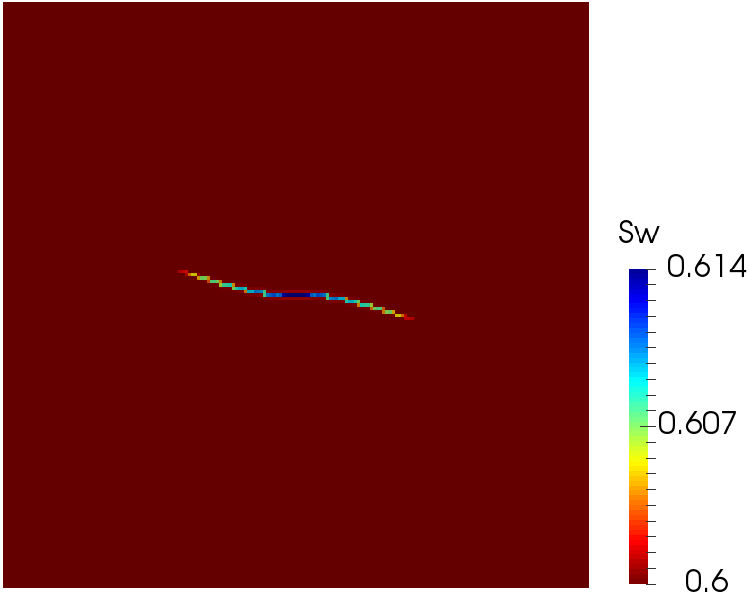}
		\caption{187$\times$173}
	\end{subfigure}

	\caption{pore pressure (top, unit: Pa) and saturation (bottom) distribution at the end of simulation 50 seconds under MMT-1 FP.}
	\label{p_s_mixedmode}
\end{figure}

Figure~\ref{p_s_mixedmode} presents snapshots of the pore pressure and saturation fields under MMT-1. The pressure fields show a high pressure band in the matrix cut through by the hydraulic fracture. The band diffuses along the normal direction. Lower fluid pressure happens behind the fracture tip region due to high tensile stresses and the resultant dilation in the corresponding matrix rock. Overall, in these cases, only a small mount of water leaks into the rock matrix as a result of the low matrix permeability as well as the scale contrast between the fracture and matrix.

\subsection{Variations in permeability and Biot parameter.}
\begin{figure}[!ht]
	\centering
	\begin{subfigure}[b]{0.6\textwidth}
		\centering
		\includegraphics[width=0.6\textwidth]{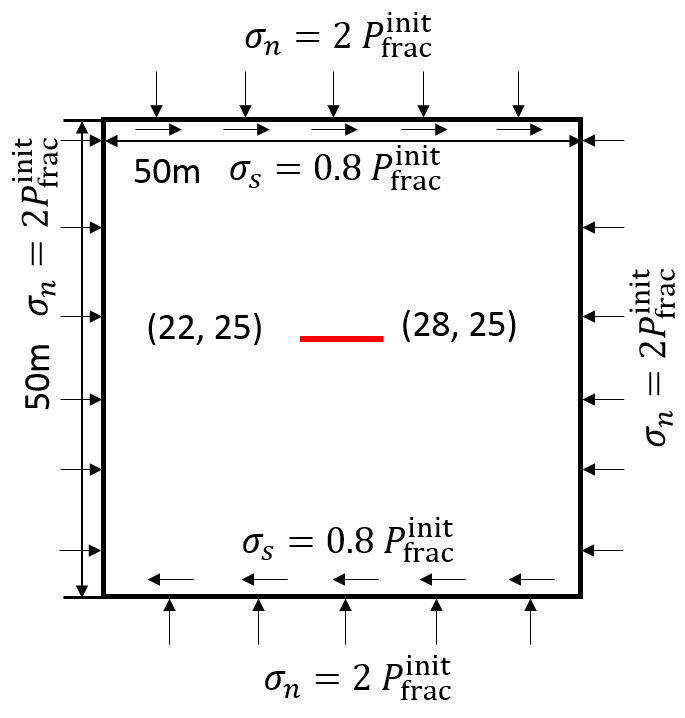}
		\caption{domain schematics}
		\label{case5schematics}
	\end{subfigure}
	\begin{subfigure}[b]{0.49\textwidth}
		\centering
		\includegraphics[width=0.8\textwidth]{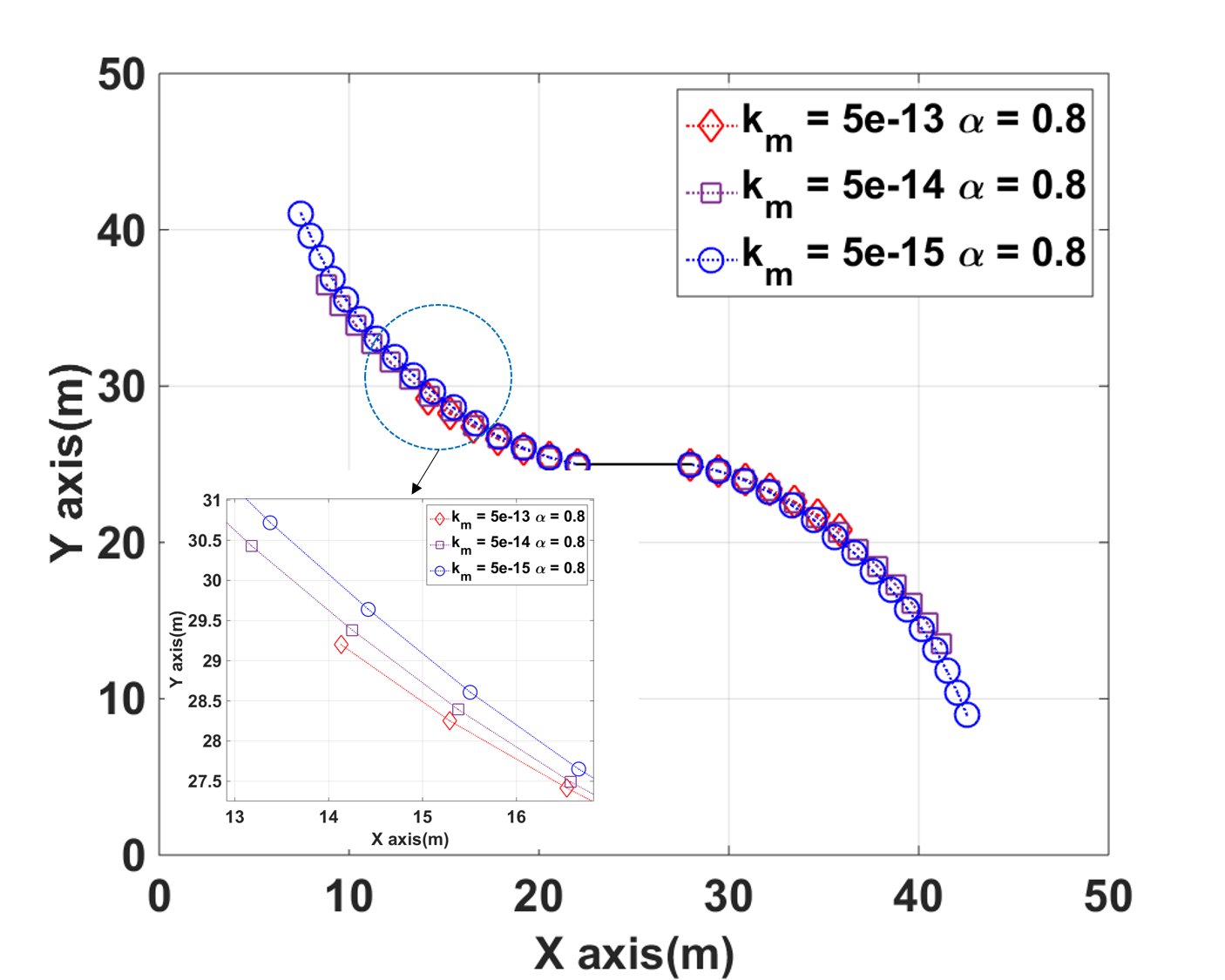}
		\caption{permeability effects}
		\label{permFP}
	\end{subfigure}
	\begin{subfigure}[b]{0.49\textwidth}
		\centering
		\includegraphics[width=0.8\textwidth]{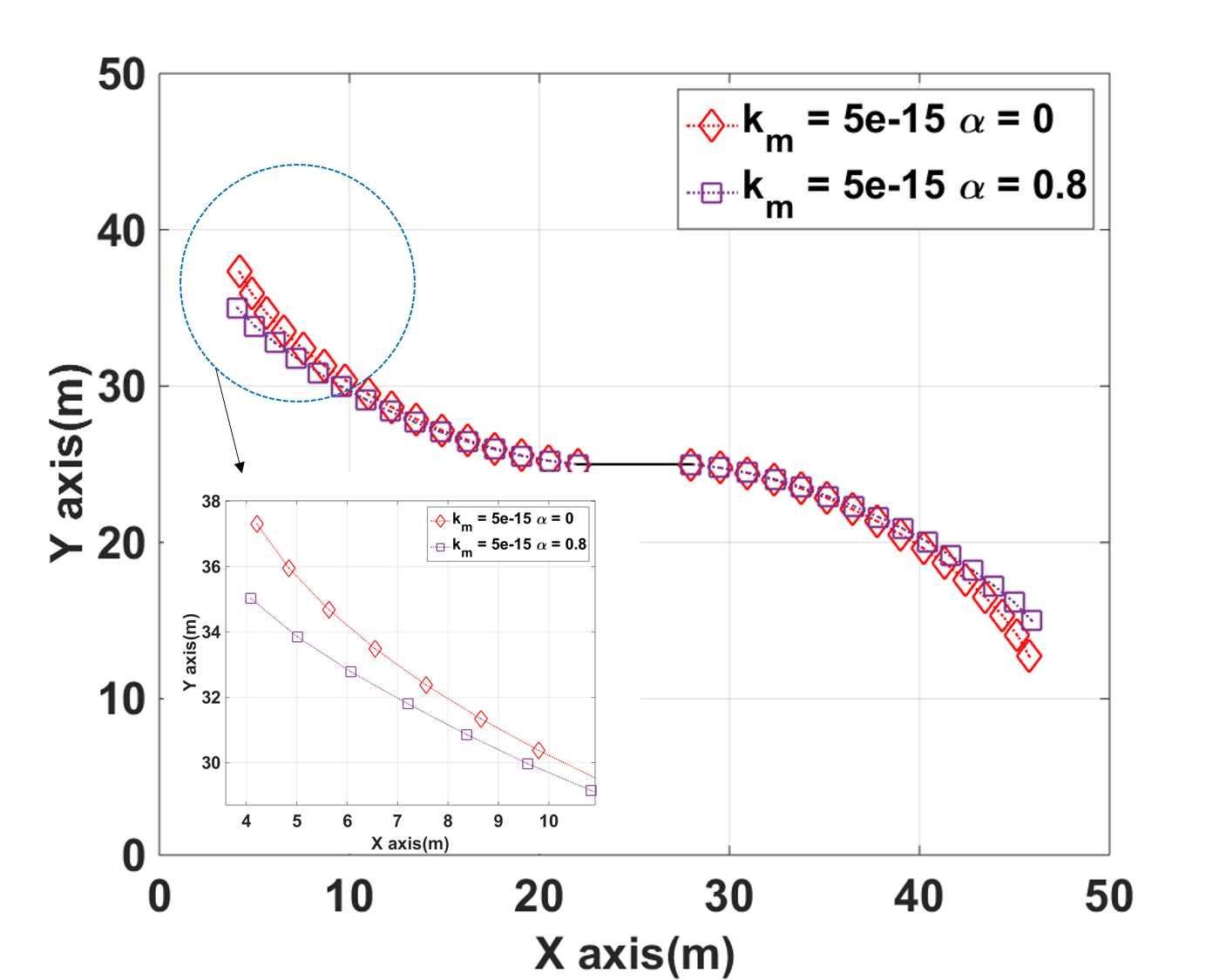}
		\caption{Biot coefficient effects}
		\label{biotFP}
	\end{subfigure}
	\caption{Poroelastic influences on FP in porous media}
	\label{poroFP}
\end{figure}

Using the model problem illustrated in Figure \ref{case5schematics}, the influences of the matrix permeability and Biot parameters on predicted simulations are investigated. Initial reservoir pressure in all cases is 5 MPa; water saturation is 0.6; water injection rate is $0.01 \text{m}^3/\text{s}$; critical SIF $K_{c}$ is set $1.5e7 \text{Pa}\sqrt{\text{m}}$; the tolerance parameter is $\epsilon_{SIF}$ is $2\%$; and $\Delta a$ is $1.5 \text{m}$.

{\color{black}
\begin{figure}[!htb]
	\centering
	\includegraphics[width=0.5\textwidth]{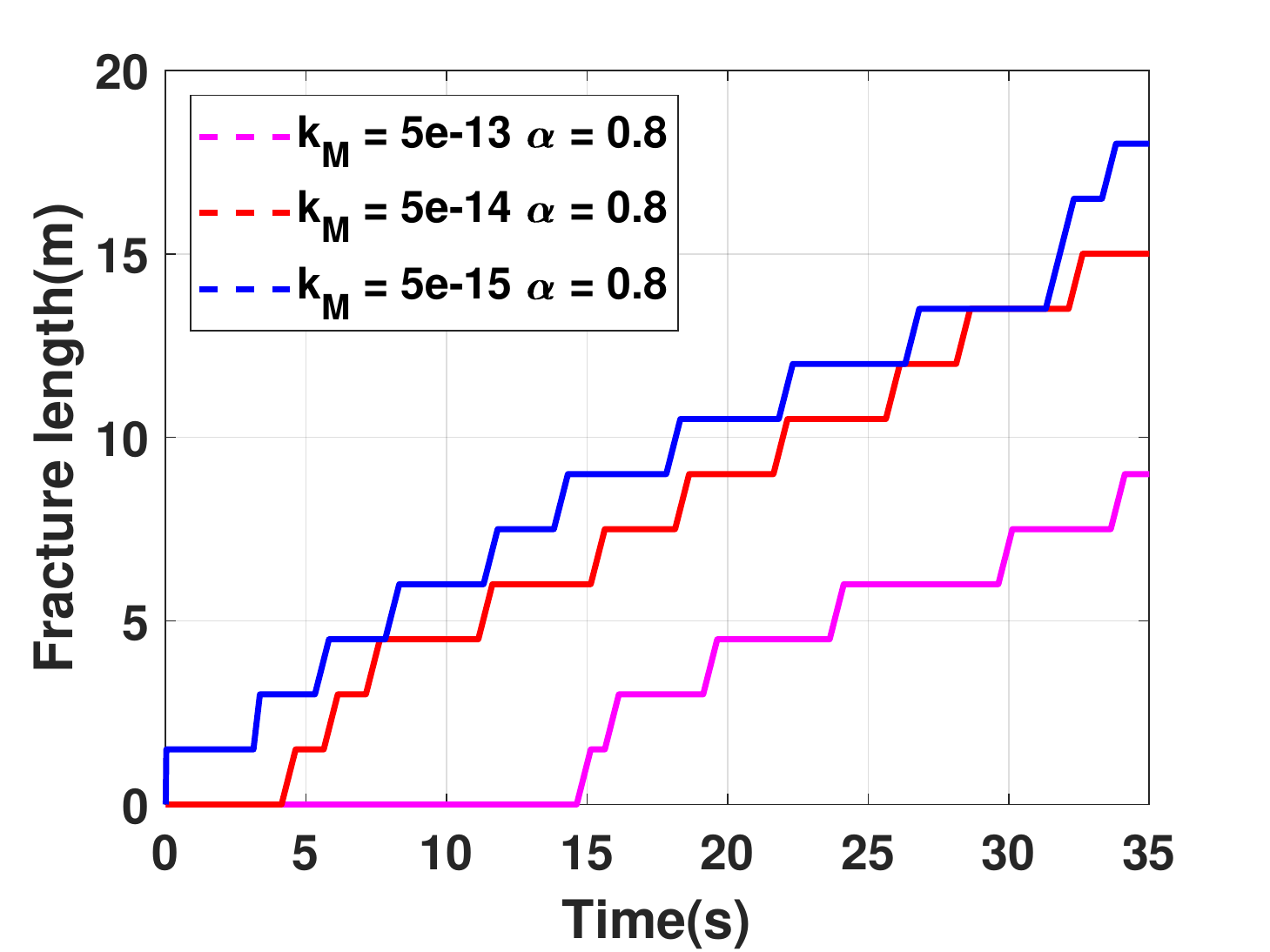}
	\caption{fracture length evolution under different $\boldsymbol{k}_{M}$}
	\label{case5_speed}
\end{figure}
By comparing the resultant fracture lengths obtained at a particular time while assuming $ \bm{k}_{M} \in \{ 5e-13, 5e-14, 5e-15 m^2 \}$ and  $\alpha=0.8$, it is apparent that lower matrix permeability leads to a higher propagation rate. Indeed, Figure \ref{case5_speed} shows that a two-order of magnitude increase in matrix permeability can lead to a $50\%$ reduction in propagation speed. This is associated with the increased fluid leak-off from fracture to matrix. Figure \ref{permFP} shows the FP direction is slightly altered by the fluid leak-off.   $K_{II}$ is a strong function of shear deformation which, in this case, is predominantly driven by the lateral forces applied to the top and bottom boundaries. The increased local matrix pressure induced by the fluid leak-off leads to a larger $K_I/K_{II}$ ratio. Subsequently, this results in reduced curvature in the propagation path observed for the high permeability field. The results agree with the observation in \cite{usui2017effect}.
}
\begin{figure}[!ht]
	\centering
	\begin{subfigure}[b]{0.49\textwidth}
		\centering
		\includegraphics[width=0.8\textwidth]{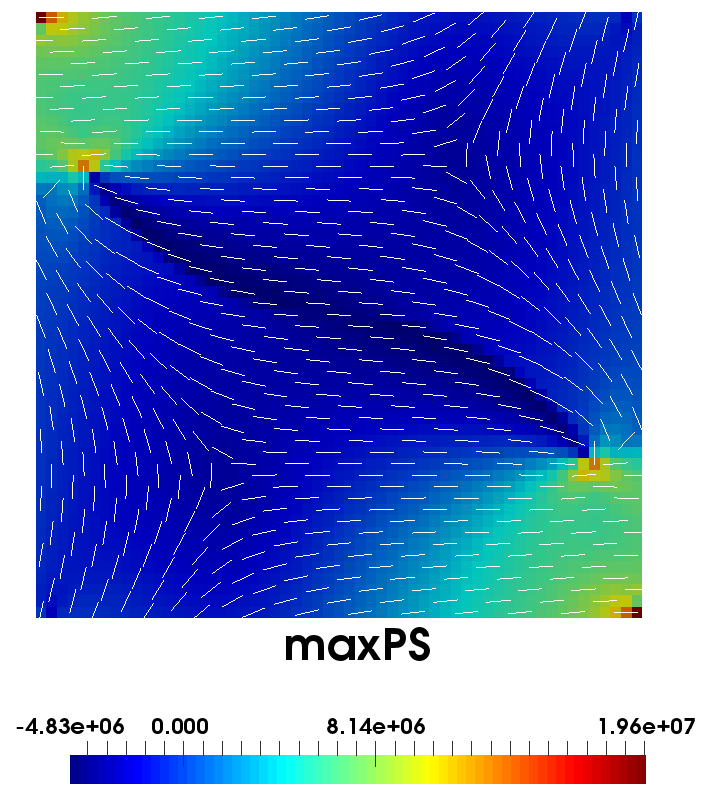}
		\caption{$\alpha = 0$}
		\label{biot1}
	\end{subfigure}
	\begin{subfigure}[b]{0.49\textwidth}
		\centering
		\includegraphics[width=0.8\textwidth]{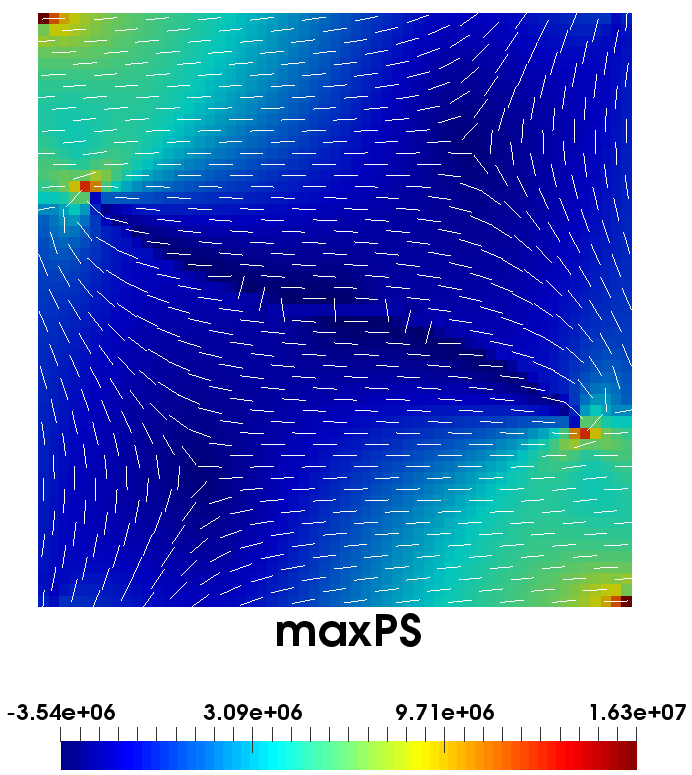}
		\caption{$\alpha = 0.8$}
		\label{biot2}
	\end{subfigure}
	\caption{Comparison of maximum principal stress (unit: Pa) and its direction (white lines) for different $\alpha$}
	\label{maxPBiot}
\end{figure}

Figure \ref{biotFP} shows results obtained while fixing permeability at $\bm{k}_{M}$ is fixed at $5e-15 \text{m}^{2}$, and selecting $\alpha$ as either $0$ or $0.8$. Figure \ref{biotFP} shows less curvature in the high $\alpha$ scenario. {\color{black} Increasing the coefficient $\alpha$ in this case causes the pore pressure to counter the compressive forces applied at the boundary and subsequently leads to an increase in the ratio $K_{I}/K_{II}$. This in turn leads to a reduction in the curvature of the resulting FP path.} The value and direction (white lines) of the maximum principal stress of two different scenarios are compared in Figure \ref{maxPBiot}. We can observe differences in the direction of the stress field near the fracture tip region that causes the deviation in the FP path. In the rest of the domain, as compared to the decoupled case ($\alpha = 0$), pore pressure also alters the direction of the maximum principal stress along the fracture body.

\subsection{Opposing, parallel fluid-driven fracture}
\begin{figure}[htb]
	\centering
	\includegraphics[width=0.3\textwidth]{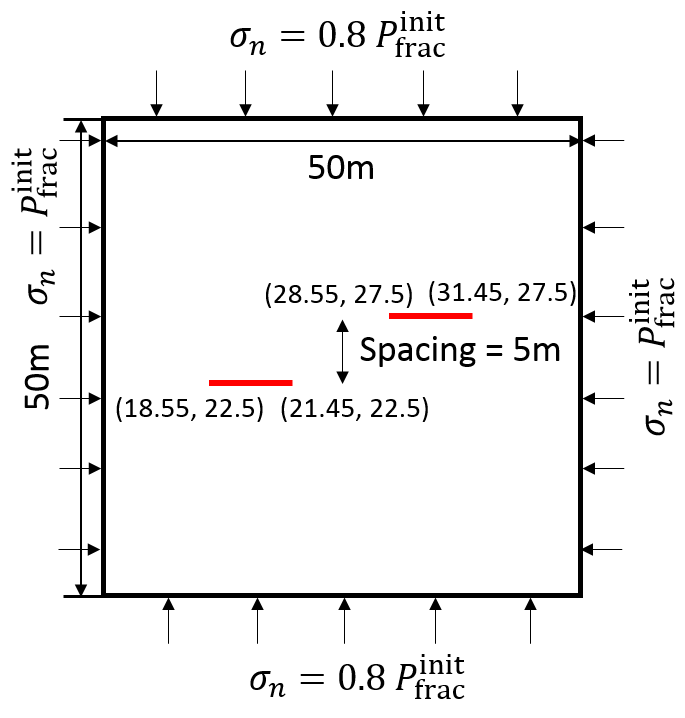}
	\caption{Initial fracture configuration and boundary conditions, where $x$ and $y$ displacement is fixed at four corners of the rectangular.}
	\label{case7_fractureConfig_boundary}
\end{figure}
The test problem is illustrated in Figure \ref{case7_fractureConfig_boundary}. A $187\times173$ mesh is applied, and water is injected into the center of either fracture at the same rate of $0.01 \text{m}^{2}/s$. The critical SIF is set at $2e7\text{Pa}\sqrt{\text{m}}$ with $5\%$ tolerance and $\Delta a$ is chosen as $0.6 \text{m}$.
\begin{figure}[!htb]
	\centering
	\begin{subfigure}[b]{0.2\textwidth}
		\centering
		\includegraphics[width=0.90\textwidth]{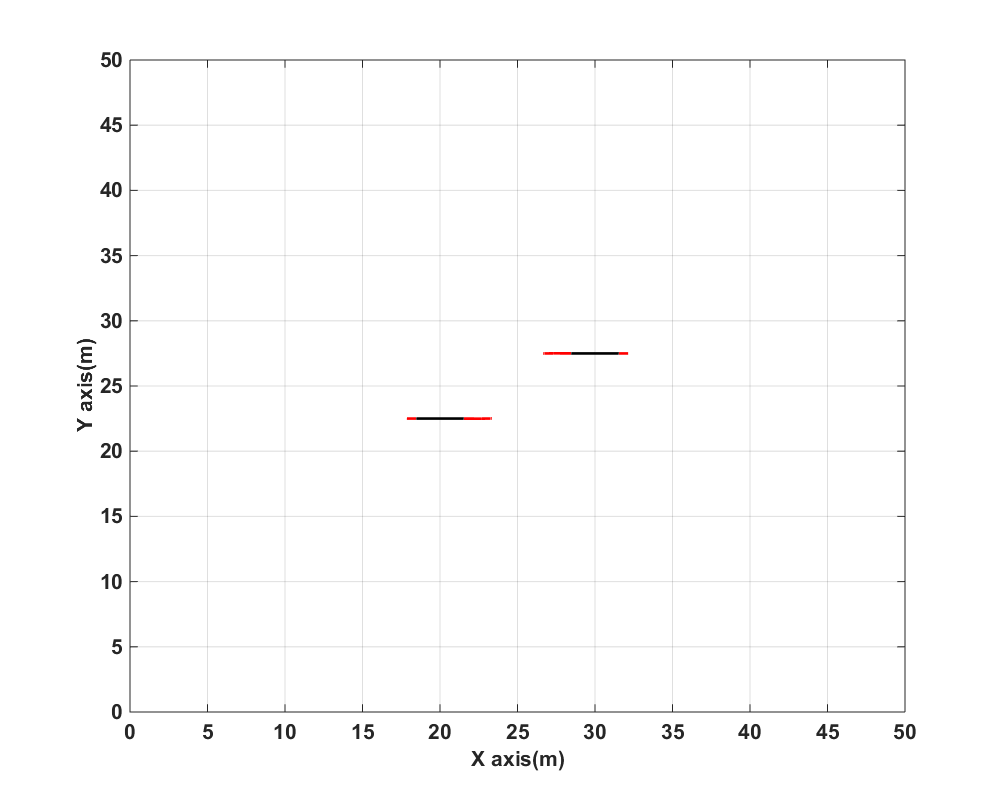}
	\end{subfigure}
	\hspace{-0.1cm}
	\begin{subfigure}[b]{0.2\textwidth}
		\centering
		\includegraphics[width=0.98\textwidth]{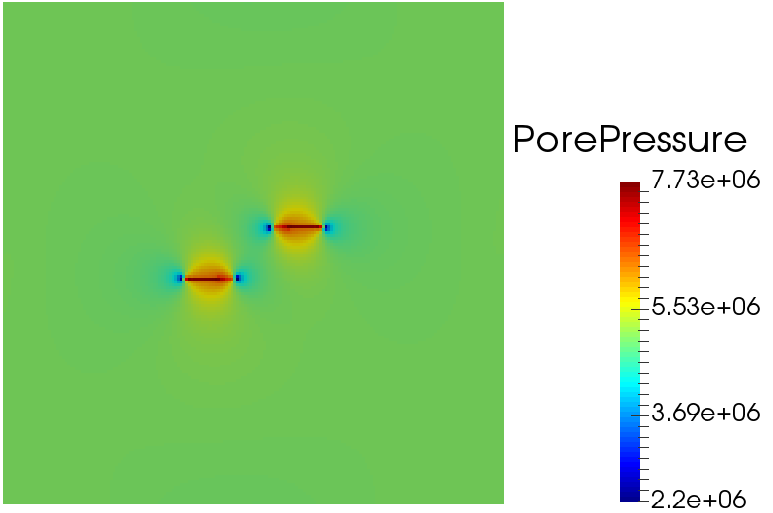}
	\end{subfigure}
	\hspace{-0.1cm}
	\begin{subfigure}[b]{0.2\textwidth}
		\centering
		\includegraphics[width=0.95\textwidth]{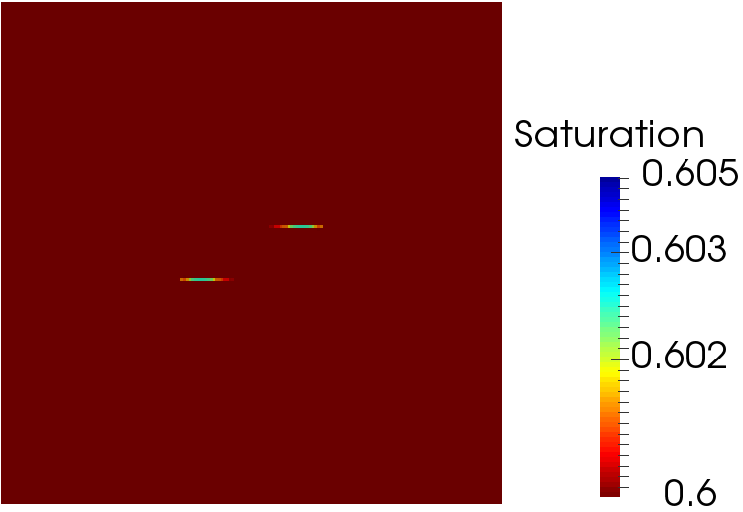}
	\end{subfigure}
	\hspace{-0.1cm}
	\begin{subfigure}[b]{0.2\textwidth}
		\centering
		\includegraphics[width=0.95\textwidth]{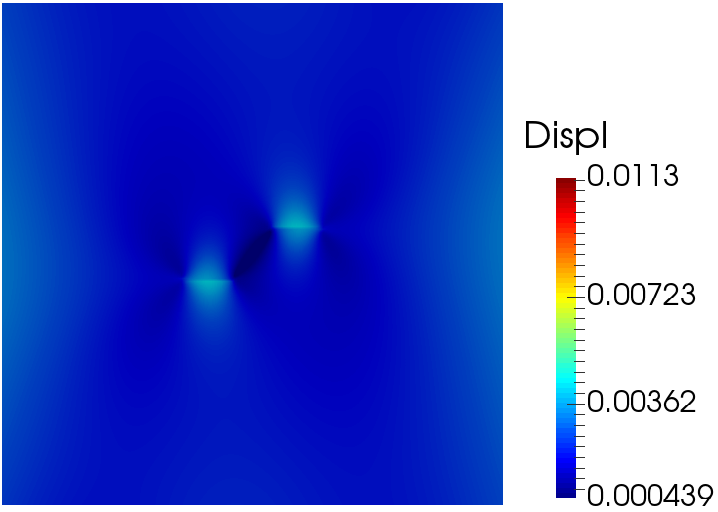}
	\end{subfigure}
	\begin{subfigure}[b]{0.2\textwidth}
		\centering
		\includegraphics[width=0.90\textwidth]{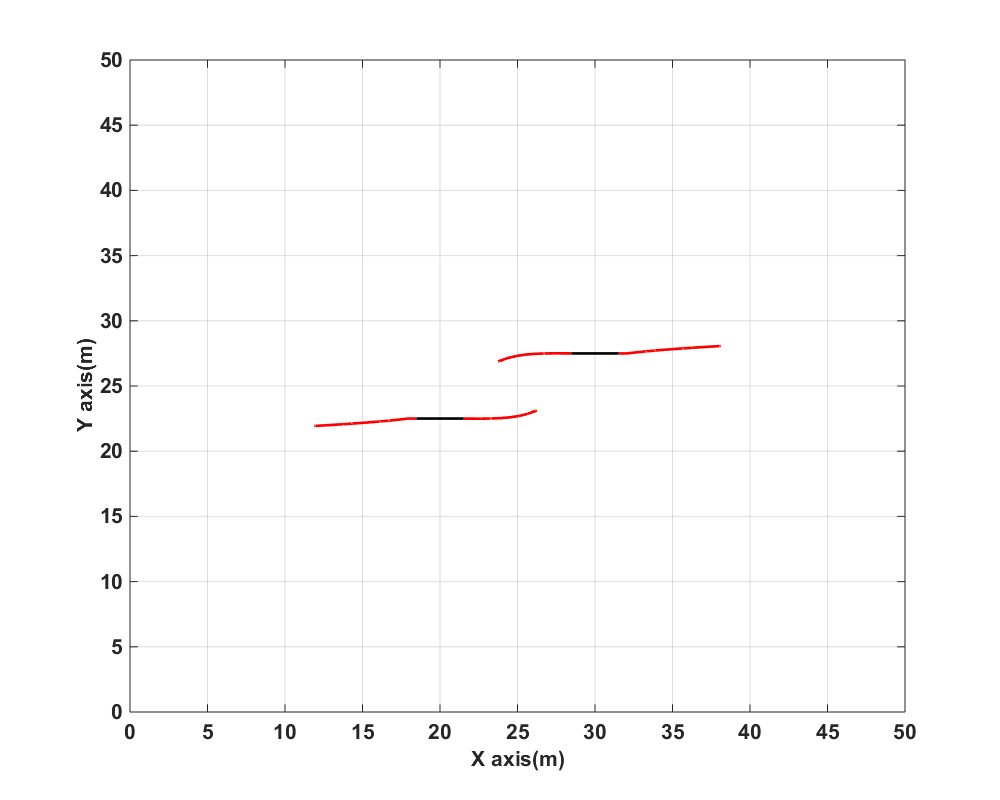}
	\end{subfigure}
	\hspace{-0.1cm}
	\begin{subfigure}[b]{0.2\textwidth}
		\centering
		\includegraphics[width=0.98\textwidth]{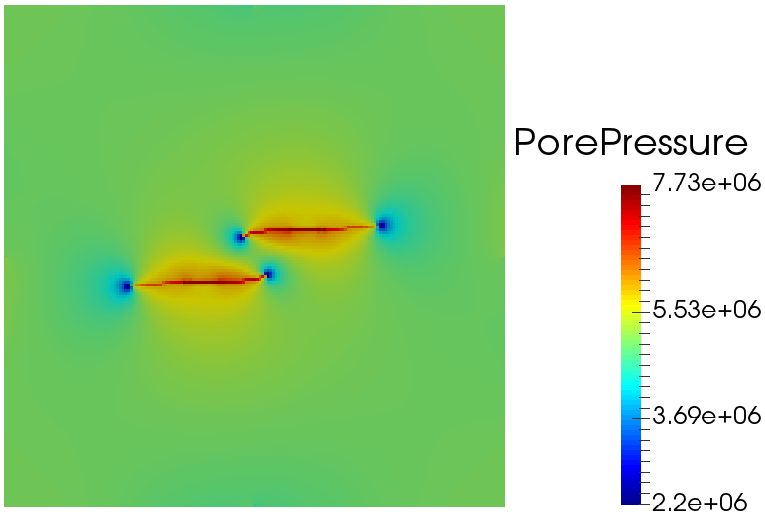}
	\end{subfigure}
	\hspace{-0.1cm}
	\begin{subfigure}[b]{0.2\textwidth}
		\centering
		\includegraphics[width=0.95\textwidth]{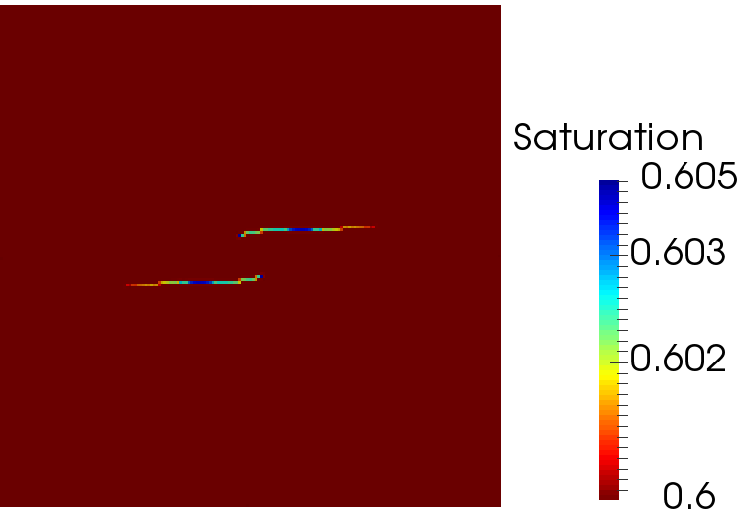}
	\end{subfigure}
	\hspace{-0.1cm}
	\begin{subfigure}[b]{0.2\textwidth}
		\centering
		\includegraphics[width=0.95\textwidth]{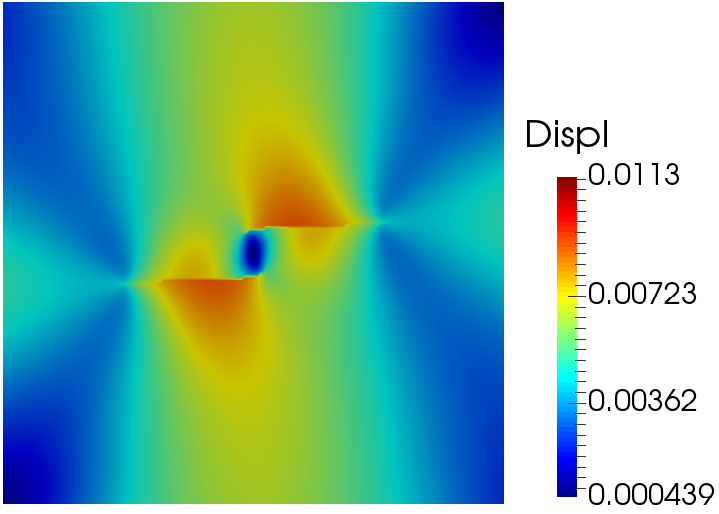}
	\end{subfigure}
	\begin{subfigure}[b]{0.2\textwidth}
		\centering
		\includegraphics[width=0.90\textwidth]{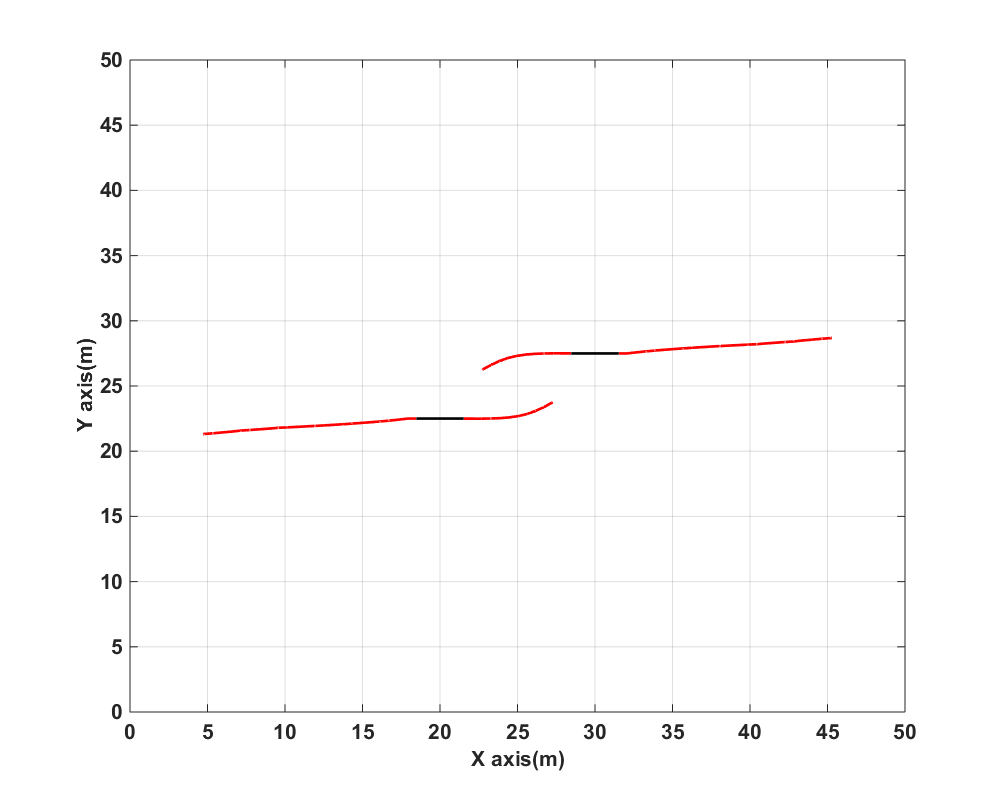}
	\end{subfigure}
	\hspace{-0.1cm}
	\begin{subfigure}[b]{0.2\textwidth}
		\centering
		\includegraphics[width=0.98\textwidth]{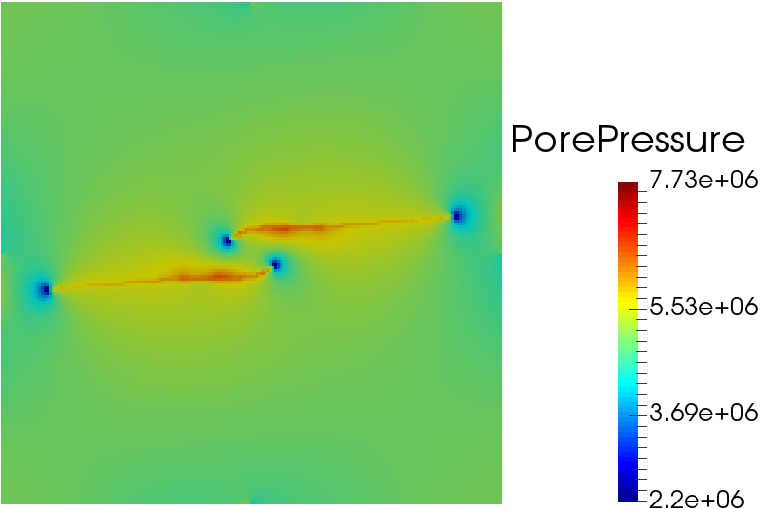}
	\end{subfigure}
	\hspace{-0.1cm}
	\begin{subfigure}[b]{0.2\textwidth}
		\centering
		\includegraphics[width=0.95\textwidth]{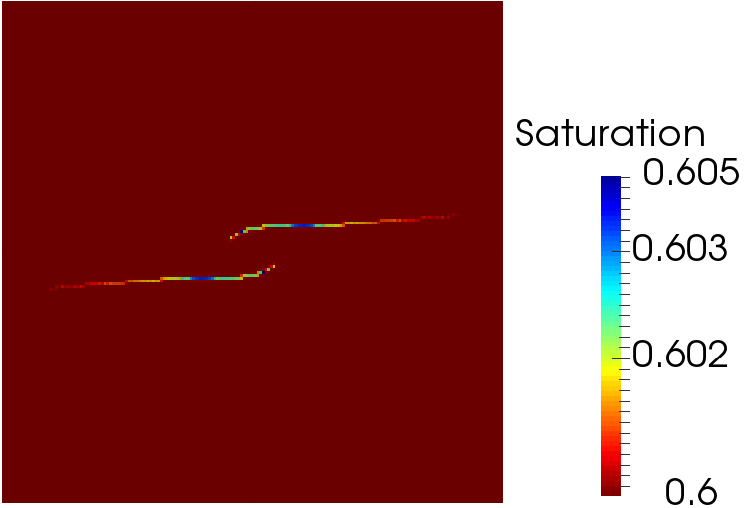}
	\end{subfigure}
	\hspace{-0.1cm}
	\begin{subfigure}[b]{0.2\textwidth}
		\centering
		\includegraphics[width=0.95\textwidth]{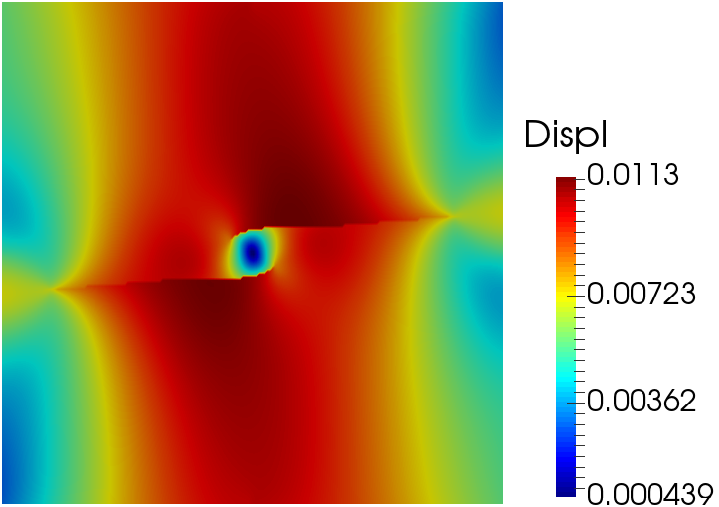}
	\end{subfigure}
	\caption{Snapshots (by column) of the FP path, pore pressure (unit: Pa), water saturation, and the displacement \textcolor{black}{(unit: m)} at $t= 3.15, 15.90, \text{ and } 34.77 \text{secs}$.}
	\label{pp}
\end{figure}

Figure~\ref{pp} shows snapshots of the fracture path, pore pressure, saturation and displacement distributions obtained for various times during the simulation. The two fractures initially propagate symmetrically until the stress field in the vicinity of the fracture ends responds to their interaction (\cite{paul20183d}). Subsequently, the localization is evident as the two tips overtake each other and arrest as they propagate towards the interior of the domain. This localization effect is similar to results reported using the phase field and XFEM methods (e.g. \cite{paul20183d, wang2016numerical}).

\subsection{Alternating periods of Hydraulic Fracturing and fluid injection and production.}\label{case 6}
\begin{figure}[htb]
	\centering
	\includegraphics[width=0.3\textwidth]{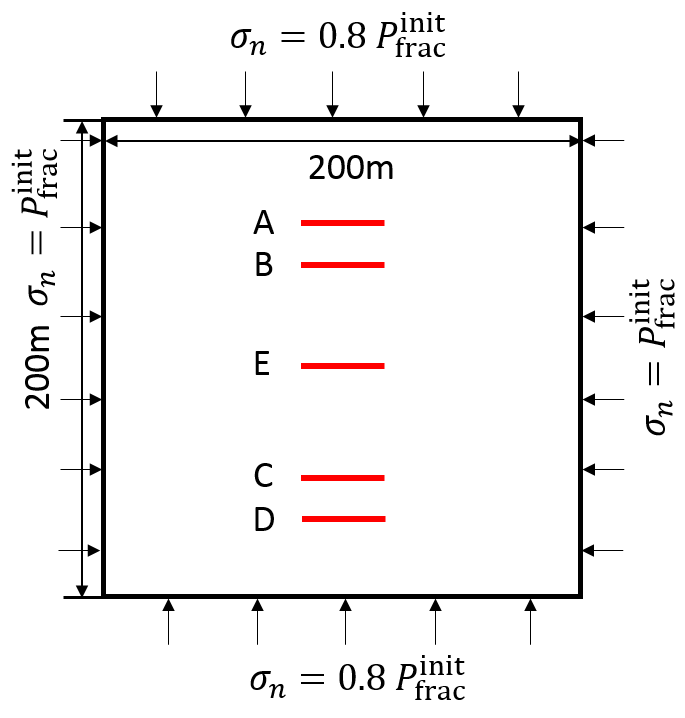}
	\caption{Initial fracture configuration and boundary conditions,  $x$ and $y$ displacement is fixed at four corners of the rectangular; fracture coordinates are (97, 160) to (103, 160), (97, 140) to (103, 140), (97, 100) to (103, 100), (97, 60) to (103, 60) , (97, 40) to (103, 40) from the top to the bottom.}
	\label{case8_fractureConfig_boundary}
\end{figure}
A key objective is to co-simulate periods of fluid flow and deformation with no propagation, and the onset and dynamics of propagation of one or more fracture. Timely engineering questions relying on such ability pertain to the design of hydraulic fracture treatments and implementation strategies in domains undergoing complex sequences of fracture, infill drilling, and production. We consider a hypothetical scenario to model such a system.

\begin{figure}[!htb]
	\centering
	\begin{subfigure}[b]{0.49\textwidth}
		\centering
		\includegraphics[width=0.9\textwidth]{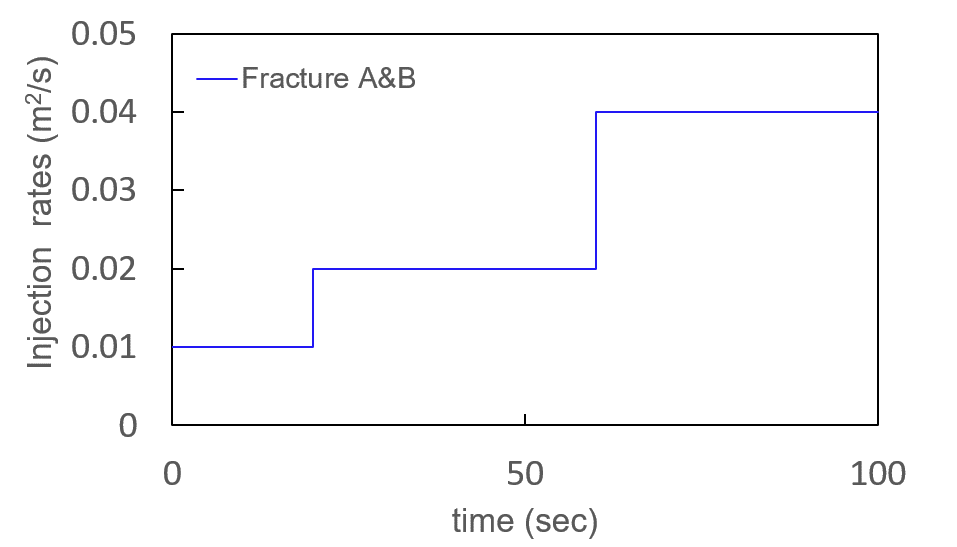}
		\caption{Injection stage 1}
		\label{twofracturetop}
	\end{subfigure}
	\begin{subfigure}[b]{0.49\textwidth}
		\centering
		\includegraphics[width=0.9\textwidth]{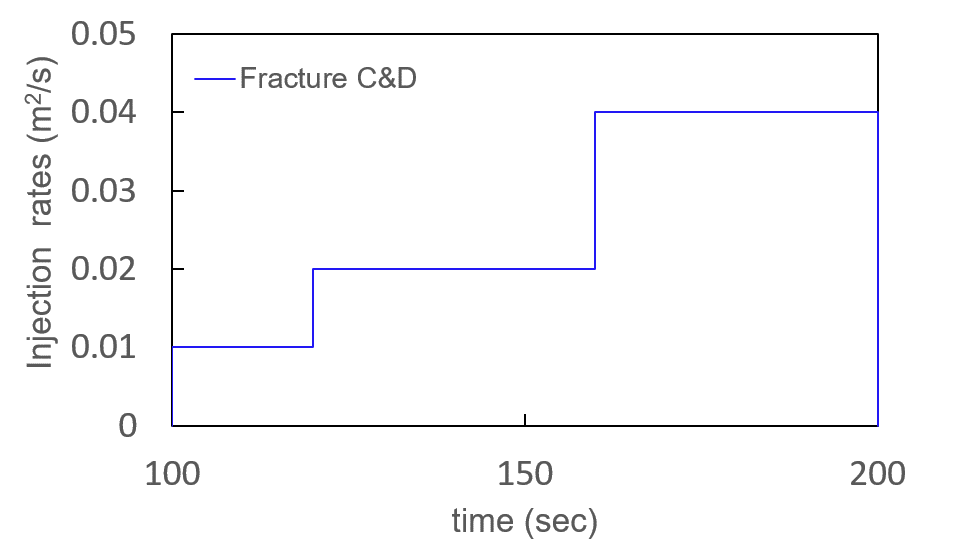}
		\caption{\textcolor{black}{Injection stage 2}}
		\label{bottomfracturetop}
	\end{subfigure}
	\begin{subfigure}[b]{0.49\textwidth}
		\centering
		\includegraphics[width=0.9\textwidth]{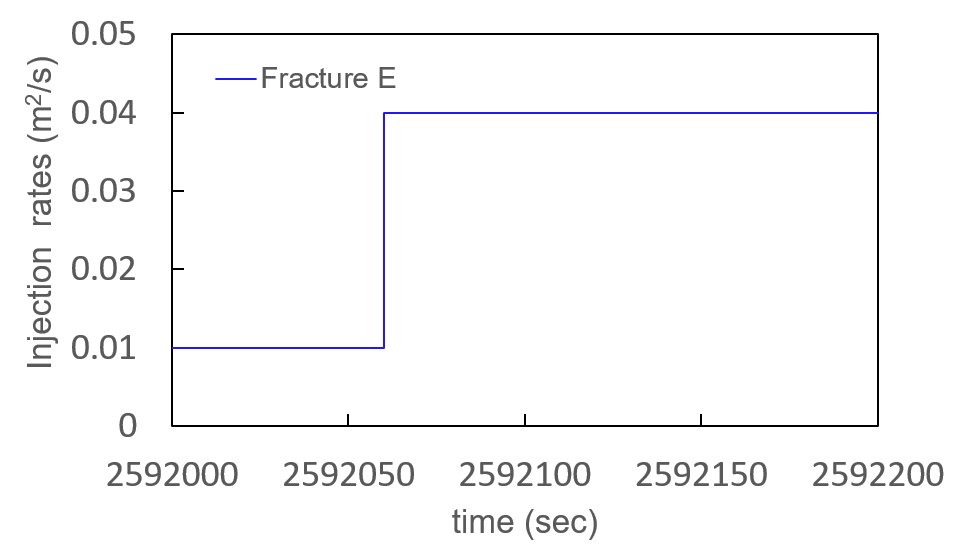}
		\caption{Injection stage 3}
		\label{middleinjectionschedule}
	\end{subfigure}
	\caption{Water rate injection schedules for three stimulation periods. Stages 1 and 2 are separated from the stage 3 by a 30 day period of fluid withdrawal.}
	\label{injectionschedule}
\end{figure}

The initial fracture pattern is showed in Figure \ref{case8_fractureConfig_boundary}. Stimulation-alternating-production is assumed to occur over four operational stages: (1) fractures $A$ and $B$ are hydraulically fractured by injection of water into their centers at the rate schedule illustrated in Figure~\ref{twofracturetop}, (2) \textcolor{black}{fractures $C$ and $D$} are hydraulically fractured by injection of water into their centers at the rate schedule illustrated in Figure~\ref{bottomfracturetop}, (3) fluid withdrawal (production) is subsequently conducted for 30 days by enforcing a constant pressure of $4.5e6$ Pa at the centers of all fractures, and (3) production is halted, and water is injected into the center of fracture $E$ using the injection rate schedule in Figure~\ref{middleinjectionschedule}. Water injection schedules for these three consecutive operations are illustrated in Figure \ref{injectionschedule}. A mesh of $137 \times 123$ is applied; $K_{c}$ is $3e7\text{Pa}\sqrt{\text{m}}$; the tolerance $\epsilon_{SIF}$ is $0.001$; the propagation length step $\Delta a$ is $3.5\text{m}$; the initial time-step is set at $0.1$ secs; the maximum step is set at $4.32e5$ secs (5 days); $\alpha_{SIF}$ is chosen as 1.0; and the Biot coefficient $\alpha$ is set to 0.4.

\begin{figure}[!htb]
	\centering
	\begin{subfigure}[b]{0.2\textwidth}
		\centering
		\includegraphics[width=0.90\textwidth]{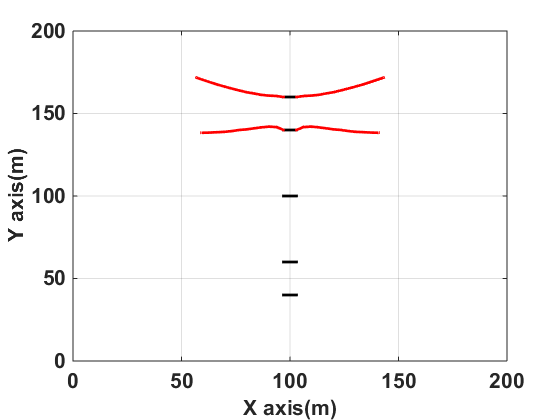}
	\end{subfigure}
	\hspace{-0.1cm}
	\begin{subfigure}[b]{0.2\textwidth}
		\centering
		\includegraphics[width=0.95\textwidth]{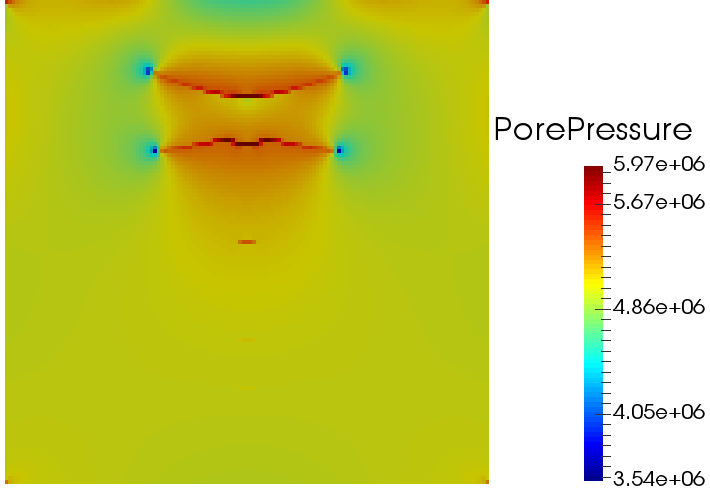}
	\end{subfigure}
	\hspace{-0.1cm}
	\begin{subfigure}[b]{0.2\textwidth}
		\centering
		\includegraphics[width=0.93\textwidth]{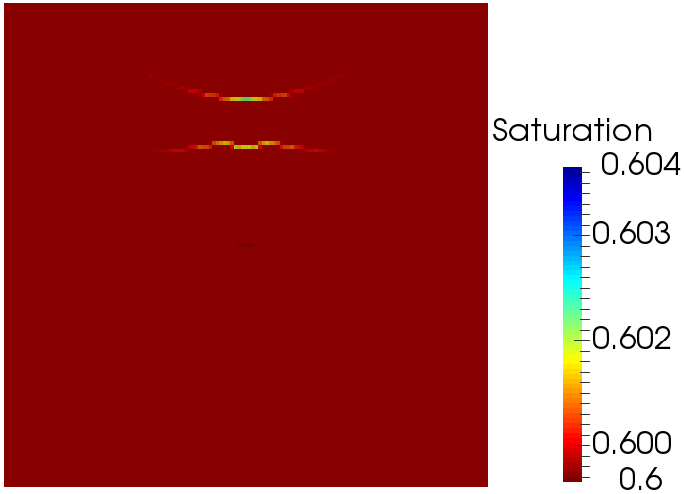}
	\end{subfigure}
	\hspace{-0.1cm}
	\begin{subfigure}[b]{0.2\textwidth}
		\centering
		\includegraphics[width=0.89\textwidth]{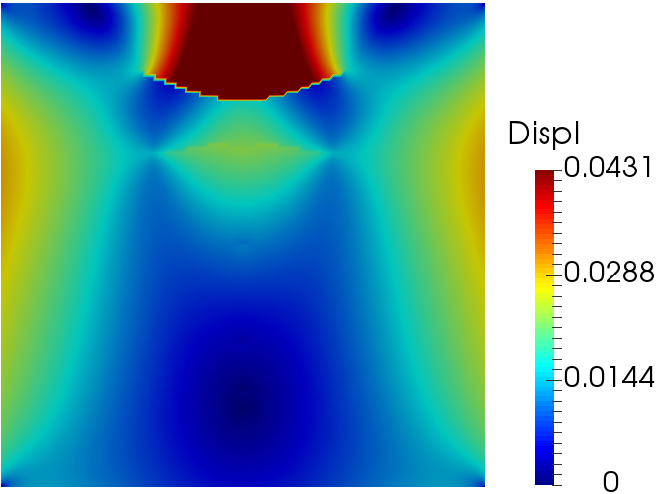}
	\end{subfigure}

	\begin{subfigure}[b]{0.2\textwidth}
		\centering
		\includegraphics[width=0.90\textwidth]{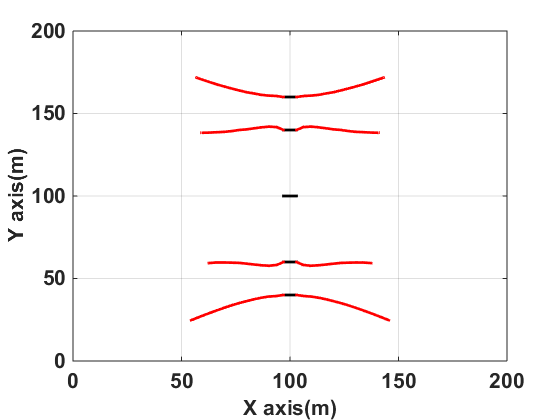}
	\end{subfigure}
	\hspace{-0.1cm}
	\begin{subfigure}[b]{0.2\textwidth}
		\centering
		\includegraphics[width=0.95\textwidth]{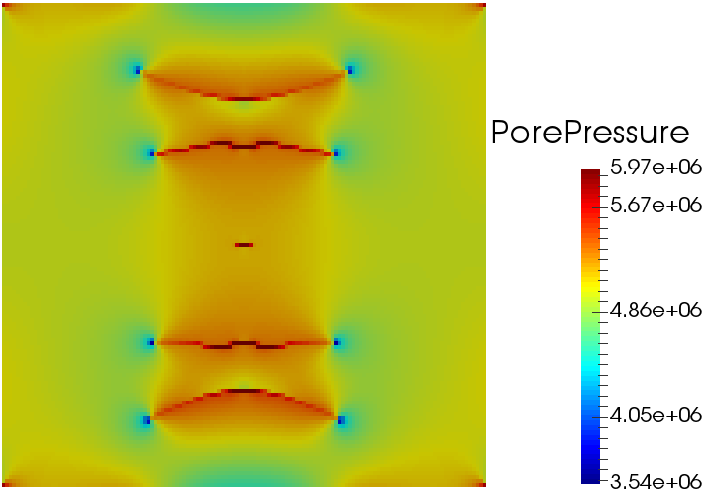}
	\end{subfigure}
	\hspace{-0.1cm}
	\begin{subfigure}[b]{0.2\textwidth}
		\centering
		\includegraphics[width=0.93\textwidth]{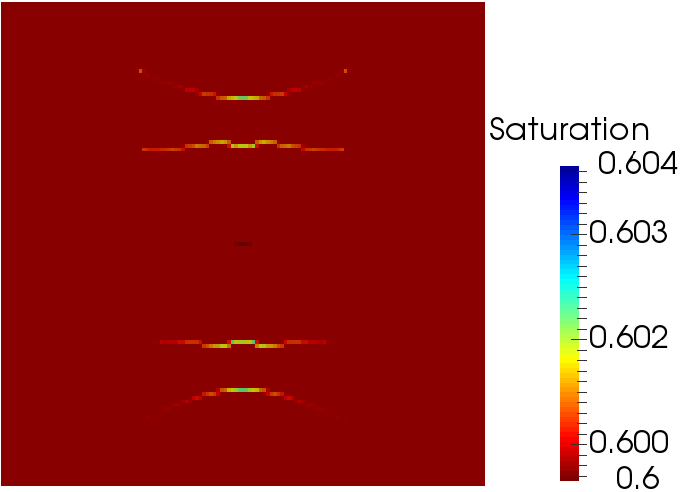}
	\end{subfigure}
	\hspace{-0.1cm}
	\begin{subfigure}[b]{0.2\textwidth}
		\centering
		\includegraphics[width=0.91\textwidth]{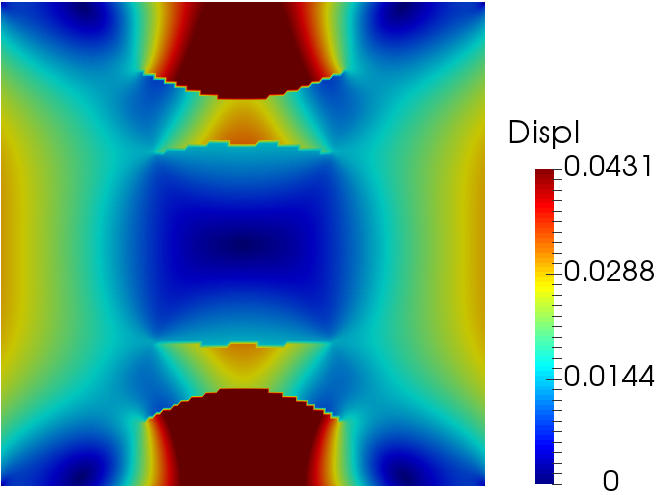}
	\end{subfigure}

	\begin{subfigure}[b]{0.2\textwidth}
	\centering
	\includegraphics[width=0.90\textwidth]{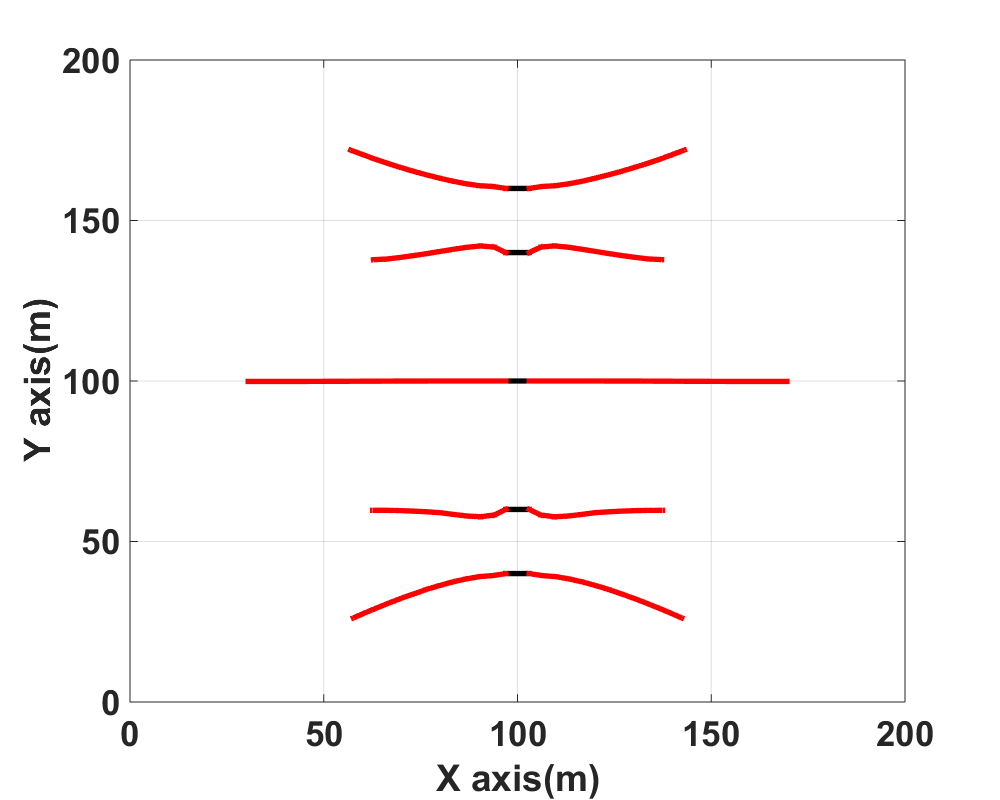}
\end{subfigure}
\hspace{-0.1cm}
\begin{subfigure}[b]{0.2\textwidth}
	\centering
	\includegraphics[width=0.95\textwidth]{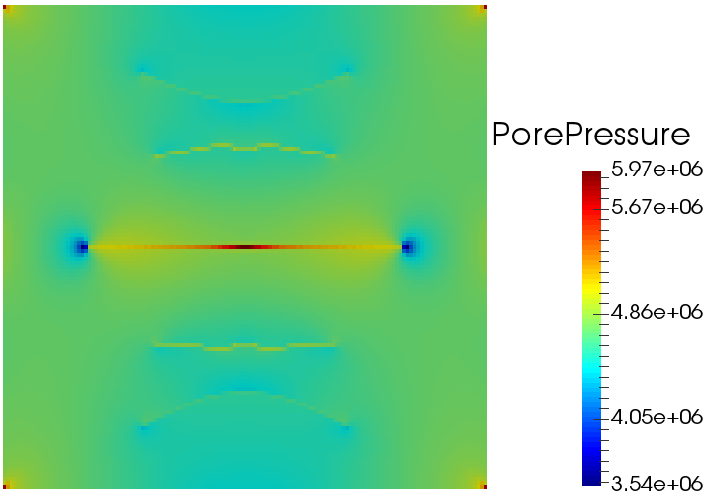}
\end{subfigure}
\hspace{-0.1cm}
\begin{subfigure}[b]{0.2\textwidth}
	\centering
	\includegraphics[width=0.93\textwidth]{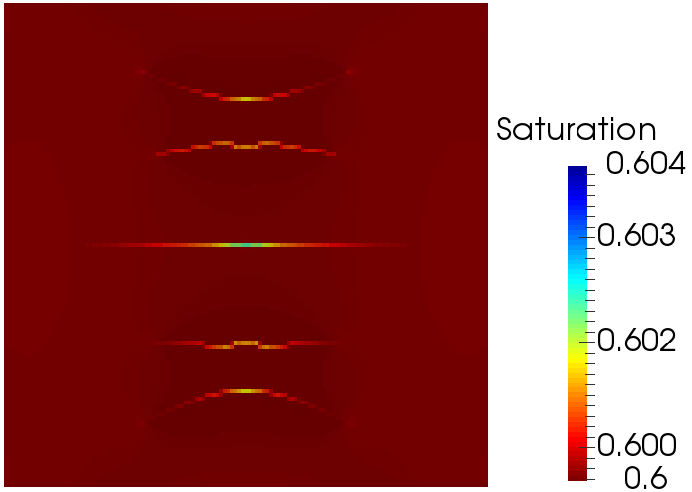}
\end{subfigure}
\hspace{-0.1cm}
\begin{subfigure}[b]{0.2\textwidth}
	\centering
	\includegraphics[width=0.91\textwidth]{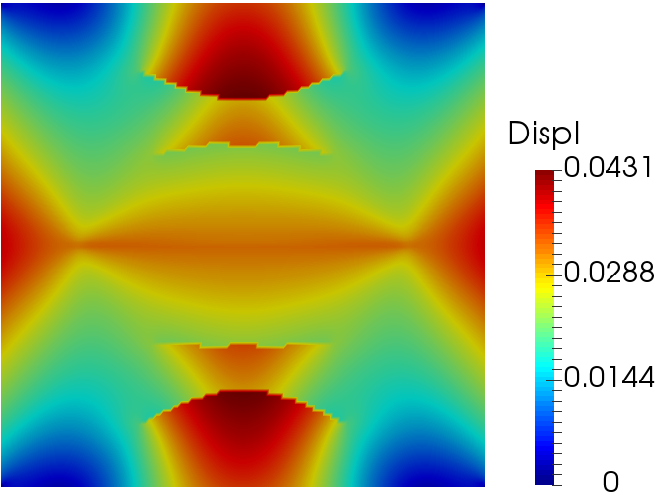}
\end{subfigure}
	\caption{Snapshots of (by column) the FP path, pore pressure (unit: Pa), water saturation, and displacement fields \textcolor{black}{(unit: m)} at the conclusion of stages (by row) 1,2, and 4.}
	\label{fractureEvolution}
\end{figure}

Figure~\ref{fractureEvolution}, shows snapshots of fracture paths, pore pressure, water saturation, and displacement fields obtained at three times during the simulation (the ends of stages 1, 2, and 4). The stress shadow effects and localization instability are evident in the paths obtained following stages 1 and 2. Moreover, Fracture E is observed to propagate in a symmetric, straight line since the stress field following symmetric production is also symmetric. Finally, while a large pressure gradient is observed in the vicinity of the fractures, saturation change due to water leak-off into the matrix is rather limited.

\begin{figure}[!htb]
	\centering
	\includegraphics[width=1\textwidth]{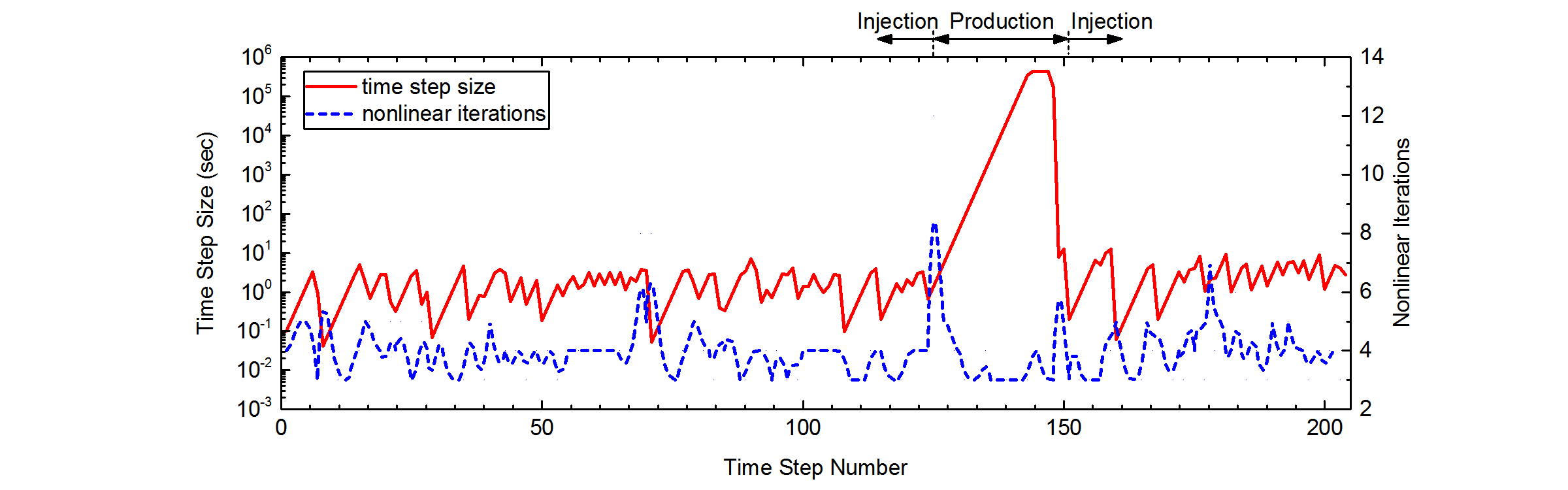}
	\caption{Nonlinear iterations and time-steps}
	\label{case8_timestep}
\end{figure}
\begin{figure}[!htb]
	\centering
	\includegraphics[width=0.4\textwidth]{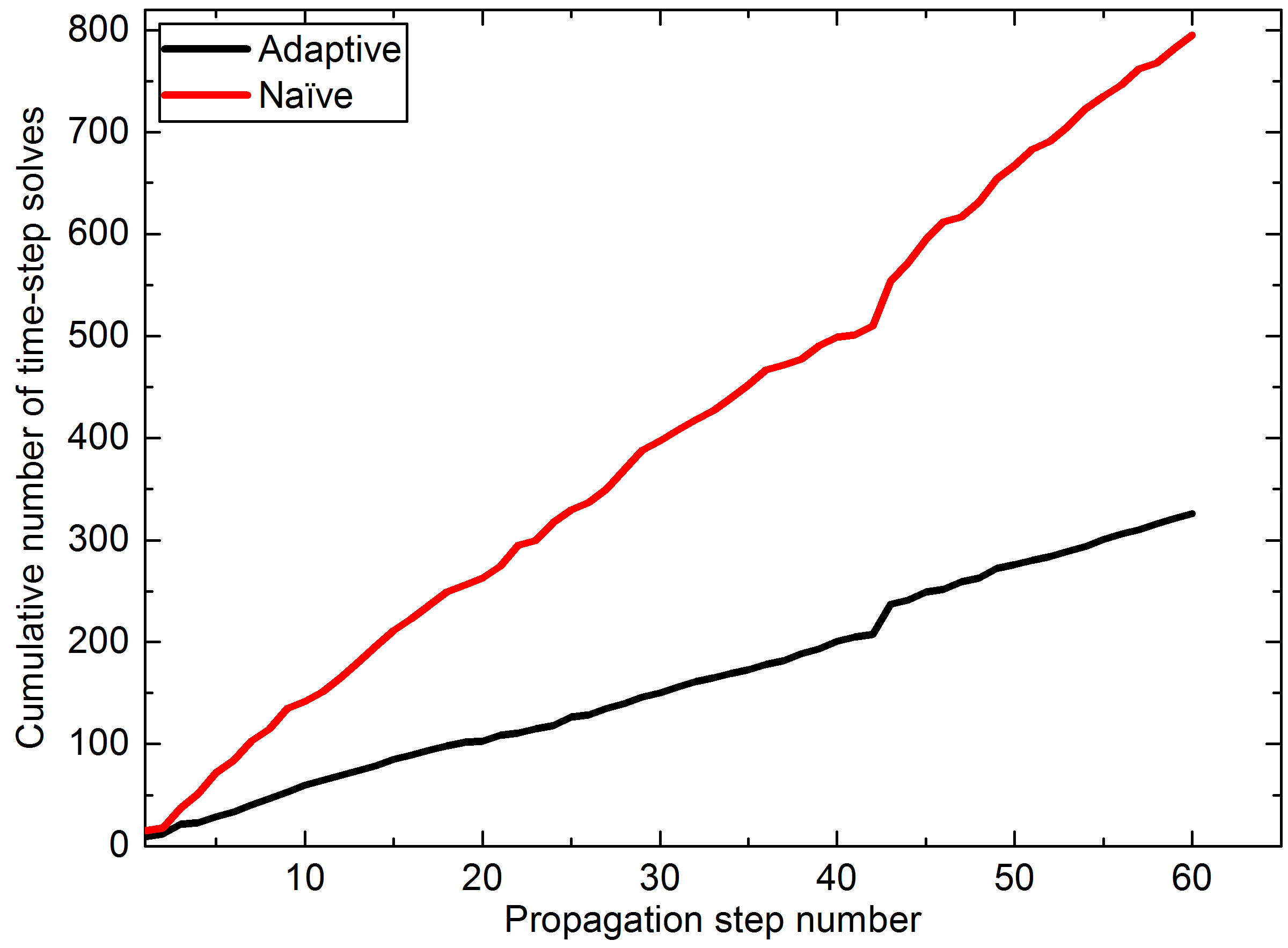}
	\caption{Cumulative time-step solves over the course of simulation}
	\label{case8_efficiency}
\end{figure}

Figure \ref{case8_timestep} presents the evolution of the nonlinear iteration count and time-step sizes over the duration of the simulation. During the hydraulic fracturing stages, time-step sizes remained below 10 secs, and no more than 10 nonlinear iterations were required for convergence. During the transition from hydraulic fracturing to production (HF-Prod) the time-step size was exponentially adapted to reach the specified maximum of $4.32e5$ secs (5 days). At the second transition from production to hydraulic fracturing (Prod-HF), the time-step size was rapidly adapted to the stimulation time scales. This is an indication of the efficacy of \cref{tscut} to prevent wasted effort due to unconverged time-step attempts. Figure~\ref{case8_efficiency} compares the evolution in cumulative nonlinear iterations computed using a naive time-step selection strategy with that of \cref{tscut}. 

\begin{figure}[H]
\begin{subfigure}[!htb]{1\textwidth}
	\centering
	\includegraphics[width=0.7\textwidth]{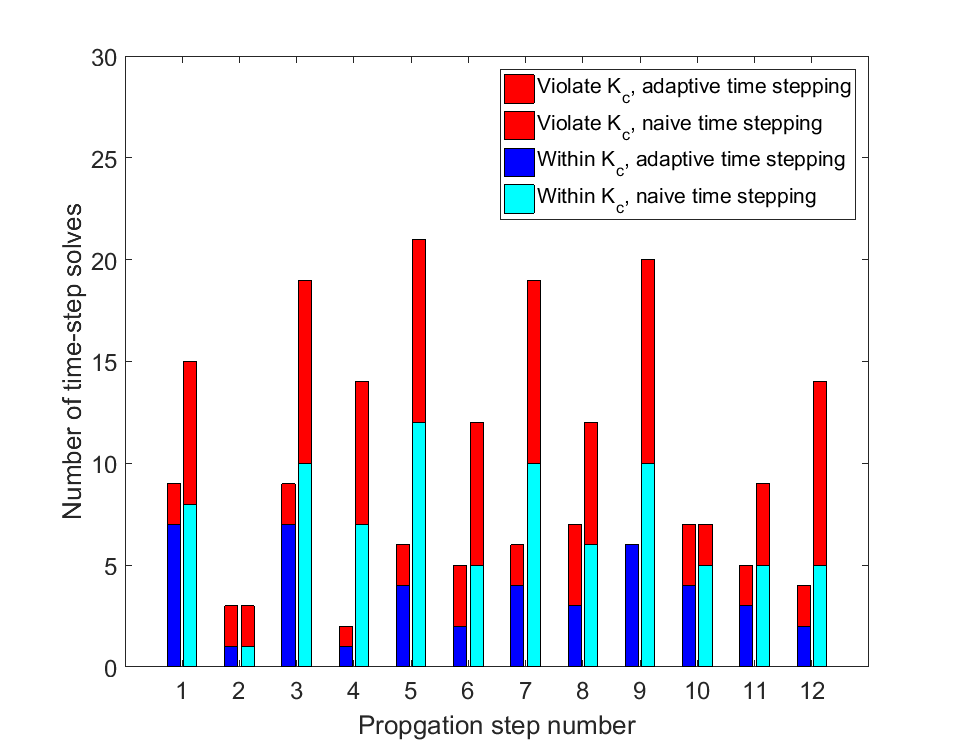}
	\caption{$\epsilon_{SIF} = 0.001$}
	\label{case8_timestepefficiency1}
\end{subfigure}
\begin{subfigure}[!htb]{1\textwidth}
	\centering
	\includegraphics[width=0.7\textwidth]{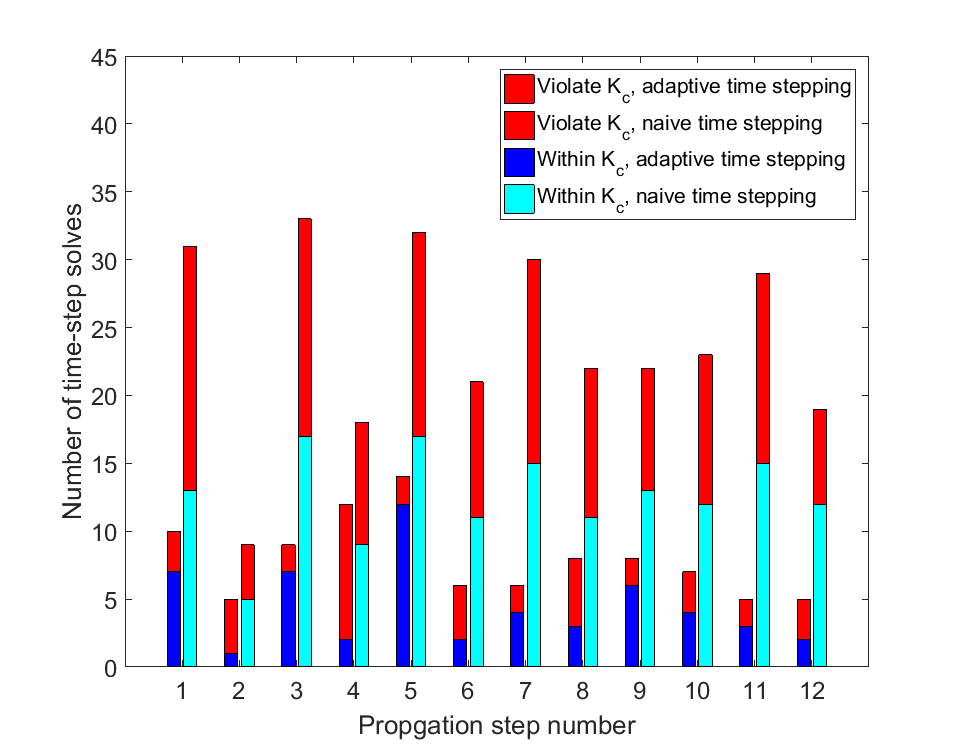}
	\caption{$\epsilon_{SIF} = 0.0001$}
	\label{case8_timestepefficiency2}
\end{subfigure}
	\caption{Number of time-step solves to reach at the FP criterion for the first 12 FP steps. }
\label{tse}
\end{figure}

There are further implications of utilizing the proposed time-step selection process. In particular, Algorithm~\ref{FPSOLVER} shows that in between accepted solutions satisfying both the propagation constraint with equality ($K_{I}^{eq}= K_{c}$) and conservation, there may be a number of sub-steps that violate the constraint ($K_{I}^{eq}\leq K_{c}$) and over which the effective SIF is to build-up. Another possibility is for states that overshoot the constraint ($K_{I}^{eq}\geq K_{c}$), and lead to a time-step cut. Figures~\ref{case8_timestepefficiency1} and~\ref{case8_timestepefficiency1} present the numbers of intermediate time-steps (blue) and failed time-steps (red) for twelve accepted propagation solutions using SIF tolerances of $1e-3$ and $1e -4$ respectively. Comparing the na\"ive and the proposed strategies, both the total number of intermediate time-steps, and failed steps can be reduced significantly. The more strict tolerance criterion leads to a larger number of intermediate and steps, and show a more drastic difference between the approaches. 
\clearpage

\begin{figure}[!htb]
	\centering
	\begin{subfigure}[b]{0.45\textwidth}
		\centering
		\includegraphics[width=1\textwidth]{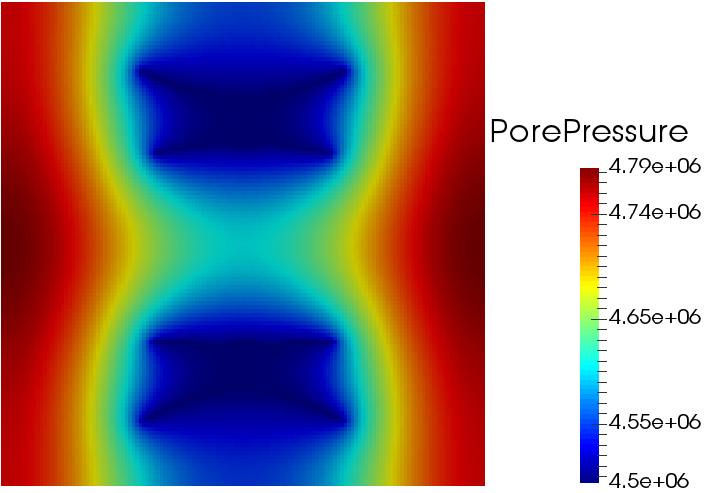}
				\caption{}
	\end{subfigure}
	\begin{subfigure}[b]{0.45\textwidth}
		\centering
		\includegraphics[width=1\textwidth]{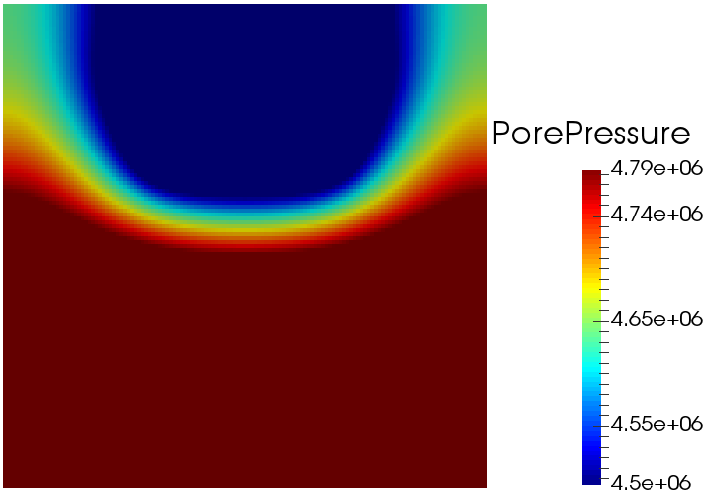}
				\caption{}
	\end{subfigure}
	\begin{subfigure}[b]{0.45\textwidth}
		\centering
		\includegraphics[width=1\textwidth]{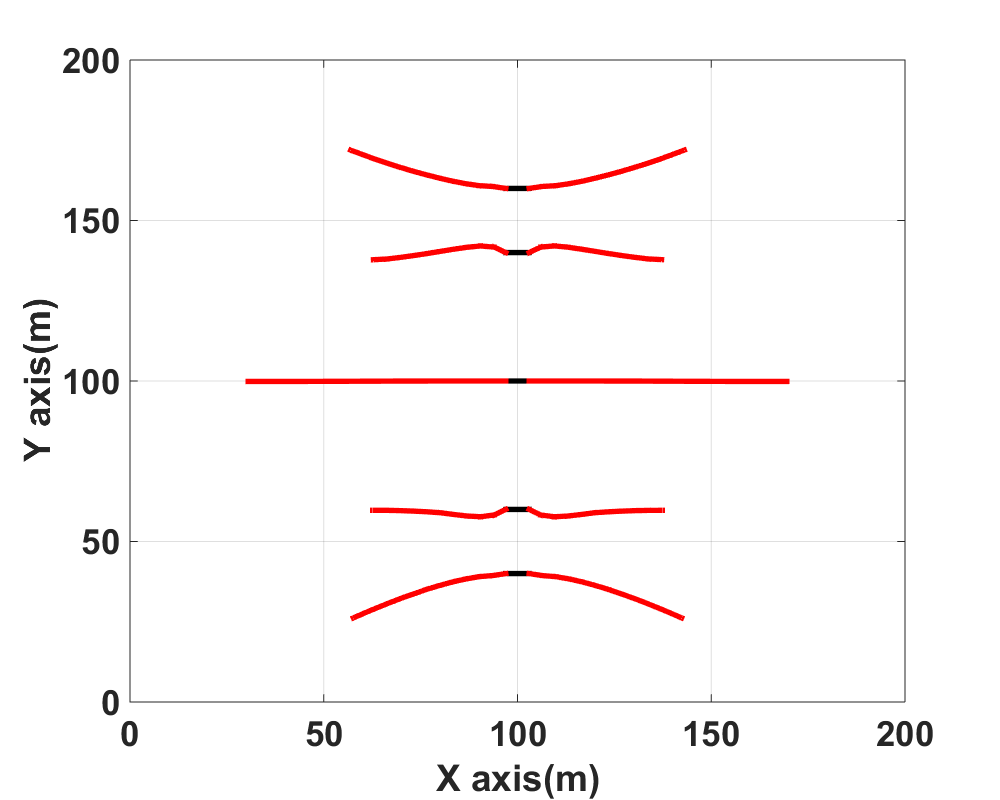}
		\caption{}
	\end{subfigure}
	\begin{subfigure}[b]{0.45\textwidth}
		\centering
		\includegraphics[width=1\textwidth]{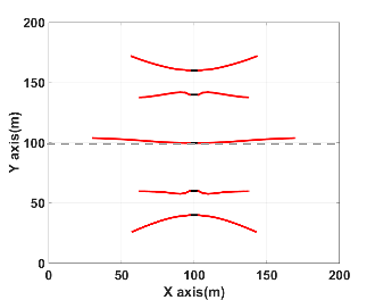}
		\caption{}
	\end{subfigure}
\caption{Comparison of pore pressure fields (top row, unit: Pa) and fracture paths (bottom row) obtained using the symmetric production stage (left column) and the asymmetric plan (right column).}
\label{case_10_Pressure}
\end{figure}

Finally, to emphasize the implications to systems involving intermittent fracturing and withdrawal operations, the simulation is repeated after slightly modifying the third operational (production) stage. In particular, during this period, production is only allowed to occur through fractures A and B instead of A, B, C, and D. Figure~\ref{case_10_Pressure} presents the pore pressure snapshots at the end of the production period and the final fracture patterns obtained for the two scenarios. The result demonstrates the utility of the integrated model to enable co-design of such operations. 

\section{Infill well fracturing}
\begin{figure}[!htb]
	\centering
	\includegraphics[width=0.6\textwidth]{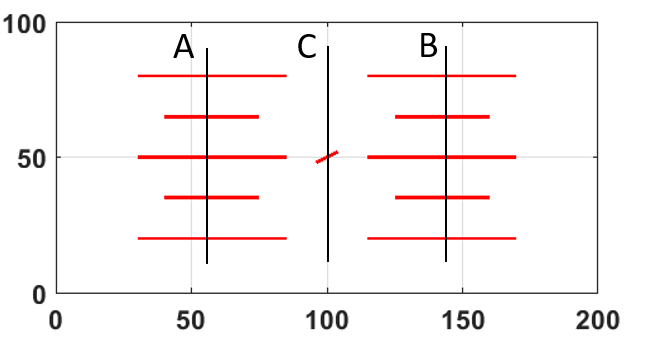}
	\caption{\textcolor{black}{Initial domain configuration.}}
	\label{infillIDC}
\end{figure}

\begin{figure}[H]
	\centering
	\includegraphics[width=0.6\textwidth]{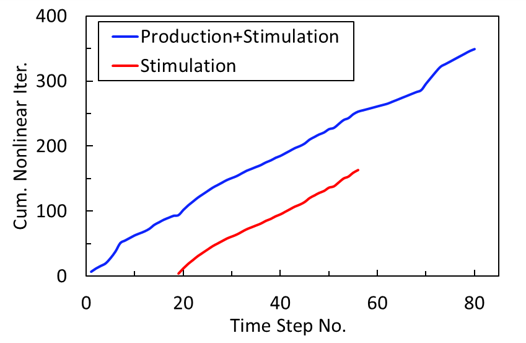}
	\caption{\textcolor{black}{Nonlinear solver performance of the poroelastic scenario.}}
	\label{NSPPS}
\end{figure}

{\color{black} In this case, simultaneous and interspersed periods of production and stimulation are studied to demonstrate the significant influence of poroelasticity on fracture propagation. The scale of the model is $200 \text{ m} \times 100 \text{ m}$ with $137 \times 77$ grid blocks. The preexisting fractures are shown in \cref{infillIDC} and are located at wells $A$ and $B$. The matrix permeability is $10^{-19} \text{ m}^{2}$; Porosity $\phi$ is $0.1$; Biot poroelastic coefficient $\alpha$ is $0.8$; Fluid viscosity of oil and water is identical, $0.1$ cp; Compressibility of oil and water is $4.3e-9 \text{Pa}^{-1}$; Young's modulus $\text{E}$ is $8.3 \text{ GPa}$ and Poisson's ratio $\nu$ is $0.3$; $K_{c}$ is $5 \text{ MPa}\sqrt{\text{m}}$, $\Delta a = 3 \text{ m}$, $\epsilon_{SIF} =0.01$. The stress at the left and right boundaries is 28 MPa while at the top and bottom boundaries it is 26 MPa. The well schedules are as follows:
\begin{itemize}
	\item[] Wells $A$ and $B$ are set to produce fluids for 5 years, and the well pressures are fixed at 5 MPa;
	\item[] Well $C$ is not operated for the first three years. Then, an injection process is triggered, and hydraulic fracturing takes place over 13 minutes. Following this period of injection at constant rate, Well C is turned to produce fluids for the remainder of time in the simulation. The injection rate is controlled at 0.001 $\text{m}^{3}/s$ while production is controlled by setting the well pressure to 5 MPa, which is lower that the ambient pressure in the model.
\end{itemize}
In order to avoid the closure of fractures due to pressure depletion, proppant is assumed to be present uniformly and is modeled by application of a uniform (elastic) force to the fracture surfaces as an inner boundary condition. This force is assumed to be elastic through an inversely linear relationship with aperture.
\begin{figure}[!htb]
	\begin{subfigure}[!htb]{1\textwidth}
		\centering
		\includegraphics[width=0.6\textwidth]{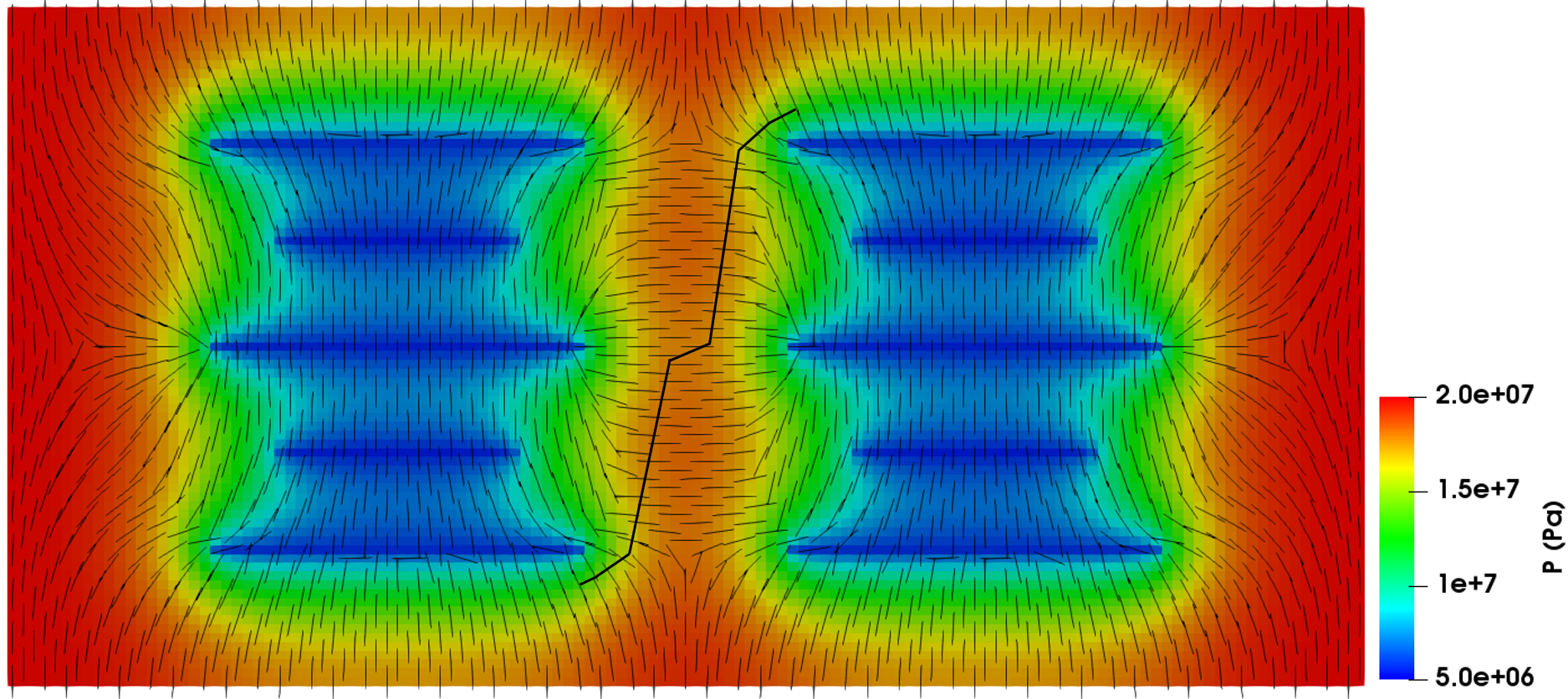}
		\caption{3 years}
		\label{poro3}
	\end{subfigure}
	\begin{subfigure}[!htb]{1\textwidth}
		\centering
		\includegraphics[width=0.6\textwidth]{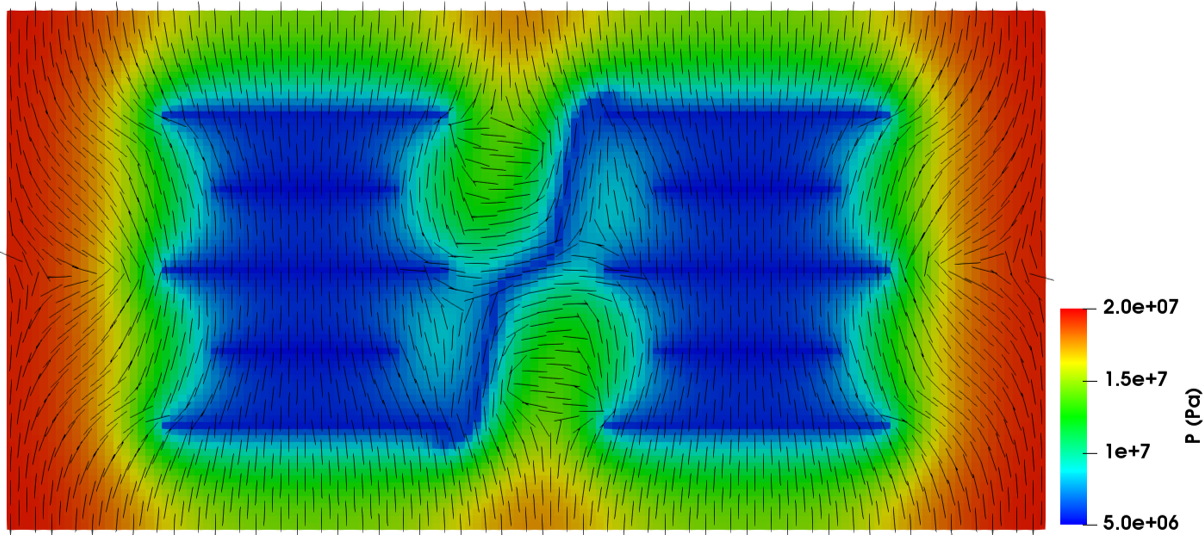}
		\caption{5 years}
		\label{poro5}
	\end{subfigure}
	\caption{\textcolor{black}{Pore pressure distribution of the poroelastic case; Short black lines are the minimum principal stress; Long black lines are the propagation path.}}
	\label{poro}
\end{figure}

\cref{poro3} shows the pressure field at three years, prior to the 13 minute fracturing process that is to occur in Well C. In the figure, the quivered black lines depict the minimum principal stress field at this time. Moreover, the solid black line depicts the fracture path that will ultimately be formed at the end of the 13 minute injection period. Due to the production and resulting pressure depletion drive by Wells $A$ and $B$, the direction of the minimum principal stress field is rotated to 90 degrees in the infill region, and becomes perpendicular to the minimum horizontal stress that is applied at the boundary. During fracture propagation, the path will follow the direction perpendicular to the minimum principal stress and hence, extend vertically through the infill region. \cref{poro5} presents the pressure field at the end of the simulation period (five years).

\begin{figure}[!htb]
	\begin{subfigure}[!htb]{1\textwidth}
		\centering
		\includegraphics[width=0.6\textwidth]{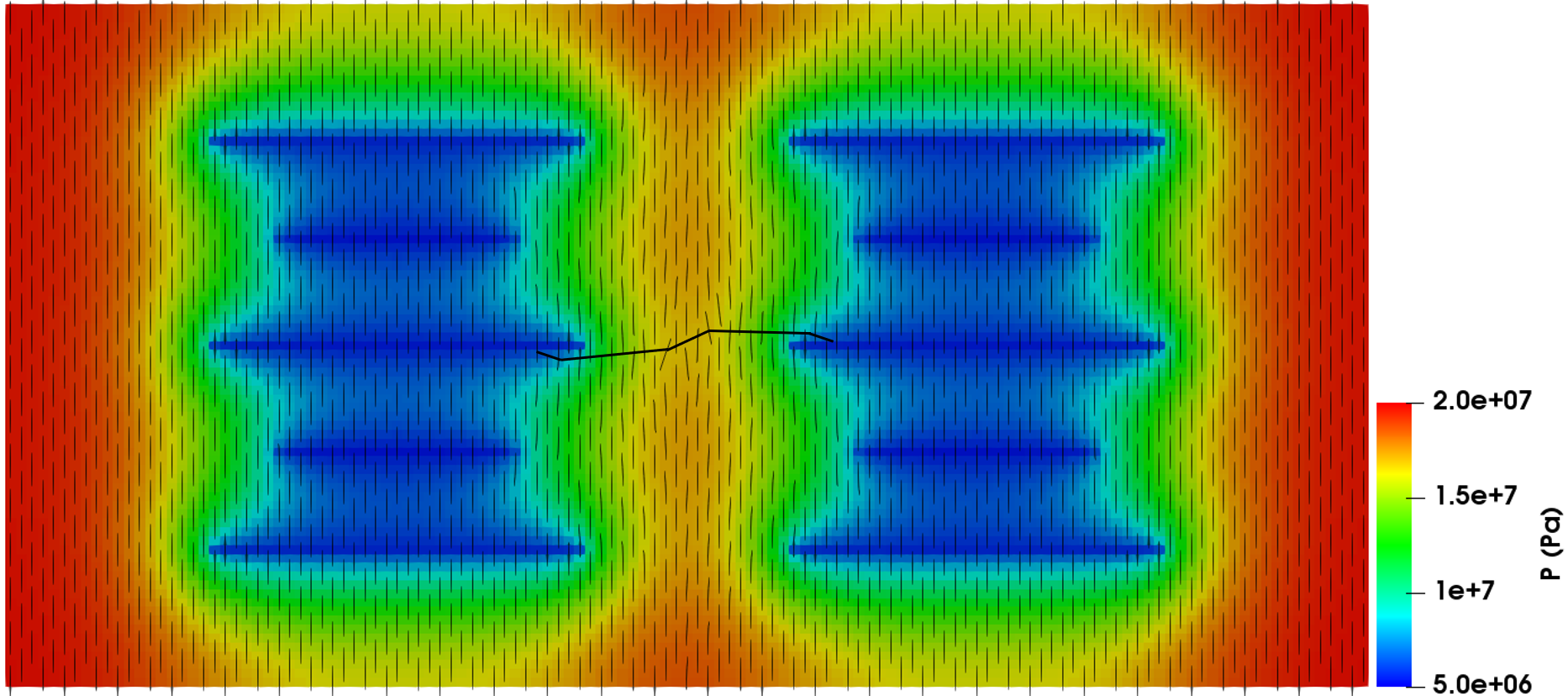}
		\caption{3 years}
		\label{elas3}
	\end{subfigure}
	\begin{subfigure}[!htb]{1\textwidth}
		\centering
		\includegraphics[width=0.6\textwidth]{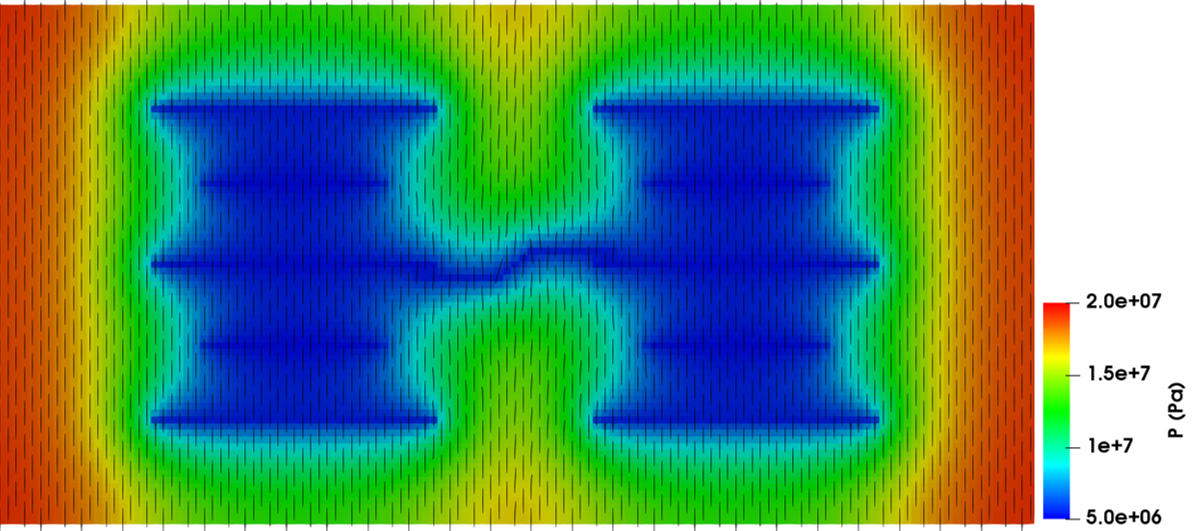}
		\caption{5 years}
		\label{elas5}
	\end{subfigure}
	\caption{\textcolor{black}{Pore pressure distribution of the elastic case; Short black lines are the minimum principal stress; Long black lines are the propagation path.}}
	\label{elas}
\end{figure}

The above simulation is repeated while ignoring the pore pressure effects on effective stress by setting the Biot coefficient to $\alpha=0$ and retaining all other conditions. The results corresponding to conditions in \cref{poro} but for this case are shown in \cref{elas}. Without poroelastic effects, the stress reorientation in the infill region is not observed, and the minimum principal stress field remains vertical and aligned with the minimum horizontal stress. The fracture propagation path is horizontal and hits the preexisting fractures at Wells $A$ and $B$. This effect is significant to the hydrodynamics as well: \cref{ComparisonCumOil} shows the cumulative evolution of fluid volumes that are produced by Well $C$ in either case, and we can observe a difference of 67\% over the five period between the elastic and poroelastic models. 

\begin{figure}[H]
	\centering
	\includegraphics[width=0.6\textwidth]{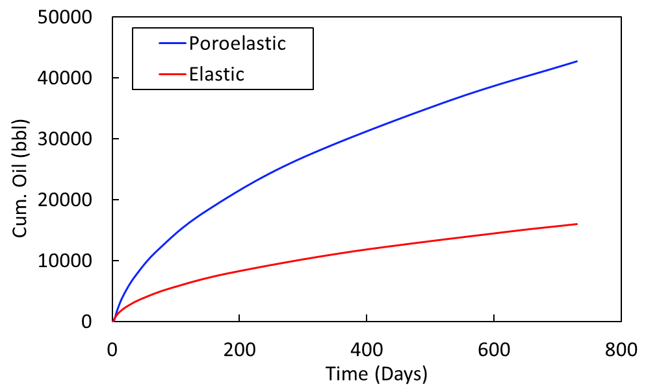}
	\caption{\textcolor{black}{Comparison of the cumulative oil production of poroelastic and elastic scenarios.}}
	\label{ComparisonCumOil}
\end{figure}

The summary of the nonlinear solver performance for the poroelastic case is shown in \cref{NSPPS}. The average nonlinear iterations per time step over the course of the simulation are 4.36 while the stimulation period takes around 46.7\% of the total nonlinear iterations.}

\section{Conclusions}\label{conclusion}
A coupled multiphase flow and mechanical model is proposed allowing the co-simulation of fluid-driven fracturing as well as injection and depletion processes in porous media using embedded meshes. The proposed mixed discrete fracture and matrix EDFM-XFEM discretization is augmented with an adaptive time-step controller, extended J-integral computation for poromechanics, and state initializations for propagated segments. The computational results show:
\begin{enumerate} 
\item The extension of the J integral estimation to poromechanics provides sufficient accuracy and/or leads to consistent solutions.
\item The proposed state-initialization strategy in newly propagated segments leads to improved nonlinear solver performance.
\item The proposed time-step controller is effective in automatically adapting to transitions, leading to computational efficiency. 
\item {
\color{black} 
Using the proposed unified model to simulate intermittent production and fracture requires computational effort that is on par with that of tying separate propagation and hydromechanical models (e.g. \cite{guo2019numerical,rezaei2019parameters}). This is facilitated by the time-step adaptivity to factor propagation onset.
}
\end{enumerate}

{\color{black}Extension of the proposed methodology to three-dimensions relies primarily on the availability of efficient and stable three-dimensional computational geometry infrastructure. In particular, this would be required to extract of geometric information such as fracture piece-wise planar intersections with the background mesh, fracture leading-edge locations, and fracture intersection. Nevertheless, the functional forms remain the same; for example the J-integral estimates presented here for SIF computations need to be performed in three-dimensions.
}
\section{Acknowledgements}
This material is based upon work supported by the U.S. Department of Energy under Award Number DE-FE-0031777. The authors also acknowledge partial funding from the members of the  TU Future Reservoir Simulation Systems \& Technology (FuRSST) Industry-University Consortium.

\bibliographystyle{unsrt}  
\bibliography{refs}
\appendix
\section{Analytical solutions for viscosity and toughness dominated solutions}
\label{appendix:analytical}
The K vertex solution is well known problem of a uniformly pressurized crack also called as Griffith's crack. $E'$ is the plain strain stress tensor, $\mu'$ is the $12\mu$. $Q$ is the flow rate into two fracture wings. The analytical expressions for the crack length and fracture aperture at the wellbore with time are,
\begin{equation}
	l = \big( \frac{E'Qt}{\sqrt{\pi}K_{C}} \big) ^{2/3}
\end{equation}

\begin{equation}
	\omega_{c}(0,t) = \frac{4K_{C}}{\sqrt{\pi}E'}l^{0.5}
\end{equation}
The M vertex solution assumes a zero toughness fracture and all of the energy overcome the flow in the channel. We adopt the 10th order solution from \cite{adachi2001},
\begin{equation}
	l = 0.6152\big(\frac{E'Q^{3}t^{4}}{\mu'}\big)^{1/6}
\end{equation}
\begin{equation}
	\omega_{c}(0,t) = 1.1260\big(\frac{\mu'}{E't} \big)^{1/3} \big(\frac{E'Q^{3}t^{4}}{\mu'}\big)^{1/6}
\end{equation}
The self similar solutions for aperture and pressure profiles for both M and K vertex solutions are also derived in the literature and not repeated here. Please refer to the original paper (\cite{adachi2001}) for more details.

{\color{black}
\section{J integral derivation}\label{app:jint}
The derivation for the \cref{area_Jint} is shown in the section. The analytical expression of the J integral is
\begin{equation}
J = \lim_{\Gamma_{I}\rightarrow 0}\int_{\Gamma_{I}}\left[ -\sigma_{ij}\frac{\partial u_{i}}{\partial x_{1}} + \sigma_{ij}\varepsilon_{ij}\delta_{1j} \right] n_{j}d\Gamma,
\end{equation}
where the effective stress is applied here and the superscript is neglected for the conciseness. Following the divergence theorem, a contour integral is casted into a volume and a line integral in 2 dimension (\cite{shih1986energy}),
\begin{equation}
\label{Jareaform}
J = \int_{\mathcal{B}_\rho}\left(\sigma_{ij}\frac{\partial u_{i}}{\partial x_{1}} - \sigma_{ij}\varepsilon_{ij}\delta_{1j} \right)\frac{\partial q}{\partial x_{j}}  + \partial_{x_{j}}\left( \sigma_{ij}\frac{\partial u_{i}}{\partial x_{1}} - \sigma_{ij}\varepsilon_{ij}\delta_{1j} \right)q d\bm{x} - \int_{\Gamma_{L}}t_{j}q\frac{\partial u_{j}}{\partial x_{1}}d\Gamma,
\end{equation}
where $t_{j}$ is the traction applied to the fracture surface. To extract the SIF, interaction integral is usually applied. By superimposing actual equilibrium fields $\left(1\right)$ and auxiliary fields  $(2)$ from the analytical solution, the superimposed state $\bar{J}$ derived from \cref{Jareaform} is
\begin{equation}
\label{Jareaform2}
\begin{split}
\bar{J} &= \int_{\mathcal{B}_\rho}\left[ \left(\sigma_{ij}^{(1)}+\sigma_{ij}^{(2)}\right)\frac{\partial \left(u_{i}^{(1)} + u_{i}^{(2)}\right)}{\partial x_{1}} - \frac{1}{2}\left(\sigma_{ij}^{(1)} + \sigma_{ij}^{(2)}\right)\left(\varepsilon_{ij}^{(1)} + \varepsilon_{ij}^{(2)}\right)\delta_{1j} \right]\frac{\partial q}{\partial x_{j}}+  \\& \partial_{x_{j}}\left[ \left(\sigma_{ij}^{(1)}+\sigma_{ij}^{(2)}\right)\frac{\partial \left(u_{i}^{(1)} + u_{i}^{(2)}\right)}{\partial x_{1}} - \frac{1}{2}\left(\sigma_{ij}^{(1)} + \sigma_{ij}^{(2)}\right)\left(\varepsilon_{ij}^{(1)} + \varepsilon_{ij}^{(2)}\right)\delta_{1j} \right]q d\bm{x} \\&- \int_{\Gamma_{L}}\left(t_{j}^{(1)} + t_{j}^{(2)}\right)\left(\frac{\partial u^{(1)}_{j} +\partial u^{(2)}_{j} }{\partial x_{1}}\right) qd\Gamma.
\end{split}
\end{equation}
The equation above can be separated into three parts 
\begin{equation}
\bar{J} = J^{(1)} + J^{(2)} + I,
\end{equation}
where $J^{(1)}$ is the domain integral of the actual state, $J^{(2)}$ is the domain integral of the auxiliary state and $I$ is the integral of mixing state:
\begin{equation}
\label{Jareaform3}
\begin{split}
I = &\int_{\mathcal{B}_\rho}\left(\sigma_{ij}^{(1)}\frac{\partial u_{i}^{(2)}}{\partial x_{1}} + \sigma^{(2)}_{ij}\frac{\partial u^{(1)}_{i}}{\partial x_{1}} - \frac{1}{2}\sigma_{ij}^{(1)}\varepsilon^{(2)}_{ij}\delta_{1j} - \frac{1}{2}\sigma^{(2)}_{ij}\varepsilon^{(1)}_{ij}\delta_{1j} \right)\frac{\partial q}{\partial x_{j}} + \\& \partial_{x_{j}}\left( \sigma^{(1)}_{ij}\frac{\partial u^{(2)}_{i}}{\partial x_{1}} + \sigma^{(2)}_{ij}\frac{\partial u^{(1)}_{i}}{\partial x_{1}}- \frac{1}{2}\sigma^{(1)}_{ij}\varepsilon^{(2)}_{ij}\delta_{1j} -\frac{1}{2}\sigma^{(2)}_{ij}\varepsilon^{(1)}_{ij}\delta_{1j}\right)q d\bm{x} \\&- \int_{\Gamma_{L}}t^{(1)}_{j}\frac{\partial u^{(2)}_{j}}{\partial x_{1}}q + t^{(2)}_{j}\frac{\partial u^{(1)}_{j}}{\partial x_{1}}q d\Gamma,
\end{split}
\end{equation}
where $\sigma_{ij}^{(1)}\varepsilon^{(2)}_{ij} = D_{ijkl}\varepsilon_{kl}^{(1)}\varepsilon_{kl}^{(2)} =\sigma_{ij}^{(2)}\epsilon^{(1)}_{ij}$. For straight fracture in homogeneous material, the second term in the areal integral vanishes (\cite{walters2005interaction}). And the $t^{(2)}$ is assumed to be zero in the auxiliary state. \cref{Jareaform3} can be further simplified into:
\begin{equation}
\label{Jareaform4}
\begin{split}
I = &\int_{\mathcal{B}_\rho}\left(\sigma_{ij}^{(1)}\frac{\partial u_{i}^{(2)}}{\partial x_{1}} + \sigma^{(2)}_{ij}\frac{\partial u^{(1)}_{i}}{\partial x_{1}} - \sigma_{ij}^{(1)}\varepsilon^{(2)}_{ij}\delta_{1j} \right)\frac{\partial q}{\partial x_{j}}d\bm{x}- \int_{\Gamma_{L}}t^{(1)}_{j}\frac{\partial u^{(2)}_{j}}{\partial x_{1}}q d\Gamma.
\end{split}
\end{equation}
where $t^{(1)}_{j}$ is equal to $\alpha p_{M} - p_{F}$.
}
\end{document}